\documentclass[12pt,twoside,english,british]{uqthesis}
\usepackage[T1]{fontenc}
\usepackage[latin9]{inputenc}
\usepackage{geometry}
\usepackage{array}
\usepackage{verbatim}
\usepackage{longtable}
\usepackage{varioref}
\usepackage{amsmath}
\usepackage{color}
\usepackage{makeidx}
\makeindex
\usepackage{graphicx}
\usepackage{setspace}
\usepackage{amssymb}
\usepackage[numbers]{natbib}

\makeatletter

\providecommand{\tabularnewline}{\\}

\usepackage{url}
\usepackage{latexsym}	

\usepackage{fancyhdr}
\let\oldmarginpar\marginpar
\renewcommand\marginpar[1]{\-\oldmarginpar[\raggedleft\scriptsize\sf #1]{\raggedright\scriptsize\sf #1}}
\usepackage{ifthen}
\newboolean{fullpaper}
\setboolean{fullpaper}{false} 

\usepackage{indentfirst}
\usepackage{balance}
\usepackage{multicol}

\AtBeginDocument{

}

\makeatother

\usepackage{babel}

\begin{document}
\thispagestyle{empty} 

\pagenumbering{roman}

\begin{center}
\textbf{\LARGE ~}
\par\end{center}{\LARGE \par}

\begin{center}
\textbf{\LARGE ~}\\
\textbf{\textsc{\LARGE Reality, locality and all that:}}
\par\end{center}{\LARGE \par}

\begin{center}
\textbf{\textsc{\textcolor{black}{{}``experimental metaphysics''
and the quantum foundations}}}
\par\end{center}

\begin{center}
\bigskip{}

\par\end{center}

\begin{center}
\textbf{\LARGE ~}\bigskip{}

\par\end{center}

\begin{center}
\bigskip{}

\par\end{center}

\begin{center}
\textbf{\LARGE ~}\bigskip{}
\bigskip{}

\par\end{center}

\begin{center}
\textbf{\LARGE ~}
\par\end{center}{\LARGE \par}

\begin{center}

\par\end{center}

\begin{center}
\textsc{A thesis submitted for the degree of Doctor of Philosophy}
\par\end{center}

\begin{center}
\textsc{to The University of Queensland}
\par\end{center}

\begin{center}
\textsc{October 2007}
\par\end{center}

\smallskip{}

\noindent \begin{center}
\smallskip{}

\par\end{center}

\begin{center}
~
\par\end{center}

\begin{center}
{\large Eric G. Cavalcanti}
\par\end{center}{\large \par}

\smallskip{}

\begin{center}
{\large Department of Physics, The University of Queensland}
\par\end{center}{\large \par}

\newpage{}

\newpage \fancyhead[RE,LO]{QUOTATIONS}\index{quotes}\label{quotes}

\thispagestyle{empty} 

\emph{\char`\"{}If one asks what, irrespective of quantum mechanics,
is characteristic of the world of ideas of physics, one is first of
all struck by the following: the concepts of physics relate to a real
outside world.\char`\"{} - Einstein}

\emph{\char`\"{}It is wrong to think that the task of physics is to
find out how Nature is. Physics concerns what we can say about Nature.\char`\"{}
- Bohr}

\emph{\char`\"{}I think there are professional problems... When I
look at quantum mechanics I see that it's a dirty theory... You have
a theory which is fundamentally ambiguous.\char`\"{} - Bell}

\emph{\char`\"{}How wonderful that we have met with a paradox. Now
we have some hope of making progress.\char`\"{} - Bohr}

%
{}

\vfill{}

\noindent \begin{center}
\copyright \  Eric G. Cavalcanti, 2007
\par\end{center}

\newpage{}

~\thispagestyle{empty} 

\noindent \begin{center}
~
\par\end{center}

\noindent \begin{center}
~
\par\end{center}

\noindent \begin{center}
~
\par\end{center}

\noindent \begin{center}
~
\par\end{center}

\noindent \begin{center}
~
\par\end{center}

\noindent \begin{center}
To Shana, who always believed, for her love and support.\newpage{}
\par\end{center}

%
{}\fancyhead[RE,LO]{}\textbf{\huge ~}{\huge \par}

\textbf{\huge ~}\thispagestyle{empty} 

\textbf{\huge Statement of Originality}{\huge \par}

~

I declare that the work presented in the thesis is, to the best of
my knowledge and belief, original and my own work, except as acknowledged
in the text or in the Statement of Contribution to Jointly-Published
Work below, and that the material has not been submitted, either in
whole or in part, for a degree at this or any other university.

\textbf{\large Statement of Contribution to Jointly-Published Work}{\large \par}

Some of the content of this Thesis is adapted from papers published
or submitted for publication jointly with other authors. The extent
of contribution of other authors is detailed in the following.

Section 3.4 is adapted from reference 5 of the List of Publications.
The work reproduced in this section was accomplished by me under the
guidance and supervision of Dr. Margaret Reid.

Chapter 4 is a reproduction, with some adaptations, of reference 3.
Most of the work in this chapter was accomplished by me with the guidance
and supervision of Prof. Peter Drummond. The detector inefficiency
calculation in 4.4.2 was done by Margaret Reid. The proofs of Section
4.5 were performed in collaboration by Chris Foster and me.

Chapter 5 is a reproduction, with some adaptations, of reference 6.
Most of the work in this chapter was accomplished by me with the guidance
and supervision of Dr. Margaret Reid. The work in Section 5.7 and
figures 5.1-5.7 and 5.9 were done by Margaret Reid.

~

~

. . . . . . . . . . . . . . . . . . . . . . . . . . . . . . . . .
. . . . . . . . . . . . . . . . . . . .

(Eric G. Cavalcanti, Candidate)

~

~

. . . . . . . . . . . . . . . . . . . . . . . . . . . . . . . . .
. . . . . . . . . . . . . . . . . . . .

(Peter D. Drummond, Principal Advisor)

\newpage{}

\textbf{\huge ~}\thispagestyle{empty} 

\textbf{\huge ~}{\huge \par}

\textbf{\huge Acknowledgements}{\huge \par}

\addcontentsline{toc}{chapter}{Acknowledgements}

\textbf{~}

So many people contributed, directly or indirectly, to the realisation
of this thesis that it is a daunting task to acknowledge them all.
So before going any further, let me first say that the following is
not at all an exhaustive attempt.

First of all I would like to thank Halina Rubinsztein-Dunlop. If it
was not for her I would not be here today, as she was my first contact
in the University of Queensland and officially my initial supervisor.
Halina supported my decision to change from experiment to theory and
helped me find a suitable supervisor, always with my best interest
at heart. I am deeply thankful for that.

A good thesis supervisor must find a fine balance between suggesting
problems to and motivating their students and allowing them to pursue
their own interests. It was with great satisfaction that I found that
my thesis supervisors, Peter Drummond, Margaret Reid and Karén Kheruntsyan,
have found just such perfect balance. I never had a lack of interesting
and stimulating problems to work on, due to their input and suggestions,
but also felt that I always had enough time to think about other problems
and complement my studies with intellectual freedom. This had a great
impact in how my career and interests have developed and I can't thank
Peter, Margaret and Karén enough for that.

Another group of people had an impact on my PhD just as important
as that of my supervisors, and certainly more frequent. Those were
my office colleagues and now friends (in alphabetical order so none
of them feels less valued), the ``usual suspects'' Andy, Chris, Paulo
and Yeong-Cherng. They were always ready to listen to my philosophical
ramblings and keen to help with my mathematical difficulties (some
of these even lead to collaborations). I will sincerely miss the friendly,
supportive and relaxed enviroment we maintained over these years.

Beyond that office there were many people, among friends, colleagues
and helpful staff, who made my PhD the rich and memorable experience
that it was. I would particularly like to thank Howard Wiseman for
sharing his knowledge, reading this manuscript and suggesting corrections,
and for giving me a job. But to name them all would take much more
space than I can fit in this page. You know who you are, and I thank
you all wholeheartedly for being there.

\newpage{}

\textbf{\huge ~}\thispagestyle{empty} 

\textbf{\huge ~}{\huge \par}

\textbf{\huge List of Publications}{\huge \par}

\addcontentsline{toc}{chapter}{List of Publications}

The following is a list of the authors' publications which are related
to the theme of this thesis, in chronological order.

~

\begin{enumerate}
\item M. D. Reid and E. G. Cavalcanti, \emph{Macroscopic quantum Schrodinger
and Einstein-Podolsky-Rosen paradoxes}, Journal of Modern Optics \textbf{52},
2245 (2005). 
\item E. G. Cavalcanti and M. D. Reid, \emph{Signatures for generalized
macroscopic superpositions,} Physical Review Letters \textbf{97},
170405 (2006).
\item E. G. Cavalcanti, C. J. Foster, M. D. Reid and P. D. Drummond, \emph{Bell
inequalities for continuous-variables correlations}, Physical Review
Letters \textbf{99}, 210405 (2007).
\item E. G. Cavalcanti and M. D. Reid, \emph{Uncertainty relations for the
realization of macroscopic quantum superpositions and EPR paradoxes},
Journal of Modern Optics \textbf{54}, 2373 (2007).
\item E. G. Cavalcanti and M.D. Reid, \emph{Criteria for generalized macroscopic
and S-scopic quantum superpositions}, accepted for publication on
Physical Review A (2008).
\item E. G. Cavalcanti, M. D. Reid, P. D. Drummond and H. A. Bachor, \emph{Unambiguous
signatures of entanglement and Bohm's spin EPR paradox}, arXiv:0711.3798.
\end{enumerate}
\newpage{}

\textbf{\huge ~}\thispagestyle{empty} 

\textbf{\huge ~}{\huge \par}

\textbf{\huge Abstract}{\huge \par}

~

In recent decades there has been a resurge of interest in the foundations
of quantum theory, partly motivated by new experimental techniques,
partly by the emerging field of quantum information science. Old questions,
asked since the seminal article by Einstein, Podolsky and Rosen (EPR),
are being revisited. The work of John Bell has changed the direction
of investigation by recognising that those fundamental philosophical
questions can have, after all, input from experiment. Abner Shimony
has aptly termed this new field of enquiry \emph{experimental metaphysics.}
The objective of this Thesis is to contribute to that body of research,
by formalising old concepts, proposing new ones, and finding new results
in well-studied areas. Without losing from sight that the appeal of
experimental metaphysics comes from the adjective, every major result
is followed by clear experimental proposals with detailed analysis
of feasibility for quantum-atom optical setups.

After setting the appropriate terminology and the basic concepts,
we will start by analysing the original argument of Einstein, Podolsky
and Rosen. We propose a general mathematical form for the assumptions
behind the EPR argument, namely those of \emph{local causality} and
\emph{completeness}. That formalisation entails what was termed a
Local Hidden State model by Wiseman \emph{et al., }which was proposed
as a formalisation of the concept of \emph{steering }first introduced
by Schrödinger in a reply to the EPR paper. Violation of any consequences
that can be derived from the assumption of that model therefore implies
a demonstration of the EPR paradox. We will show how one can then
re-derive the well-known EPR-Reid criterion for continuous-variables
correlations, and derive new ones applicable to the spin setting considered
by Bohm.

The spin set-up of the EPR-Bohm paradox was used by Bell to derive
his now famous theorem demonstrating the incompatibility of the assumption
of local causality and the predictions of quantum mechanics. The inequalities
which bear his name can be derived for any number of discrete outcomes,
but so far there has been no derivation which can be directly applied
to the continuous-variables case of the original EPR paradox. We close
the circle by deriving a class of inequalities which make no explicit
mention about the number of outcomes of the experiments involved,
and can therefore be used in continuous-variables measurements with
no need for binning the continuous results into discrete ones. Apart
from that intrinsic interest, these inequalities could prove important
as a means to perform an unambiguous test of Bell inequalities since
optical homodyne detection can be performed with high detection efficiency.
The technique, which is based on a simple variance inequality, can
also be used to re-derive a large class of well-known Bell-type inequalities
and at the same time find their quantum bound, making explicit from
a formal point of view that the non-commutativity of the local operators
is at the heart of the quantum violations.

Finally, we address the issue of macroscopic superpositions originally
sparked by the infamous \char`\"{}cat paradox\char`\"{} of Schrödinger.
We consider macroscopic, mesoscopic and `\emph{S}-scopic' quantum
superpositions of eigenstates of an observable, and develop some signatures
for their existence. We define the extent, or size \emph{S} of a (pure-state)
superposition, with respect to an observable \emph{X}, as being the
maximum difference in the outcomes of \emph{X} predicted by that superposition.
Such superpositions are referred to as generalised \emph{S}-scopic
superpositions to distinguish them from the extreme superpositions
that superpose only the two states that have a difference \emph{S}
in their prediction for the observable. We also consider generalised
\emph{S}-scopic superpositions of coherent states. We explore the
constraints that are placed on the statistics if we suppose a system
to be described by mixtures of superpositions that are restricted
in size. In this way we arrive at experimental criteria that are sufficient
to deduce the existence of a generalised \emph{S}-scopic superposition.
The signatures developed are useful where one is able to demonstrate
a degree of squeezing.

\tableofcontents{}

\cleardoublepage

\addcontentsline{toc}{chapter}{List of Figures}\listoffigures

\chapter*{List of Abbreviations and Acronyms}

\label{sec:List-of-Abbreviations}\index{abbreviations}

\addcontentsline{toc}{chapter}{List of Abreviations}

\begin{longtable}{rp{0.9\textwidth}}
MRRF & Minimal Realist-Relativistic Framework\tabularnewline
OT & Operational Theory\tabularnewline
HVM & Hidden Variable Model\tabularnewline
LC & Local Causality\tabularnewline
D & Determinism\tabularnewline
P & Predictability\tabularnewline
IU & Irreducible Unpredictability\tabularnewline
L & Locality\tabularnewline
OI & Outcome Independence\tabularnewline
LD & Local Determinism\tabularnewline
SL & Signal Locality\tabularnewline
PS & Phenomenon Space\tabularnewline
vSL & Set of phenomena violating Signal Locality\tabularnewline
vL & Set of phenomena violating Locality\tabularnewline
vLC & Set of phenomena violating Local Causality\tabularnewline
vLD & Set of phenomena violating Local Determinism\tabularnewline
vD & Set of phenomena violating Determinism\tabularnewline
vP & Set of phenomena violating Predictability\tabularnewline
SL & Signal Locality\tabularnewline
OQT & Operational Quantum Theory\tabularnewline
EPR & Einstein, Podolsky and Rosen (paradox/argument) \cite{Einstein1935}\tabularnewline
HUP & Heisenberg's Uncertainty Principle\tabularnewline
LHS & (i) Local Hidden State (model); (ii) left hand side (of an equation)\tabularnewline
RHS & Right hand side (of an equation)\tabularnewline
LHV & Local Hidden Variables (model)\tabularnewline
CV & Continuous Variables\tabularnewline
CHSH & Clauser, Horne, Shimony and Holt (inequality)\cite{Clauser1969}\tabularnewline
MABK & Mermin, Ardehali, Belinskii and Klyshko (inequalities) \cite{Mermin1990,Ardehali1992,Belinskii1993}\tabularnewline
GHZ & Greenberger, Horne and Zeilinger (state) \cite{Gisin1998}\tabularnewline
 & \tabularnewline
\end{longtable}

\newpage{}

\cleardoublepage

\pagenumbering{arabic}

\fancyhead[RE]{\leftmark} \fancyhead[LO]{\rightmark}

\chapter{\label{cha:Introduction}Introduction}

In a recent book by Lee Smolin, that author proposed a list of the
5 greatest problems in contemporary physics. In second place, just
after the problem of quantum gravity, were \emph{the foundational
problems of quantum mechanics. }Below this were the problems of unification
of particles and forces, of explaining the free constants of the standard
model and the problem of dark matter and dark energy.

The prominent position may sound quaint for those who have been taught
that the problems in the quantum foundations were all solved many
years ago by Bohr, Heisenberg, von Neumann and the other founders
of the theory. That impression is especially understandable given
the enormous empirical success of the theory. However, a large part
of the community is starting to recognise that the problems that Einstein,
Schrödinger and others have raised since the theory's beginnings are
as relevant and urgent as ever.

Two reasons may be advanced as prime contributors to this increased
interest in the quantum foundations. Firstly, the emergence of the
field of quantum information and computation, which aims to harness
the quantum nature of the world for previously impossible tasks, has
raised physicists' awareness for the foundational problems by exposing
a larger audience to the bizarre nature of quantum phenomena.

Secondly, some theorists such as Smolin are starting to suspect that
the failures to find a quantum theory of gravity may be related to
our failure of understanding quantum mechanics. Success in the first
of the above problems, to those authors, will have to come hand in
hand with success in the second.

This does not at all mean that these authors advocate a return to
Einstein's dream of a local and realist theory. Since Bell's famous
1964 theorem, and the many experiments that confirm violation of Bell
inequalities, we know this is a hopeless goal\label{loopholes}%
\footnote{Although, strictly speaking, there are some open loopholes in all
violations of Bell's inequalities \cite{Clauser1978,Gisin2007}. It
seems unlikely that local realism will be restored when those loopholes
are closed, but it is of fundamental importance to be able to settle
the issue once and for all. We will have something to say about that
in Chapter \ref{cha:CV-Bell}.%
}. But following the quote from the same Bell on page \pageref{quotes},
there are \emph{professional} problems with quantum mechanics. Those
like Bell who point out the contradictions within the theory, most
notably those arising out of the so-called \emph{measurement problem},
are not merely indicating that quantum theory is not locally realistic
--- no-one was more aware of that than Bell! Those theorists are pointing
out that we ought to have a coherent picture of the whole of reality,
not just of experimentalists' laboratory fiddlings, even if that picture
turns out to be quite distinct from any classical one.

The return to a more professional attitude, in my opinion, will have
to include a more careful attention to \emph{philosophical} issues.
Philosophers have since long battled with the conceptual quandaries
which quantum mechanics forces us to face. A professional attitude
towards the quantum questions cannot avoid using terms such as \emph{ontology
}and \emph{epistemology. }Even if Bohr is right and quantum mechanics
regards not Nature herself, but regards what \emph{we} \emph{can say
}(and therefore what we can know)\emph{ }about Nature, we should be
able to say clearly what \emph{we }in this sentence means, and we
should be able to understand how our knowledge seems to follow well-defined
physical laws.

The present situation with quantum mechanics could be compared to
the situation of the Special Theory of Relativity before Einstein
interpreted the Lorentz transformations. Eintein's revolution was
one of interpretation, and it lead to a revolution in how we would
come to understand and use the theory of relativity. It is also interesting
to conjecture about what would happen with the General Theory if the
first breakthrough of interpretation achieved by the Special Theory
were not laid down. With historical hindsight it is easy to see the
value of Einstein's interpretational leap. However, before 1905 it
was unthinkable that the solution to the problem of the electrodynamics
of moving bodies would lead to such a deep restructuring of our basic
fundamental notions about space and time.

Similarly, one could argue that we are in a similar pre-revolutionary
phase with respect to quantum mechanics (not necessarily in the sense
that the revolutionary leap is imminent, but that it is yet to come).
And pointing out the current empirical successes of the theory is
even more of a reason to pursue its foundational problems. If the
present quagmire of postulates and quantisation rules is so successful,
one can only dream of what we could achieve with a satisfactory understanding.

\section{{}``Experimental Metaphysics''}

To understand the source of the conflicts in the foundations of Quantum
Mechanics, it is essential to know where and how our classical models
and intuitions start to fail to describe a quantum world. This is
the subject of \emph{experimental metaphysics}. The term was originally
coined by Abner Shimony\textbf{ }\cite{Shimony1989}\textbf{ }to describe
the new area of enquiry opened by Bell in 1964 when he recognised
the existence of experimentally testable implications that could be
derived from some general metaphysical assumptions --- namely, those
that go under the rubric of local realism%
\footnote{To use standard terminology. A more careful nomenclature will be introduced
in Chapter \ref{cha:Concepts}.%
}. For the first time, it was clearly recognised that very general
philosophical theses could have input from experiment.

At the time of Bell those questions were not part of the concerns
of most physicists, but today we have learned to perceive the nonlocality
evidenced by Bell as a \emph{resource. }The fields of quantum information
and computation rely on these counter-intuitive features of Quantum
Mechanics for speeding up computational tasks or achieving results
--- such as unconditionally secure quantum cryptography --- impossible
to achieve before. It therefore becomes an important task to map those
resources and recognise how exactly they are distinct from classical
resources. This is another problem towards which Experimental Metaphysics
can contribute.

The purpose of this thesis is to contribute to that body of research,
by formalising old concepts, proposing new ones, and finding new results
in well-studied areas. Without losing from sight that the appeal of
experimental metaphysics comes from the adjective, every major result
is followed by clear experimental proposals with detailed analysis
of feasibility for quantum-atom optical setups.

\section{Outline of the Thesis}

In Chapter \ref{cha:Concepts} we set up the appropriate terminology
and the basic concepts. Most of it will be simply careful definitions
of standard concepts, but some definitions may be new and some results
and consequences may not have been fully appreciated before. In particular,
I present a new result on a relation between signal locality and the
irreducible unpredictability of Nature.

Chapter \ref{cha:EPR-paradox} is related to publications 4 and 6
of the List of Publications. In that chapter, we analyse the original
argument of Einstein, Podolsky and Rosen (EPR) \cite{Einstein1935},
and propose a general mathematical form for the assumptions behind
that argument, namely those of \emph{local causality} and \emph{completeness}
\emph{of quantum theory}. That will entail what was termed a Local
Hidden State model by Wiseman \emph{et al. }\cite{Wiseman2007}\emph{,
}which was proposed as a formalisation of the concept of \emph{steering
}first introduced by Schrödinger \cite{Schroedinger1935} in a reply
to the EPR paper. Violation of any consequences that can be derived
from the assumption of that model therefore implies a demonstration
of the EPR paradox. We will show how one can then re-derive the well-known
EPR-Reid criterion \cite{Reid1989} for continuous-variables correlations,
and derive new ones applicable to the spin setting considered by Bohm
\cite{Bohm1951}.

The spin set-up of the EPR-Bohm paradox was used by Bell \cite{Bell1964}
to derive his now famous theorem demonstrating the incompatibility
of the assumption of local causality and the predictions of quantum
mechanics. The inequalities which bear his name can be derived for
any number of discrete outcomes, but so far there has been no derivation
which can be directly applied to the continuous-variables case of
the original EPR paradox. In Chapter \ref{cha:CV-Bell}, related to
publication 3, we close the circle by deriving a class of inequalities
which make no explicit mention about the number of outcomes of the
experiments involved, and can therefore be used in continuous-variables
measurements with no need for binning the continuous results into
discrete ones. Apart from that intrinsic interest, these inequalities
could prove important as a means to perform an unambiguous test of
Bell inequalities which does not suffer from the logical loopholes\textbf{
}\cite{Clauser1978,Gisin2007} that plague all experimental demonstrations
so far, since optical homodyne detection can be performed with high
detection efficiency. The technique, which is based on a simple variance
inequality, can also be used to re-derive a large class of well-known
Bell-type inequalities and at the same time find their quantum bound,
making explicit from a formal point of view that the non-commutativity
of the local operators is at the heart of the quantum violations.

Finally, in Chapter \ref{cha:Macro-Super}, related to publications
1, 2, 4 and 5, we address the issue of macroscopic superpositions
originally sparked by the infamous \char`\"{}cat paradox\char`\"{}
of Schrödinger \cite{Schroedinger1935}, presented in the same seminal
paper where he coined the terms \emph{entanglement }and \emph{steering}.
We consider macroscopic, mesoscopic and `\emph{S}-scopic' quantum
superpositions of eigenstates of an observable, and develop some signatures
for their existence. We define the extent, or size \emph{S} of a (pure-state)
superposition, with respect to an observable \emph{X}, as being the
maximum difference in the outcomes of \emph{X} predicted by that superposition.
Such superpositions are referred to as generalised \emph{S}-scopic
superpositions to distinguish them from the extreme superpositions
that superpose only the two states that have a difference \emph{S}
in their prediction for the observable. We also consider generalised
\emph{S}-scopic superpositions of coherent states. We explore the
constraints that are placed on the statistics if we suppose a system
to be described by mixtures of superpositions that are restricted
in size. In this way we arrive at experimental criteria that are sufficient
to deduce the existence of a generalised \emph{S}-scopic superposition.
The signatures developed are useful where one is able to demonstrate
a degree of squeezing.

\chapter{\label{cha:Concepts}Concepts of Experimental Metaphysics}

In June 2007, in a conference on Quantum Foundations in the charming
little town of Växjö, Sweden, dedicated to the 80$^{\text{th}}$anniversary
of the Copenhagen Interpretation, I have noticed an unexpectedly large
number of debates about \emph{what} experimental violations of Bell
inequalities prove. I was definitely expecting debates about, say,
how to make sense of a world where Bell inequalities are violated,
but not as much about what they \emph{mean} in the first place. It
became clear to me that the reason behind many of the disagreements
(though I wouldn't say all of them) was the lack of common ground
in definitions of terms such as \emph{locality} or \emph{realism}
and in the distinction between models and the phenomena they predict.

\emph{Local realism} is the catch-all term that is usually employed
to represent the set of assumptions which Bell's theorem shows to
be incompatible with the quantum mechanical predictions and (up to
some open loopholes) violated by Nature. Even though the final mathematical
form of the constraints imposed by local realism is quite uncontroversial,
there are numerous authors who debate what exactly the underlying
assumptions correspond to and what features of our world view must
be modified to accommodate the violation of Bell inequalities. See
for example the collection of papers edited by J. T. Cushing and E.
McMullin in \cite{Cushing1989} and the book of Tim Maudlin \cite{Maudlin1994}
for some in-depth discussion of these issues. I won't attempt to go
as deeply into the myriad questions that can be addressed in the philosophical
surroundings of Bell's theorem. My main purpose is to establish as
clearly as possible the terminology I will use. 

That said, in this chapter I will introduce some new usage of terms
and some implications which may not have been fully recognised before.
While some of it will be my own work, much of my current understanding
of these concepts is due to insights gained from discussions with
Howard Wiseman, to whom I am grateful. The presentation style and
most definitions were influenced by notes from that author. 

The most important new result of my own will be an interesting connection
between the assumption of signal locality, or no-faster-than-light-signalling,
and the notion of \emph{predictabilty. }I will show that the assumption
of signal locality -- which must be satisfied if one assumes relativistic
invariance -- together with the experimental observation of violation
of Bell inequalities, lead to the conclusion that Nature is irreducibly
unpredictable, quite independently of anything from the formalism
of quantum mechanics. This establishes a deep connection between two
of the main puzzles of quantum mechanics: Bell-nonlocality and the
Uncertainty Principle.

\section{The Minimal Realist-Relativistic Framework}

Many of the debates around the meaning of the Bell theorem regard
the status of the word `realism'\emph{ }in `local realism'. In a recent
analysis, Norsen \cite{Norsen2007a} has argued that there's no such
assumption among those that go into a derivation of a Bell inequality,
or at least that any such assumption is so fundamental that no scientific
theory can be built without it. I would not go as far as saying that
there's no such assumption, although I agree with that author that
there are misconceptions around the term (and I will try to clear
some of them here) and that it is important to recognise that the
assumption of realism is part of an underlying framework without which
Bell's concept of local causality cannot even be expressed. It is
not possible, as we will discuss in more detail, to \emph{maintain}
\emph{local causality while rejecting realism}. There are, however,
other possible usages of the word `locality' which are possible to
be maintained even in light of Bell's theorem, most notably the concept
of no faster-than-light signalling or \emph{signal locality}. However,
those are emphatically \emph{not} the local causality of Bell%
\footnote{Besides, the concept of signalling is not entirely without conceptual
difficulties as we'll see later in this chapter.%
}, and one does not need to reject realism to keep signal locality.

So let us attempt to understand what `realism' in `local realism'
can possibly mean. What better source for that than the most notable
supporter of realism in the 20$^{\text{th}}$ century? Einstein's
concept of Reality is clearly expressed in the following passage:

\begin{quote}
\char`\"{}If one asks what, irrespective of quantum mechanics, is
characteristic of the world of ideas of physics, one is first of all
struck by the following: the concepts of physics relate to a real
outside world... It is further characteristic of these physical objects
that they are thought of as arranged in a space-time continuum. An
essential aspect of this arrangement of things in physics is that
they lay claim, at a certain time, to an existence independent of
one another, provided these objects `are situated in different parts
of space'.\char`\"{} (\cite{Born1971}, pg. 168)
\end{quote}
I will encapsulate in the following axiom a weakened form of the concept
of Reality expressed by Einstein in the above quote.

\begin{verse}
\textbf{Axiom 1 (Reality):} \emph{Events occur independently of observers
or frames of reference}.
\end{verse}
That is, this is the assumption that \emph{the very fact that events
occur} is something independent of observers or reference frames.
A pebble being splashed by water on a deserted beach, a photo-detector
signalling the arrival of a photon, the Big Bang, Axiom 1 states that
the reality of those events is independent of anyone's description
or observation.

That's not to say that different observers can't give different accounts
about \emph{where }or \emph{when} those events occurred; it's just
that they will all agree, if Axiom 1 is true, that they \emph{in fact
}occurred%
\footnote{This is actually not a sufficient condition. It is possible that no
observers disagree about the reality of events, and yet Axiom 1 to
be false. This would only require that observers only be able to communicate
with just those observers who would agree with them.%
}. That does not mean, of course, that all physical quantities are
independent of observers or reference frames -- some physical quantities
such as velocity are patently relative in that way. 

Axiom 1 does not necessarily imply that there are hidden variables
underlying quantum properties either; one could regard the interactions
between quantum systems and their evolutions before measurement simply
as not corresponding to \emph{bona-fide }`events' in the above sense
(maybe `virtual events' would be an appropriate term for such interactions,
if they are regarded simply as mathematical devices). In fact this
seems to be an accurate representation of the view most commonly espoused
by physicists in the orthodox camp, clearly expressed in Wheeler's
famous maxim {}``No phenomenon is a phenomenon until it is an observed
phenomenon'' or by Peres' {}``Unperformed experiments have no results''
\cite{Peres1978}. Neither does this axiom imply that all events are
human-scale observable events; it is important to allow for the possibility
that a theory will make claims about some events even if we can't
directly know them. We need to emphasise, however, that one can conceive
of frameworks in which this axiom does not hold, and I will return
to this point in \ref{sub:Reality}.

The other important assumption in Einstein's concept is that \char`\"{}these
physical objects (...) are thought of as arranged in a space-time
continuum\char`\"{}, which I represent in the following axiom.

\begin{verse}
\textbf{Axiom 2 (Space-time): }\emph{All events can be embedded as
points in a single relativistic space-time, where the concepts of
space-like separation, light cones, and other standard relativistic
concepts can be applied unambiguously.}
\end{verse}
The conjunction of Axioms 1 and 2 constitute what I will call a \emph{Minimal
Realist-Relativistic Framework\index{Minimal Realist-Relativistic Framework}
(MRRF)}\label{ver:MRRF}. All of the subsequent notions and theorems
will assume this framework.

While the above was stated as an ontological framework (i.e., as assumptions
about how the world \emph{is}), one can re-frame it as an epistemological
framework (i.e., as assumptions about what one can \emph{know}). So
Axiom 1 could be rephrased as saying that all observers agree about
which events they see and Axiom 2 as stating that all observers can
describe those events consistently as corresponding to points in a
single space-time. Any metaphysical realist (one who believes that
there exists a world independent of observers) would take the fact
that all observers \emph{agree} about which events occur as evidence
that they \emph{really occur}, independently of any observers. However,
even a metaphysical \emph{anti-}realist (one who does\emph{ not} believe
that there exists a world independent of observers) could accept those
axioms in this weaker epistemic form, and that's the reason I include
them here --- so that anti-realists don't feel completely secure against
the consequences of Bell's theorem.

\section{Causation and Bell's local causality}

Maybe surprisingly to most physicists, causation is a problematic
concept in philosophy even for reasons independent of Bell. I won't
delve into those difficulties here%
\footnote{For a thorough philosophical treatment of causation see \cite{Dowe2000}.%
}, but will stick to the concepts which are relevant for the current
purposes. 

That Bell takes his theorem as implying a conflict between quantum
mechanics and relativity is evident in the following%
\footnote{For a more in-depth analysis of Bell's concept than I will present
here, see \cite{Norsen2007b}\textbf{.}%
}:

\begin{quote}
\char`\"{}For me then this is the real problem with quantum theory:
the apparently essential conflict between any sharp formulation and
fundamental relativity. That is to say, we have an apparent incompatibility,
at the deepest level, between the two fundamental pillars of contemporary
theory...\char`\"{} (\cite{Bell1987}, pg. 172)\textbf{ }
\end{quote}
We will come soon to a derivation of this \char`\"{}essential conflict\char`\"{}.
But for that we need to understand the concept of local causality
Bell takes relativity to imply. In \char`\"{}La nouvelle cuisine\char`\"{},
Bell starts with an amusing quote from H.B.G. Casimir: 

\begin{quote}
\char`\"{}I want to boil and egg. I put the egg into boiling water
and I set an alarm for five minutes. Five minutes later the alarm
rings and the egg is done. Now the alarm clock has been running according
to the laws of classical mechanics uninfluenced by what happened to
the egg. And the egg is coagulating according to laws of physical
chemistry and is uninfluenced by the running of the clock. Yet the
coincidence of these two unrelated causal happenings is meaningful,
because, I, the great chef, imposed a structure in my kitchen.\char`\"{}
(\cite{Bell1987}, pg. 232)
\end{quote}
This passage is to illustrate a principle that has always been a hallmark
of the scientific enterprise: whenever correlations between events
occur, either one event causes the other or they share a common cause.
This is known in philosophy of science as the \emph{principle of common
cause}, which was first formulated in a clear manner by Hans Reichenbach
\cite{Reichenbach}. Bell is thinking along very similar lines when
he describes the relativistic principle of local causality as (with
reference to Figure \ref{fig:Bell1}):

\begin{figure}
\begin{centering}
\includegraphics[width=13cm]{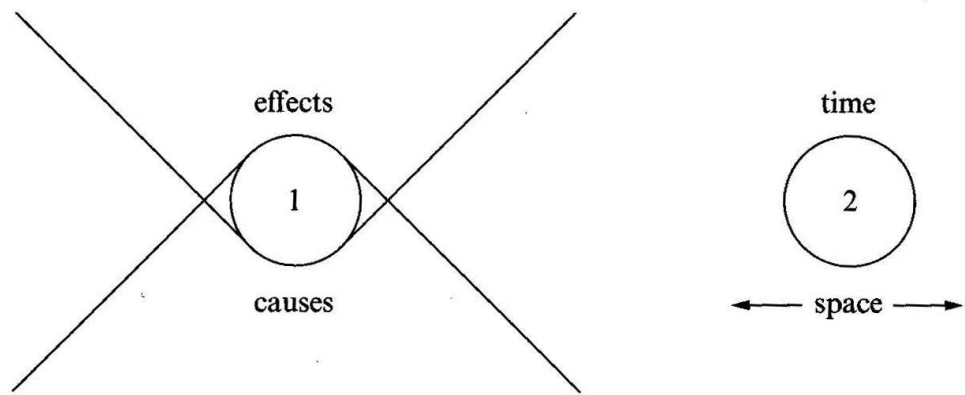}
\par\end{centering}

\caption{\label{fig:Bell1}\char`\"{}Space-time location of causes and effects
of events in region 1.\char`\"{} (reproduced with permission from
\cite{Bell1987}, pg. 239. Copyright Cambridge University Press. ) }

\end{figure}

\begin{figure}
\begin{centering}
\includegraphics[width=14cm]{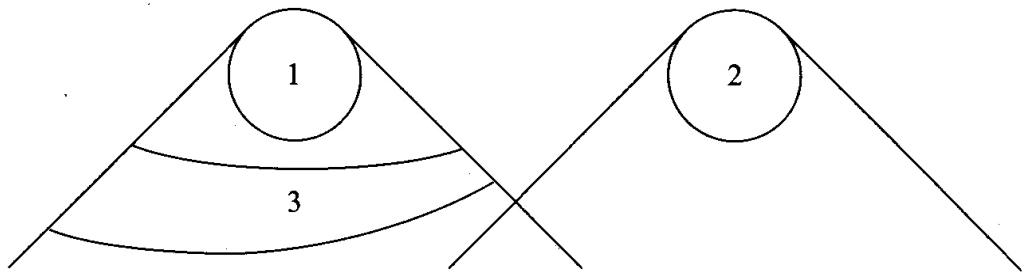}
\par\end{centering}

\caption{\label{fig:Bell2}\char`\"{}Full specification of what happens in
3 makes events in 2 irrelevant for predictions about 1 in a locally
causal theory.\char`\"{} (reproduced with permission from \cite{Bell1987},
pg. 240. Copyright Cambridge University Press.) }

\end{figure}

\begin{quote}
\char`\"{}The direct causes (and effects) of events are near by, and
even the indirect causes (and effects) are no further away than permitted
by the velocity of light. Thus for events in a space-time region 1
(...) we would look for causes in the backward light cone, and for
effects in the future light cone. In a region like 2, space-like separated
from 1, we would seek neither causes nor effects of events in 1. Of
course this does not mean that events in 1 and 2 might not be correlated,
as are the ringing of Professor Casimir's alarm and the readiness
of his egg. They are two separate results of his previous actions.\char`\"{}
(\cite{Bell1987}, pg. 239)
\end{quote}
The above principle, Bell admits, \char`\"{}is not yet sufficiently
sharp and clean for mathematics.\char`\"{} He then considers the following
as an implication of the principle above (with reference to Figure
\ref{fig:Bell2}):

\begin{quote}
\char`\"{}A theory will be said to be locally causal if the probabilities
attached to values of local beables in a space-time region 1 are unaltered
by specification of values of local beables in a space-like separated
region 2, when what happens in the backward light cone of 1 is already
sufficiently specified, for example by a full specification of local
beables in a space-time region 3...\char`\"{} (\cite{Bell1987}, pg.
239)
\end{quote}

The concept of \char`\"{}beable\char`\"{} was invented by Bell to
contrast with the notion of \char`\"{}observable\char`\"{} that is
fundamental in orthodox quantum mechanics:

\begin{quote}
\char`\"{}The beables of the theory are those elements which might
correspond to elements of reality, to things which exist. Their existence
does not depend on `observation'. Indeed observation and observers
must be made out of beables.\char`\"{} (\cite{Bell1987}, pg. 174)
\end{quote}
The above and the specification in Bell's formulation of local causality
that the relevant beables are \char`\"{}local beables\char`\"{} are
evidence that what he means by `local beables' is precisely what I
meant by `events' in the previous section. I will stick to the latter
terminology as it seems very non-problematic and familiar to physicists,
who encounter the term with precisely this meaning in any undergraduate
course in relativity.

We are now ready to formulate the principle mathematically:

\begin{verse}
\textbf{Definition 1 (Local causality):} \emph{A theory will be said
to be locally causal iff for every pair of events $E_{1},\, E_{2}$
respectively contained in space-like separated regions 1 and 2, the
probability, posited by the theory, of occurrence of event $E_{1}$
is independent of $E_{2}$, given the specification of some sufficient
set of events $\mathcal{E}_{p1}$ in the past light cone of 1, i.e.,
}\begin{equation}
P(E_{1}|E_{2},\mathcal{E}_{p1})=P(E_{1}|\mathcal{E}_{p1}).\label{eq:LocalCausality}\end{equation}

\end{verse}
It's important to emphasise the need only for a \emph{sufficient}
set of events (and not necessarily the full specification of \emph{all}
events in the past light cone), for a reason which will be clear in
the next section. What counts as a sufficient set of events will,
of course, depend on the theory. The set of events in region 3 of
Figure \ref{fig:Bell2} would be sufficient in a theory where all
causal chains are continuous. However, one can envisage theories in
which that does not occur%
\footnote{For example, if a theory (as is the case in orthodox quantum theory)
considers as \emph{bona-fide} events the preparation of a quantum
system in a region $R$ and the outcomes of measurements done on that
system in a region $1$ in the future light cone of $R$, but regards
the intermediate evolution of the system as not corresponding to events
in the sense of Axiom 1, one could setup a case in which the set of
events in region $3$ of Figure \ref{fig:Bell2}\textbf{ }would not
be sufficient to screen off events in $1$ from events in $2$ even
in the absence of entanglement. That, however, should arguably not
be considered as a failure of local causality.%
}. This matter will not be too important for Bell's theorem as the
purpose of that will be to show that \emph{no such set can exist anywhere
in the past light cone of $E_{1}$} and therefore that Local Causality
fails.

\section{Other general definitions\label{sec:Other-general-definitions}}

\begin{figure}
\begin{centering}
\includegraphics[width=14cm]{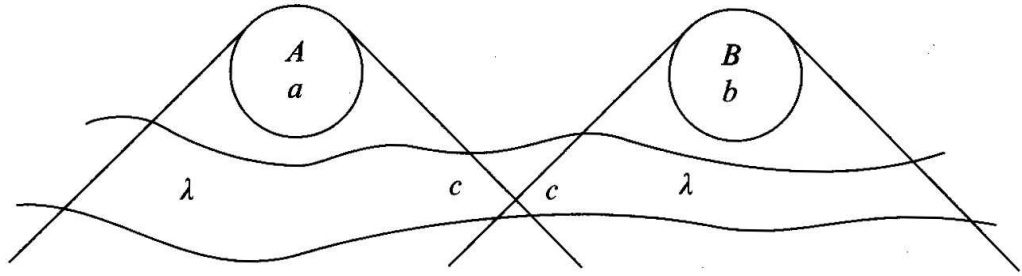}
\par\end{centering}

\caption{\label{fig:Bell3}The bipartite experimental setup used for discussion
of the concepts of this chapter. (reproduced with permission from
\cite{Bell1987}, pg. 242. Copyright Cambridge University Press.)}

\end{figure}

The previous Axioms and definitions were quite independent of any
particular setup. To derive specific consequences of local causality
we need to introduce some concrete experimental situation. The most
common setting used in discussions of Bell's theorem is constituted
by two parties, traditionally called Alice and Bob, and this is the
scheme of Figure \ref{fig:Bell3}. The definitions and theorems of
this chapter will make use of that case only, but most of them could
be generalised in an obvious way for arbitrary parties.

\begin{itemize}
\item Alice and Bob are two spatially separated observers who can perform
a number of measurements and observe their outcomes. 
\item For each pair of systems they perform measurements upon, the choices
of measurement settings and their respective outcomes occur in regions
which are space-like separated from each other, so that no signal
travelling at a speed less than or equal to that of light could connect
any two of them;
\item For each pair of systems, we will denote by $a$ and $b$ Alice's
and Bob's respective measurement settings, and by $A$ and $B$ their
corresponding observed outcomes. 
\item Each pair of systems is prepared by an agreed-upon reproducible procedure
$c$. The events corresponding to this preparation procedure are necessarily
in the intersection of the past light cones of the measurements yielding
$A$ and \textbf{$B$;}
\item $a$, \textbf{$b$}, $A$, $B$ and $c$ represent events in the sense
of the MRRF;
\item $\lambda$ represents any further variables (in addition to $a$,
$b$ and $c$) that may be relevant to the outcomes of the measurements
considered, that is, conditioned upon which the probabilities of the
experimental outcomes may be further specified. They are not fully
determined by the preparation procedure $c$, and as such may be deemed
\char`\"{}hidden variables\char`\"{}. They are necessarily not known
in advance, since any knowledge of additional variables could be assimilated
into the preparation $c$. Conversely, if some of the variables associated
with the preparation procedure are unknown, they must be assimilated
into $\lambda$. In other words, the distinction between $c$ and
$\lambda$ is merely epistemic, i.e., $c$ represents the set of \emph{known}
relevant variables and $\lambda$ represents the set of \emph{unknown}
(but not necessarily \emph{unknowable}) relevant variables. Strictly
speaking $\lambda$ describes a set of \emph{events} in the union
of the past light cones of the experiments as needed by Definition
1. However, one can also think of them as physical variables which
are specified by those events;
\item When an equation involving variables appears, it is to be understood
that the equality holds for all values of those variables.
\end{itemize}
We can now specify the third axiom needed for Bell's theorem.

\begin{verse}
\textbf{Axiom 3 (External Conditionalisation}, or\textbf{ Free will},
or \textbf{No backwards-causality):} \emph{The choices of experiment
$a$, $b$, can be conditioned on free variables uncorrelated with
$\lambda$, such that knowledge of those choices does not provide
any further information about the hidden variables, i.e., it does
not change their probability distribution. Formally, \begin{equation}
P(\lambda|a,b,c)=P(\lambda|c).\label{eq:freewill}\end{equation}
}
\end{verse}
In other words, the choices can be freely made, independently of any
relevant variables that influence the outcomes of the measurements
under study. This is a fundamental requirement of any theory in which
it is possible to separate the world into the system of interest and
the \emph{rest of the world}, where the rest of the world can be ignored
as irrelevant to the evolution of the system. Some people may picture
that as an allowance for the experimenters to make those choices at
their own free will. This picture introduces the inconvenience of
requiring us to explain what we mean by \char`\"{}free will\char`\"{},
which is completely besides the point. Even if the world is completely
deterministic and human free will is an illusion, External Conditionalisation
would be an almost unavoidable assumption of any physical theory.

To make that clear, one can imagine that those choices are made by
pointing photo-detectors at opposite parts of the sky and deciding
based on fluctuations of the cosmic microwave background radiation;
or that they depend on the output of a pseudo-random number generator;
or that they are decided at the whim of an experimentalist; or by
some further random quantum process, say, a measurement on a (presumably)
uncorrelated spin-1/2 particle; or by any combination of those processes.
This is where we see the importance of being clear about the need
for the specification of only a \emph{sufficient }set of events \emph{$\mathcal{E}_{p1}$
}in the Definition 1. Of course the choices of experiment will depend
on \emph{some }events in the past light cone of the events under study,
but a\emph{ }\char`\"{}superdeterministic\char`\"{} theory would be
needed to entertain the possibility that the factors on which those
choices depend can influence the evolution of the system under study.
In such theory, \emph{any possible variable} which one chooses to
conditionalise the choices of experiment on would be statistically
correlated with the set of hidden variables which are relevant to
the experiments of Alice and Bob. And they would need to be conspiratorially
correlated in such a way as to fool us into believing that local causality
is violated by the correlations between those experiments while in
reality the world is strictly locally causal. 

That said, a possible way in which equation \eqref{eq:freewill} could
be violated is by way of backwards-in-time causality or \emph{retrocausality.
}Some authors \cite{Price1996} have claimed that in fact such backwards
causality should be expected from the fact that fundamental physics
is (mostly) time-symmetric, and could be the source of quantum nonlocality.
This view, however, is far from being generally accepted.

\begin{verse}
\textbf{Definition 2: }\emph{A }\textbf{phenomenon}\emph{ is defined,
for a given preparation procedure $c$, by the relative frequencies\begin{equation}
f(A,B|a,b,c).\label{eq:phenomenon}\end{equation}
for all measurements $a$, $b$, and corresponding outcomes $A$,
$B$.}
\end{verse}
The use of \emph{frequencies} in that definition, instead of \emph{probabilities},
is motivated by the fact that it is rather uncontroversial that frequencies
are the things we observe, at least to an arbitrarily good approximation.
Whether or not frequencies directly correspond to probabilities will
depend on one's interpretation of probabilities. Here we will make
the distinction between the observable frequencies (the phenomenon)
and the probabilities which are employed to predict or explain the
phenomenon (the model). In general, models can make use of unobserved
or even unobservable variables. Nevertheless, those variables can
be taken to have an existence independent of observers, i.e., they
can be taken to have an ontological status. Thus one can refer to
such models as ontological models. The following is the most general
kind of ontological model for the phenomena under consideration, in
which the observed phenomena can be explained as arising out of our
ignorance of underlying variables, where our ignorance is accounted
for with standard probability theory, and where Axiom 3 holds.

\begin{verse}
\textbf{Definition 3:} \emph{An }\textbf{ontological model}\emph{
(or }\textbf{model}\textbf{\emph{ }}\emph{in short) for a phenomenon
consists of the set $\Lambda$ of values of $\lambda$, together with
a probability distribution $P(\lambda|c)$ for every preparation procedure
$c$ and a specification of }\begin{equation}
P(A,B|a,b,c,\lambda)\label{eq:modelprob}\end{equation}
\emph{which predicts the phenomenon\begin{equation}
\sum_{\lambda\in\Lambda}\, P(\lambda|c)\, P(A,B|a,b,c,\lambda)=f(A,B|a,b,c).\label{eq:modelfreq}\end{equation}
}
\end{verse}
Nothing in the following discussions hangs on whether the $\lambda$
are discrete or continuous, so for simplicity I use discrete hidden
variables. The above definition can be extended to a continuous set
of hidden variables in the standard way.

\begin{verse}
\textbf{Definition 4:} \emph{An }\textbf{operational theory}\emph{
(OT) is the class of }\textbf{trivial models}\emph{, i.e., the class
of models for which \begin{equation}
P(A,B,|a,b,c,\lambda)=f(A,B|a,b,c).\label{eq:OT}\end{equation}
}
\end{verse}
That is, in operational theories no hidden variables further specify
the probabilities. One could argue that an operational theory contains
no $\lambda$'s, but this definition, as a class of models with trivial
dependences on $\lambda$, allows an operational theorist to talk
about the $\lambda$'s even if they are not operationally meaningful.

\begin{verse}
\textbf{Definition 5:} \emph{A }\textbf{hidden variable model }\emph{(HVM)
is any model that is not trivial.}
\end{verse}
We will now determine what the Definition 1 of local causality implies
to this experimental situation. Since $a$ and $A$ are space-like
separated from $b$ and $B$, and the set of hidden variables completely
specifies all relevant events in the past light cone of $a,\, A$
and $b,\, B$, we obtain

\begin{verse}
\textbf{Corollary 1:}\emph{ A model is }\textbf{locally causal}\emph{,
i.e., a model satisfies }\textbf{local causality }\emph{(LC) iff}\begin{equation}
P(A|a,b,B,c,\lambda)=P(A|a,c,\lambda),\label{eq:LC}\end{equation}
plus the corresponding equations for $B$.

\textbf{Definition 6:}\textbf{\emph{ }}\emph{A model is said to be
}\textbf{deterministic}\emph{, or to satisfy }\textbf{determinism}\emph{
(D), iff \begin{equation}
P(A,B|a,b,c,\lambda)\in\{0,1\}.\label{eq:determinism}\end{equation}
}
\end{verse}
This implies that $A$ and $B$ are functions of the variables which
condition those probabilities, i.e., \begin{equation}
A=A(a,b,c,\lambda),\; B=B(a,b,c,\lambda).\label{eq:functions}\end{equation}

\begin{verse}
\textbf{Definition 7:}\textbf{\emph{ }}\emph{A model is said to be
}\textbf{predictable}\emph{, or to satisfy }\textbf{predictability}\emph{
(P) iff \begin{equation}
P(A,B|a,b,c,\lambda)=P(A,B|a,b,c)\in\{0,1\}.\label{eq:predictability}\end{equation}
}
\end{verse}
That is, a model is predictable iff it is trivial and deterministic.
This implies that $A$ and $B$ are functions of \begin{equation}
A=A(a,b,c),\; B=B(a,b,c).\label{eq:predictablefunctions}\end{equation}
This definition is motivated by the fact that the variables represented
by $c$ are known. The former definition (determinism) is \emph{ontological
}(about how the world \emph{is})\emph{, }while this definition (predictability)
is \emph{epistemic }(about what one can know).

\begin{verse}
\textbf{Definition 8:}\textbf{\emph{ }}\emph{A phenomenon $\Phi_{1}$
associated with preparation $c_{1}$ is said to be }\textbf{irreducibly
unpredictable}\emph{, or to satisfy }\textbf{irreducible unpredictability}\emph{
(IU) iff it has no predictable models and there is no phenomenon $\Phi_{2}$
associated with a preparation $c_{2}=c_{1}\cup c'$ (for all $c'$)
which has a predictable model and such that the frequencies $f_{1}$
of $\Phi_{1}$ are given by}

\emph{\begin{equation}
f_{1}(A,B|a,b,c_{1})=\sum_{c'}P(c'|c_{1})P(A,B|a,b,c_{1},c'),\label{eq:IU}\end{equation}
where $P(A,B|a,b,c_{1},c')\in\{0,1\}$.}
\end{verse}
In other words, a phenomenon is irreducibly unpredictable iff it has
no predictable model and cannot be rendered predictable by knowledge
of further variables. The reason for \eqref{eq:IU} is that the only
way in which an unpredictable phenomenon could be rendered predictable
\emph{without fundamentally changing the phenomenon }would be if there
were a deterministic hidden variable model which predicted the phenomenon,
but for which the \char`\"{}hidden variables\char`\"{} could be in
principle known in advance. But if the hidden variables were known
in advance they could (in fact they should, since the only distinction
between $\lambda$ and $c$, as mentioned before, is that the latter
are known) be incorporated into the preparation variables. That is
why I use the notation $c'$ for those further knowable variables.

\begin{verse}
\textbf{Definition 9:}\textbf{\emph{ }}\emph{A model is said to satisfy
}\textbf{locality }\emph{(L) iff \begin{equation}
P(A|a,b,c,\lambda)=P(A|a,c,\lambda),\label{eq:L}\end{equation}
}plus the corresponding equation for $B$.
\end{verse}
This was precisely the meaning that Bell intended for this term in
his original 1964 paper \cite{Bell1964}, although without formal
definition. Shimony called this \emph{parameter independence }(\cite{Cushing1989},
pg. 25).

\begin{verse}
\textbf{Definition 10:}\emph{ A model is said to satisfy }\textbf{outcome
independence }\emph{(OI) iff}\begin{equation}
P(A|a,b,B,c,\lambda)=P(A|a,b,c,\lambda),\label{eq:OI}\end{equation}
plus the corresponding equation for $B$.
\end{verse}
This concept was introduced by Jarrett \cite{Jarrett1984}, under
the misleading name of \emph{completeness. }The present terminology
is due to Shimony (\cite{Cushing1989}, pg. 25). One could also consider
calling it causality, so that local causality would be the conjunction
of locality and causality. This choice would be justified if one takes
causality to be a weakened form of determinism, in which the outcomes
depend (maybe stochastically) only on the experimental settings, but
not on the distant outcomes. In other words, in this view of causality
the effects (the outcomes, things which are not controllable) at Alice's
can depend only on the local or distant causes (the settings, things
which are controllable) but not on the distant effects. This terminology,
however, would seem to rule out a common {}``explanation'' of the
quantum correlations: that the measurement on Bob's system \emph{causes}
the quantum state to collapse which \emph{causes} the outcomes at
Alice to be what they are. This (non-relativistically invariant) causal
picture would violate outcome independence since the collapsed quantum
state depends on Bob's outcome.

\begin{verse}
\textbf{Definition 11:}\textbf{\emph{ }}\emph{A model is said to be
}\textbf{locally deterministic}\emph{, or to satisfy }\textbf{local
determinism}\emph{ (LD) iff it satisfies locality and determinism.}

\textbf{Definition 12:}\textbf{\emph{ }}\emph{A }\textbf{model}\emph{
is said to }\textbf{violate}\emph{ p iff it lacks that property.}

\textbf{Definition 13:}\textbf{\emph{ }}\emph{An }\textbf{operational
theory}\emph{ is said to }\textbf{violate}\emph{ p iff all trivial
models violate p.}

\textbf{Definition 14:}\textbf{\emph{ }}\emph{A }\textbf{phenomenon}\emph{
is said to }\textbf{violate}\emph{ p iff all models violate p.}

\textbf{Definition 15:}\textbf{\emph{ }}\textbf{Nature}\emph{ is said
to }\textbf{violate}\emph{ p iff a phenomenon violating p is observed.}
\end{verse}

\section{General results}

We now present the general results and relations (that is, those which
are not specific to quantum mechanics) concerning the definitions
of the previous section. 

By the definition of conditional probability $P(A,B|a,b,c,\lambda)=P(A|B,a,b,c,\lambda)P(B|a,b,c,\lambda)$.
Using Corollary 1 we arrive at

\begin{verse}
\textbf{Theorem 1:}\emph{ Local causality is equivalent to }\textbf{factorisability}\emph{,
i.e.,}\begin{equation}
P(A,B|a,b,c,\lambda)=P(A|a,c,\lambda)P(B|b,c,\lambda).\label{eq:factorizability}\end{equation}

\end{verse}
Jarrett \cite{Jarrett1984} showed that:

\begin{verse}
\textbf{Theorem 2 (Jarrett 1984a):}\textbf{\emph{ }}\emph{Local causality
is the conjunction of locality and outcome independence.}
\end{verse}
The purpose of that decomposition was to argue that outcome independence
was the concept to be blamed, while locality was the real consequence
of relativity\emph{ }and therefore ought to be maintained\emph{. }We'll
return to this important point later.

\begin{verse}
\textbf{Theorem 3:}\textbf{\emph{ }}\emph{Local causality is strictly
stronger than locality.}
\end{verse}
That is, every locally causal model satisfies locality but not vice-versa.
The proof is simple. Theorem 1 implies the first statement, and orthodox
quantum mechanics provides an example of a model which satisfies locality
but violates local causality.

\begin{verse}
\textbf{Theorem 4:}\textbf{\emph{ }}\emph{Determinism is strictly
stronger than outcome independence.}
\end{verse}
If determinism holds, the outcome $A$ is fully determined by $(a,b,c,\lambda)$.
Therefore knowledge of $B$ cannot change the probability of $A$
if $(a,b,c,\lambda)$ are specified, which is just the statement of
outcome independence. To see the failure of the converse, just note
that orthodox quantum mechanics for separable states violates determinism
but satisfies outcome independence.

\begin{verse}
\textbf{Corollary 2:}\textbf{\emph{ }}\emph{Local determinism is strictly
stronger than local causality.}
\end{verse}
This follows from the definition of local determinism and Theorems
2 and 4.

\begin{verse}
\textbf{Theorem 5:}\textbf{\emph{ }}\emph{Predictability is strictly
stronger than determinism.}
\end{verse}
Every predictable model is obviously deterministic, but some deterministic
models are not predictable. A trivial example is that of a deterministic
model in which one does not know all relevant variables $\lambda$
(or does not know them precisely enough) but could know them in principle,
(e.g. classical statistical mechanics), but there are models in which
one cannot know all $\lambda$ even in principle (e.g. Bohmian mechanics).

\begin{verse}
\textbf{Theorem 6:}\textbf{\emph{ }}\emph{Some phenomena violate predictability.}
\end{verse}
Since the distinction between the hidden variables $(\lambda)$ and
the variables that specify the preparation $(c)$ is an epistemic
one, and since a phenomenon is \emph{defined} partly by $c$, there
are trivial examples of unpredictable phenomena --- one just needs
to ignore some relevant variables. 

This does not imply that there are irreducibly unpredictable phenomena.
I'll return to this question later.

\begin{verse}
\textbf{Theorem 7:}\textbf{\emph{ }}\emph{No phenomenon violates determinism.}
\end{verse}
That doesn't mean that no model violates determinism, but that every
phenomenon has a possible model which satisfies determinism. To see
that, define $\lambda_{0}$ and $\lambda_{1}$ by $P(A,B|a,b,c,\lambda_{0})=0$
and $P(A,B|a,b,c,\lambda_{1})=1$. Substituting in \eqref{eq:modelfreq}
we obtain $f(A,B|a,b,c)=P(\lambda_{1}|c)$, resulting in that every
possible frequency can be modelled in this form. Of course, that leaves
open the question of whether there \emph{really exist} such events
corresponding to the variables to play the role of $\lambda_{0}$
and $\lambda_{1}$, so one could read Theorem 7 as saying that it
is impossible to \emph{prove} that any phenomenon violates determinism.

\begin{verse}
\textbf{Corollary 3:}\emph{ No phenomenon violates outcome independence.}

\textbf{Theorem 8 (Fine 1982):}\textbf{\emph{ }}\emph{A phenomenon
violates local determinism iff it violates local causality.}
\end{verse}
That is, any phenomenon that has a locally deterministic model has
a locally causal model and vice versa. This theorem is due originally
to Fine \cite{Fine1982}. Note that this does \emph{not }mean that
all locally causal models are locally deterministic and vice versa
(which in fact is false as per Corollary 2). 

\textbf{Proof.} Any phenomenon that has a locally deterministic model
automatically has a locally causal model, since all LD models are
LC. To see the converse, remember that if there exists a locally causal
model, the frequencies can be written as \begin{equation}
f(A,B|a,b,c)=\sum_{\lambda}P_{LC}(\lambda|c)P_{LC}(A|a,c,\lambda)P_{LC}(B|b,c,\lambda).\label{eq:prooffinetheorem1}\end{equation}
We now add extra hidden variables for each of the factors on the right
hand side, in a similar fashion as for the proof of Theorem 7. We
decompose \begin{equation}
P_{LC}(A|a,c,\lambda)=\sum_{\lambda_{A}}P'(\lambda_{A}|c,\lambda)P_{LD}(A|a,c,\lambda_{A}),\label{eq:prooffinetheorem2}\end{equation}
where $P{}_{LD}(A|a,c,\lambda_{A})=\lambda_{A}\in\{0,1\}$. With a
similar decomposition for $B$, we obtain

\begin{eqnarray}
f(A,B|a,b,c) & = & \sum_{\lambda}P_{LC}(\lambda|c)P_{LC}(A|a,c,\lambda)P_{LC}(B|b,c,\lambda).\nonumber \\
 & = & \sum_{\lambda,\lambda_{A},\lambda_{B}}P_{LC}(\lambda|c)P'(\lambda_{A},\lambda_{B}|c,\lambda)P_{LD}(A|a,c,\lambda_{A})P_{LD}(B|b,c,\lambda_{B})\label{eq:prooffinetheorem3}\\
 & = & \sum_{\lambda'}P_{LD}(\lambda'|c)P_{LD}(A|a,c,\lambda')P_{LD}(B|b,c,\lambda'),\nonumber \end{eqnarray}
where we define $\lambda'\equiv(\lambda,\lambda_{A},\lambda_{B})$
and $P_{LD}(\lambda'|c)\equiv P_{LC}(\lambda|c)P'(\lambda_{A},\lambda_{B}|c,\lambda)$,
obtaining a locally deterministic model as desired.

\subsection{Local causality and signalling}

Bell was adamant in stressing that his concept of local causality
was quite distinct from the concept of no faster than light signalling.
One of the reasons for Bell's rejection of the importance of the concept
of signalling was that he understood that it was hard to talk about
signalling without using anthropocentric terms like `information'
and `controllability':

\begin{quote}
\char`\"{}Suppose we are finally obliged to accept the existence of
these correlations at long range, (...). Can \emph{we} then signal
faster than light? To answer this we need at least a schematic theory
of what \emph{we} can do, a fragment of a theory of human beings.
Suppose we can control variables like $a$ and $b$ above, but not
those like $A$ and $B$. I do not quite know what `like' means here,
but suppose the beables somehow fall into two classes, `controllables'
and `uncontrollables'. The latter are no use for \emph{sending} signals,
but can be used for \emph{reception.}\char`\"{} (\cite{Bell1987},
pg.60)

But he rejects the idea that signal locality be taken as the fundamental
limitation imposed by relativity:

\char`\"{}Do we have to fall back on `no signalling faster than light'
as the expression of the fundamental causal structure of contemporary
theoretical physics? That is hard for me to accept. For one thing
we have lost the idea that correlations can be explained, or at least
this idea awaits reformulation. More importantly, the `no signalling...'
notion rests on concepts which are desperately vague, or vaguely applicable.
The assertion that `we cannot signal faster than light' immediately
provokes the question:
\begin{quote}
Who do we think \emph{we }are?
\end{quote}
\emph{We} who can make `measurements', \emph{we} who can manipulate
`external fields', \emph{we }who can signal at all, even if not faster
than light? Do \emph{we} include chemists, or only physicists, plants,
or only animals, pocket calculators, or only mainframe computers?\char`\"{}
(\cite{Bell1987}, pg.245)
\end{quote}
That is one of the reasons I have first presented the definition of
local causality in a general context without the operational definitions
used from Section \ref{sec:Other-general-definitions} onwards. In
Definition 1 no mention was made of measurements and outcomes, but
only of events and space-time structure. We have then translated the
consequences of that into our operational definitions with Corollary
1. But to even start talking about signalling, we need to have the
operational model set up and explicit mention what are the controllable,
uncontrollable and observable variables within it. This is precisely
why Bell rejects the idea that this is a fundamental notion. I agree
with Bell that if one wants to entertain the idea that signal locality
is the fundamental restriction from relativity, one needs to put fundamental
weight on epistemological terms. However, as long as one clearly understands
that, I have no qualms with that position, in fact I find it an interesting
possibility to pursue, and in this chapter it will lead to a nice
new result.

Given the assumptions that the variables $a$ and $b$ are controllable
(an assumption, in fact, already made in Axiom 3) we can formulate
the concept of signal locality as follows.

\begin{verse}
\textbf{Definition 16:}\textbf{\emph{ }}\emph{A }\textbf{phenomenon}\emph{
is said to satisfy }\textbf{signal locality}\emph{ (SL) iff \begin{equation}
f(A|a,b,c)=f(A|a,c),\label{eq:SL}\end{equation}
}plus the corresponding equation for $B$.
\end{verse}
The reason is straightforward. If the phenomenon violates signal locality,
then there exist at least two possible choices of setting $b$, $b'$
such that $f(A|a,b,c)\neq f(A|a,b',c)$. Therefore by looking at the
frequency of outcomes of $A$ in a large enough ensemble (and in principle
it is always possible for Alice to make all of the measurements in
her ensemble space-like separated from all measurements in Bob's ensemble),
Alice can determine with arbitrary accuracy what setting Bob has chosen.
Conditionalising this choice on a source of information, Bob can thereby
send signals to Alice.

\begin{verse}
\textbf{Definition 17:}\textbf{\emph{ }}\emph{A }\textbf{model}\emph{
is said to satisfy }\textbf{signal locality}\emph{ (SL) iff the phenomena
it predicts satisfies SL.}
\end{verse}
The reason behind Jarrett's preference for locality (mentioned below
Theorem 2) over outcome independence is that Jarrett believed that
locality was equivalent to signal locality. However, that is an unwarranted
assumption as argued by Maudlin \cite{Maudlin1994}. The main counter-example
is Bohmian mechanics. That theory violates L but not SL. The reason
is that any attempt of controlling the distant outcome by the local
choice of setting is thwarted by an unavoidable lack of knowledge
of the hidden variables. On the other hand, if a model satisfies L
it necessarily satisfies SL, as can be easily seen by substituting
Eq. \eqref{eq:L} in \eqref{eq:modelfreq} and summing over $B$.
In other words,

\begin{verse}
\textbf{Theorem 9:}\textbf{\emph{ }}\emph{:Locality is strictly stronger
than signal locality.}
\end{verse}
While Jarrett's rejection of models which violate locality is not
well-grounded, his repulse of violations of locality is. 

\begin{verse}
\textbf{Theorem 10:}\textbf{\emph{ }}\emph{A phenomenon violates locality
iff it violates signal locality.}
\end{verse}
As argued above, if a phenomenon has a model which satisfies locality
then the phenomena satisfies signal locality. The converse is also
true: if a phenomenon satisfies signal locality then there is a model
which satisfies locality: one example is the trivial model that corresponds
to that phenomenon.

\begin{verse}
\textbf{Theorem 11:}\textbf{\emph{ }}\emph{If relativistic invariance
is assumed, a phenomenon which violates signal locality will lead
to contradictions.}
\end{verse}
Faster than light signalling can lead to paradoxes like the famous
\char`\"{}grandfather paradox\char`\"{} of time travel --- a time
traveller goes to the past and kills his grandfather before his father
was born, paradoxically thwarting the possibility of his own existence.
Consider a scenario in which signal locality is violated by Alice's
and Bob's experimental apparatuses such that $f(A_{1}|a_{1},b_{1},c_{1})\neq f(A_{1}|a_{1},b_{1}',c_{1})$.
How exactly these frequencies differ is unimportant, as Alice can
always use an arbitrarily large ensemble such that $f(A_{1}'|b_{1})\approx1$
and $f(A_{1}'|b_{1}')\approx0$, where $A_{1}'$ is the event which
corresponds to the frequency of the outcome $A_{1}$ within the ensemble
being approximately $f(A_{1}|a_{1},b_{1},c_{1})$ and with good confidence
distinct from $f(A_{1}|a_{1},b_{1}',c_{1})$. Since $A_{1}'$ and
the choice between $b_{1}$ and $b_{1}'$ are space-like separated,
there exists a reference frame in which they occur simultaneously.
If relativistic invariance is assumed, then it must be possible to
produce, in any reference frame, a similar instantaneous signalling
setup (otherwise there would be a preferred reference frame), and
if we assume that there's no preferred spatial direction for such
signal transmission (if there were a preferred spatial direction,
that would also be a violation of the principle of relativity), we
can setup another pair of boxes where $f(B_{2}|a_{2},b_{2},c_{2})\neq f(B_{2}|a_{2}',b_{2},c_{2})$.
The choice $a_{2}/a_{2}'$ of Alice's experimental setting would be
conditioned on $A_{1}'$ (and therefore would be in the future light
cone of $A_{1}'$) and by use of a similar procedure as for the first
setup, Alice and Bob arrange things such that $f(B_{2}'|A_{1}')\approx1,\, f(B_{2}'|\neg A_{1}')\approx0$,
where $'\neg'$ denotes logical negation. Bob's outcome $B_{2}'$
is arranged to be in the past light cone of the choice between $b_{1}$
and $b_{1}'$, and Bob conditions that choice such that he will choose
\textbf{$b_{1}'$ }if the result of measurement 2 is $B_{2}'$, otherwise
he will choose $b_{1}$. So we finally obtain the contradictory set
of implications\begin{align}
b_{1} & \Rightarrow A_{1}'\Rightarrow B_{2}'\Rightarrow b_{1}'\Rightarrow\neg b_{1}\nonumber \\
b_{1}' & \Rightarrow\neg A_{1}'\Rightarrow\neg B_{2}'\Rightarrow b_{1}\Rightarrow\neg b_{1}'.\label{eq:paradox}\end{align}
That is, if Bob chooses $b_{1}$, then he does not choose $b_{1}$,
If Bob chooses $b_{1}'$, then he does not choose $b_{1}'$. So it
is fair to say that relativity seems to exclude the possibility of
violation of signal locality%
\footnote{Although there are several attempts to make sense of this kind of
paradox in the philosophical literature about time travel. However,
most attempts either violate Axiom 2, by postulating multiple universes,
or Axiom 3, by restricting the possibility of conditionalising the
choices of experiment on any variables we choose, thereby blocking
the setup of the paradox from the start. If one wants to keep those
axioms, I am not aware of any clean way around the paradox except
enforcing signal locality.%
}. Bell would say that this is not all it excludes, that relativity
implies not simply signal locality but local causality. An interesting
question is therefore whether violation of local causality by a phenomenon
which does not violate signal locality can lead to such time travel
paradoxes. If signal locality is satisfied, the above scenario cannot
be set up, and the fact that we have observed violations of local
causality (up to some loopholes) seems to point to the fact that no
contradictions can arise out of that violation (inasmuch as a contradiction
cannot be actually observed). However sensible this statement may
sound, I can't provide a proof to it. But it is interesting to mention
it as a conjecture.

\begin{verse}
\textbf{Conjecture 1:}\textbf{\emph{ }}\emph{A phenomenon that violates
local causality but not signal locality does not lead to contradictions
even if relativistic invariance is assumed.}

\textbf{Remark 1:}\textbf{\emph{ }}\emph{Nevertheless, a phenomenon
that violates local causality is a }\textbf{non-local resource}\emph{.
That is, there are tasks Alice and Bob can do using this phenomenon
that they could not do if they had access only to locally causal phenomena.}

\textbf{Theorem 12:}\textbf{\emph{ }}\emph{For a phenomenon to violate
locality it is necessary and sufficient for the operational theory
to violate locality.}
\end{verse}
This follows directly from Theorem 10 and the definitions of operational
theory and signal locality.

\begin{verse}
\textbf{Theorem 13:}\textbf{\emph{ }}\emph{For a phenomenon to violate
local causality it is necessary but not sufficient for the operational
theory to violate local causality.}
\end{verse}
The necessary part is obvious. We can show insufficiency by appealing
to a counter-example: single particle quantum mechanics restricted
to measurements of position. The operational theory violates local
causality (see Theorem below) but there's a locally causal model for
the phenomena involved: single-particle Bohmian mechanics.

\begin{verse}
\textbf{Remark 2:}\textbf{\emph{ }}\emph{Thus, unlike locality, to
say whether a phenomenon violates local causality it is necessary
to consider hidden variable models.}
\end{verse}

\section{Quantum Results}

\begin{verse}
\textbf{Definition 18:}\textbf{\emph{ }}\textbf{Orthodox quantum theory}\emph{
or }\textbf{operational quantum theory}\emph{ (OQT) is an operational
theory in which \begin{equation}
P(A,B|a,b,c,\lambda)=Tr[\hat{\Pi}_{A}\otimes\hat{\Pi}_{B}\rho_{c}],\label{eq:OQT}\end{equation}
where }$\hat{\Pi}_{A}$\emph{ is a projector onto the subspace corresponding
to outcome $A$ of observable $\hat{a}$ and similarly for $B$, while
$\rho_{c}$ is a positive unit-trace operator associated with the
preparation} $c$.

\textbf{Definition 19:}\textbf{\emph{ }}\emph{A}\textbf{ quantum model}\emph{
is a model whose predictions agree with OQT. }

\textbf{Definition 20:}\textbf{\emph{ }}\emph{A}\textbf{ quantum phenomenon}\emph{
is a phenomenon predicted by OQT.}

\textbf{Theorem 14 (Einstein 1927):}\textbf{\emph{ }}\emph{OQT violates
local causality.}
\end{verse}
This was first proved by Einstein at the 1927 Solvay conference \cite{Wick1995}.
Consider a single particle with a position wavefunction spread out
over space, and consider $a$ as a measurement of the particle position
in the region around Alice and similarly for Bob. Then the probability
that Alice finds the particle in her region is not independent of
whether Bob finds the particle in his region. And since in OQT there
are no hidden variables, \eqref{eq:LC} is not satisfied.

\begin{verse}
\textbf{Theorem 15 (Heisenberg 1930):}\textbf{\emph{ }}\emph{OQT does
not violate locality.}
\end{verse}
This is a well-known result that arises out of the fact that operators
corresponding to measurements in spatially separated regions commute.
In his 1930 book \cite{Heinsenberg1930}, Heisenberg replied to Einstein's
objection to OQT above by pointing out that the theory would not allow
faster-than-light signalling.

\begin{verse}
\textbf{Corollary 4:}\emph{ Quantum phenomena do not violate signal
locality.}

\textbf{Theorem 16 (Jarrett 1984b):}\textbf{\emph{ }}\emph{OQT violates
outcome independence.}
\end{verse}
This is a consequence of Jarrett's Theorem 2, and Theorems 14 and
15.

\subsection{The EPR argument}

As we mentioned above, in 1927 Einstein had already shown that OQT
violated local causality (the term wasn't used until Bell, but Einstein
clearly stated that quantum mechanics \char`\"{}contradicts the principles
of relativity\char`\"{} \cite{Wiseman2006}). That was always intended
by Einstein as a proof that OQT was incomplete, i.e., that there must
be further hidden variables beyond the quantum state to completely
specify the properties of physical systems.

Einstein's 1927 argument however is not as widely recognised as the
Einstein, Podolsky and Rosen (EPR) argument of 1935 \cite{Einstein1935}.
Maybe because the argument in that paper was more well articulated,
maybe because in the Solvay conference Einstein also tried to prove
quantum mechanics to be inconsistent --- not only incomplete --- by
carefully set-up thought experiments. Those attempts were thwarted
by Bohr, who argued that Einstein was not consistently using the uncertainty
principle at all levels in his analysis of the experiments. The physics
community largely takes Bohr to have triumphed over Einstein on those
arguments, and therefore Einstein's failure on that front would have
been automatically transferred to his 1927 theorem about local causality.

In a recent paper \cite{Harrigan2007a}, Harrigan and Spekkens suggest
that Einstein himself preferred the (more complicated) argument for
incompleteness using entangled states. More precisely, Einstein's
preferred argument was that used in a 1935 correspondence with Schrödinger,
not the EPR argument which --- those authors claim --- does not reflect
precisely Einstein's opinion on the matter. The reason for that preference,
they argue, is that this argument not only rules out the completeness
of quantum mechanics (if one assumes local causality, of course),
but also it provides an extra argument for the view that the quantum
states are merely epistemic in nature.

Another reason for preference of the EPR argument, mentioned in \cite{Wiseman2006}
and \cite{Harrigan2007a} over Einstein's 1935 is an experimental
one: the measurement statistics used in the 1927 argument can be trivially
simulated with a mixed state, while those of the EPR argument cannot.
The critics could evade the earlier argument by denying the coherence
of the state upon which the measurements are performed, a move that
cannot be made in the later.

But the most likely reason, in my opinion, for the community's recognition
of the EPR argument over the 1927 one is that the former also substantially
differs from the later in that it introduces entangled states. Even
if EPR failed in convincing the community of the incompleteness of
orthodox quantum mechanics, the kind of states they considered opened
the door to Bell's recognition of the failure of one of EPR's premises
instead --- local causality --- and from there lead to the multiple
applications in the modern field of quantum information science. They
also influenced Schrödinger to envisage his infamous \char`\"{}cat
paradox\char`\"{} \cite{Schroedinger1935} (and to coin the term `entanglement'
to refer to the strange kind of states EPR consider).

EPR's argument is essentially that given (i) a suitable necessary
condition for \emph{completeness} of a theory; (ii) an apparently
reasonable sufficient condition for determining when a physical variable
corresponds to an \char`\"{}element of physical reality\char`\"{};
(iii) the assumption of local causality; and (iv) some predictions
of quantum mechanics concerning entangled states; one must conclude
that quantum mechanics is incomplete, in the sense that there must
exist hidden variables to further specify physical states. I won't
go into the EPR argument in detail here, however, as it will be the
subject of Chapter \ref{cha:EPR-paradox}.

\subsection{Bell's Theorems}

We now arrive at the famous Bell theorems. By 1964 it was known that
OQT violates local causality, but one could still imagine, as EPR
did, that a more detailed description of the phenomena was possible
in which local causality was maintained. Bell's startling contribution
was to show that this hope was futile. The importance of this theorem
cannot be overemphasised. It has even been dubbed the \char`\"{}most
profound discovery of science\char`\"{} \cite{Stapp1977}. In 1964,
however, Bell did not prove the stronger theorem about local causality,
but only a weaker version:

\begin{verse}
\textbf{Theorem 17 (Bell 1964):}\textbf{\emph{ }}\emph{All deterministic
quantum models violate locality.}
\end{verse}
In other words, quantum phenomena violate local determinism. By Fine's
Theorem 8 this is equivalent to the strong form of Bell's theorem:

\begin{verse}
\textbf{Theorem 18 (Bell 1971):}\emph{ Quantum phenomena violate local
causality.}
\end{verse}
However, that was not fully recognised until much later. In 1964 Bell
was considering locality and not local causality, as is evident by
this passage:

\begin{quote}
\char`\"{}It is the requirement of locality, or more precisely that
the result of a measurement on one system be unaffected by operations
on a distant system with which it has interacted in the past, that
creates the essential difficulty.\char`\"{} (\cite{Bell1964}, pg.
14)
\end{quote}
and that he considers only deterministic hidden variables is clear
in the following discussion of the specific setup:

\begin{quote}
\char`\"{}The result $A$ (...) is then determined by $a$ and $\lambda$,
and the result $B$ (...) is determined by $b$ and $\lambda$, and
$A(a,\lambda)=\pm1,\, B(b,\lambda)=\pm1$. The vital assumption is
that the result $B$ (...) does not depend on the setting $a$, (...)
nor$A$ on $b$.\char`\"{} (\cite{Bell1964}, pg. 15)
\end{quote}
In a 1971 paper (\cite{Bell1987}, pg. 37), Bell considered a model
which allowed some \char`\"{}indeterminism with a certain local character\char`\"{}
associated with the detectors --- essentially a factorisable model
with arbitrary probabilities, which by Theorem 1 follows from local
causality --- so this could be taken as the first clear proof of Theorem
18. But he did not clearly use the term local causality with the meaning
of Definition 1 until 1976 (\cite{Bell1987}, pg. 54).

While in hindsight Bell's 1964 and 1971 theorems are logically equivalent
(by way of Theorem 8), the 1964 theorem could not be directly applied
to experimental situations, since it assumed perfect correlations
(which are obviously not observable in any real experiment). The Bell
inequality of 1971 however (essentially a version of the Clauser,
Horne, Shimony, Holt (CHSH) inequality of 1969 \cite{Clauser1969},
but which was derived from a local deterministic model) is applicable
to real experimental situations. Many experiments realised since then
strongly follow the quantum mechanical predictions, and (up to some
loopholes involving detection efficiencies and/or lack of space-like
separation) support the conclusion

\begin{verse}
\textbf{Conclusion 1:}\emph{ Nature violates local causality.}
\end{verse}

\subsection{Determinism and Predictability}

\begin{verse}
\textbf{Theorem 19 (Born):}\textbf{\emph{ }}\emph{OQT violates determinism.}
\end{verse}
This is essentially the content of Born's postulate that the modulus
square of the wave function corresponds to probabilities of outcomes
of measurements, the extension of which for mixed states is given
by \eqref{eq:OQT}. Since for any state (determined by a preparation
$c$) there is at least one measurement for which the probabilities
are different from $1$ or $0$, determinism is violated by OQT.

\begin{verse}
\textbf{Corollary 4:}\textbf{\emph{ }}\emph{OQT violates predictability.}

\textbf{Corollary 5:}\textbf{\emph{ }}\emph{Quantum phenomena violate
predictability.}
\end{verse}
That is a consequence of Corollary 4 plus the fact that the definition
of predictability implies that if a model is not predictable, then
the phenomena it predicts violates predictability. This still does
not mean that quantum phenomena are irreducibly unpredictable, that
is, it doesn't mean that there cannot be further knowable variables
such that the phenomena would be rendered predictable. That there
are no such further knowable variables is essentially the content
of Heisenberg's Uncertainty Principle (HUP), that is, 

\begin{verse}
\textbf{Theorem 20 (Heinsenberg 1926):}\textbf{\emph{ }}\emph{Quantum
phenomena are irreducibly unpredictable.}
\end{verse}
But the proof of Heisenberg's uncertainty principle, based on the
commutation relations between different observables, can only be made
within quantum mechanics. Bohmian mechanics is deterministic but also
not predictable, but again those facts are model-dependent. The definition
of quantum phenomenon is \char`\"{}a phenomenon predicted by OQT\char`\"{}.
OQT states that some phenomena are irreducibly unpredictable (by way
of HUP), but OQT could be wrong about that. However, Theorem 7 (which
states that no phenomenon violates determinism) may be taken to imply
that nothing can be said about predictability in a model-independent
way. 

This debate was in the centre of the famous Einstein-Bohr debates
starting in the Solvay conference of 1927. Einstein was then unsatisfied
with the indeterminism of quantum mechanics and attempted to show
that it was inconsistent by devising clever \emph{gedanken} experiments
aiming to break Heisenberg's Uncertainty Principle. But Bohr thwarted
every such attempt by pointing a flaw in Einstein's reasoning. History
is one the side of Bohr who is widely regarded as the victor of those
debates. Einstein's flaw was essentially to ignore the uncertainty
principle for some variables, which would allow him to obtain more
knowledge than permitted by OQT for some \emph{other} variables. Bohr's
reply was to point that if the principle was observed consistently
throughout the problem, Einstein's move would fail. However, of course,
Bohr could never \emph{prove }that OQT was correct and that the HUP\emph{
}must always hold, all he did was point out that OQT was \emph{consistent.}
There is no question about that, but could he give a better argument
to convince Einstein that OQT was \emph{correct} about that? On this
question I offer the following theorem.

\begin{verse}
\textbf{Theorem 21:}\textbf{\emph{ }}\emph{If relativistic invariance
is assumed, Nature is irreducibly unpredictable or some observed phenomena
can lead to contradictions.}
\end{verse}
The contradictions in question are just those involved in the grandfather-type
paradox considered in Theorem 11. The proof is simple. Suppose Nature
is not irreducibly unpredictable, i.e., that for any observed phenomenon
there is a possible trivial deterministic model that reproduces the
phenomenon as in \eqref{eq:IU}. By Theorem 17 there are phenomena
for which there is no local deterministic model, therefore any such
deterministic model must be nonlocal. However, since the models under
consideration are trivial, if they are nonlocal the phenomena they
predict must violate signal locality. Therefore if those phenomena
are not irreducibly unpredictable then they violate signal locality
and by Theorem 11 will lead to contradictions if relativistic invariance
is assumed. Moreover, such phenomena violating local determinism have
been observed, therefore if Nature is not irreducibly unpredictable
then some observed phenomena can lead to contradictions (if one obtains
the hitherto hidden but observable information to render the phenomena
predictable).

In the last of Einstein's thought experiments, Bohr's reply made use
of Einstein's own General Theory of Relativity to reject Einstein's
thought experiment. Of course, Einstein or Bohr were not aware of
Bell's Theorem, but interestingly, if they were, Bohr could, by the
use of Einstein's Special Theory, have convinced Einstein not only
that OQT was consistent, but that Nature is irreducibly unpredictable
regardless of quantum mechanics.

\begin{verse}
\textbf{Conclusion 2:}\textbf{\emph{ }}\emph{The above definitions
and theorems imply the following structure in }\textbf{phenomenon
space }\emph{(PS):}

\begin{eqnarray}
vSL & = & vL\subset vLD=vLC\subset PS,\nonumber \\
\{\} & = & vD\subset vP\subset PS\label{eq:PS}\end{eqnarray}
\emph{Here $\{\}$ denotes the empty set, vSL denotes the set of phenomena
violating signal locality, and similarly for vL, vLC, vLD, vD and
vP.}
\end{verse}

\section{Making sense of it}

In the beginning of this Chapter I indicated that there are, still
today, many discussions about what Bell's theorem(s) prove and what
concepts it requires us to give up. Here I'll sketch a few of those
debates and what we can conclude about them in light of the careful
definitions and theorems of this chapter.

\subsection{Locality and outcome independence}

I have already mentioned Jarrett's preference of rejecting outcome
independence and keeping locality. The motivation is essentially Theorems
9 and 10, which state that violation of locality by a phenomenon will
lead to paradoxes if relativistic invariance is assumed. However,
before rushing into conclusions, we must remember that the concept
of a phenomenon violating a property is quite distinct from that of
a model lacking that property. In fact, we have seen that it is possible
for a model to violate locality while the related phenomena do not.
So keeping locality \emph{does not }necessarily lead to paradoxes.
A desire to avoid paradoxical situations is not sufficient reason
to reject locality. 

Bell's theorem states that quantum phenomena violate local causality.
Heisenberg's Theorem 15 says that quantum phenomena do not violate
locality. One could take that to mean that quantum phenomena violate
outcome independence. But by Corollary 3, it is \emph{impossible}
for any phenomenon to violate outcome independence. The choice will
depend on the model. Some models respect locality but not OI, such
as OQT. Some respect OI but not locality, such as Bohmian mechanics.
Some respect neither, as Nelson's mechanics. Bell's theorem says that
none can respect both.

\subsection{Determinism and hidden variables \label{sub:Determinism-and-hidden}}

There's a common misconception in the literature about Bell inequalities
which maintains that what Bell's theorem tells us to give up is determinism
and/or hidden variables. That probably arises from the fact that the
original 1964 Bell theorem (and many subsequent derivations) was in
fact about the failure of local determinism, that is, the conjunction
of locality and determinism. For a reader who was already used to
the violation of determinism by orthodox quantum mechanics, and who
didn't notice the fact that violation of locality by a model did not
imply (the truly objectionable) violation of signal locality, the
choice seemed clear. And since the early models did not mention the
possibility of nondeterministic hidden variables, it is understandable
if those readers then took Bell's theorem as definitive proof of the
failure of the project of hidden variables.

Curiously, that was quite the opposite from Bell's intent. Bell was
a supporter of the hidden variable program, and his purpose was to
show that it was not fair to reject Bohmian mechanics --- the leading
contender among the hidden variable theories --- due to its nonlocality,
since that was an unavoidable feature of any hidden variable model.

In any case, we now know that what is at stake is not just local determinism,
but the weaker concept of local causality. And by Remark 2, to even
consider the concept of a phenomenon violating local causality, one
needs to consider hidden variable theories. And even if one has other
reasons to reject hidden variables, one cannot avoid the fact that
OQT violates local causality. In other words, rejecting determinism
and/or hidden variables does not make the world an entirely local
place.

A common reason for rejecting hidden variables, Bohmian mechanics
in particular, is that this theory is deterministic but does not allow
predictability. If the hidden variables exist, why can't we \emph{know
}about them? Theorem 21, while not directly answering the question,
points out to the fact that (unless relativistic invariance is violated)
\emph{Nature} is irreducibly unpredictable, so that this problem is
not exclusive of Bohmian mechanics (just as nonlocality, Bell showed,
isn't either).

With no intention of being a supporter of hidden variables, it's interesting
to point out that there are other reasons to consider them: they are
potential solutions to the measurement problem, in that they do not
make `measurement' a fundamental feature of the world like OQT, they
make sense of quantum cosmology, and they allow an ignorance interpretation
of probabilities.

\subsection{Reality\label{sub:Reality}}

Finally, some theorists defend the idea that what violations of Bell's
inequalities tell us to give up is \emph{realism. }These people take
the term 'local realism' to be a conjuntion of 'locality' and 'realism',
and favour to reject the latter. We can identify four main views in
which such a conjunction, in our terminology, can consistently be
taken. In the first, Axioms 1 to 3 are accepted implicitly and realism
is equated with hidden variables. This view, therefore, falls under
the same criticism pointed out in \ref{sub:Determinism-and-hidden}.
In the second, Axioms 1 to 3 and the possibility of HVs are accepted
implicitly, but one equates 'realism' with outcome independence or
determinism. This would make sense if one believes that the only acceptable
hidden variable models are those that are deterministic, or at least
that satisfy outcome independence. Then one can consistently reject
'realism' in this sense and keep locality. However, the holder of
this view cannot escape the fact that local causality is still violated
by OQT. In a third possibility, Axioms 1 to 3 are accepted implicilty,
all the concepts defined through use of an ontological model are put
under the label 'realism', and 'locality' is taken to mean signal
locality. In this view, therefore, 'locality' is not violated, and
one refuses to talk about anything that goes under the name 'realism'.
This, however, cannot be more than a form of operationalism, for which
the only escape from the fact that OQT violates local causality would
be to hold that local causality is not an interesting concept.

A fourth possibility is to reject Axioms 1 or 2. However, the motivation
for that cannot be simply the desire to keep local causality, because
the MRRF had to be assumed before one can even \emph{define }local
causality. So a theorist who chooses to reject realism probably has
other motivations than simply to save local causality.

That said, there are possible moves in that direction. A popular one
is Everett's (or Many-Worlds) interpretation \cite{Everett1957},
which maintains that there is in reality a single, unitarily-evolving,
wave function for the universe (or multiverse?). Therefore in a sense
all possible outcomes of all experiments (and of all particle interactions
throughout the universe for that matter) occur. Axiom 2 is clearly
violated. Axiom 1 is not necessarily violated, since events really
occur somewhere in the multiverse independently of observers. It's
just that we don't have direct access to all of them.

Yet another possibility is to maintain a relational view of quantum
states, along the lines supported for example by Rovelli in \cite{Rovelli1996}.
In this view, quantum states are always relative to an observer or
reference frame. It is possible that different observers will disagree
about whether or not some events occur, and in a relational world
there's no matter of fact about such events independently of any observer.
Axiom 1 is clearly violated. Axiom 2 may or may not be, depending
on how one pursues the idea. It is possible that relational quantum
theories of gravity (i.e., of space-time geometry) will also reject
it. To emphasise the meaning of a violation of Axiom 1, let me be
explicit: in a fully relational theory, it is possible that \emph{the
very existence of events} are not absolute facts independent of reference
frames. This would be a startling conclusion, far beyond the mere
relativity of space and time which we have come to accept and ---
dare I say --- understand. Can \emph{reality itself} be relative?
Although this idea goes dangerously close to a radical solipsism,
a more moderate reading may be possible. In special relativity, the
fact that lengths or time lapses are relative does not entail that
they \emph{don't exist} --- they just have this previously unsuspected
property. In the same vein, to say that events are relative to reference
frames does not entail that they don't exist --- again, it is just
another surprising property. However, I would say that although it
is intriguing, whether the whole idea ultimately makes sense (and
most importantly, whether it can lead to new predictions) remains
to be seen.

\chapter{The EPR paradox and Steering\label{cha:EPR-paradox}}

As outlined in Section \ref{cha:Concepts}, in a seminal 1935 paper,
Einstein, Podolsky and Rosen (EPR) \cite{Einstein1935} demonstrated
an inconsistency between the premises that go under the name of \emph{local
realism} and the notion that quantum mechanics is complete. EPR never
regarded it as a paradox, but as an argument to prove the incompleteness
of quantum mechanics. The name `paradox' was probably introduced by
those who could not believe with EPR that quantum mechanics was indeed
incomplete but could not see a flaw in the argument either. In hindsight,
we now know (since Bell) that, while the argument is sound, one of
the premises --- local causality --- is false. However, we will retain
the historically prevalent term `paradox'%
\footnote{The American Heritage Dictionary defines `paradox' as 1. A seemingly
contradictory statement that may nonetheless be true. 2. One exhibiting
inexplicable or contradictory aspects. 3. An assertion that is essentially
self-contradictory, though based on a valid deduction from acceptable
premises. 4. A statement contrary to received opinion. Our usage of
the term is therefore in accordance with definition 3 if one takes
the premises of the EPR argument to be acceptable (in which case we
arrive at a contradiction) and with definition 4 if, instead, one
takes the argument to imply the failure of one of those a priori reasonable
premises.%
}. Our reasons to study the EPR paradox are threefold.

Firstly, we aim to do historical justice to EPR and put their argument
in their correct standing, distilling its essence and formalising
it to make it clear how it relates to the notion of local causality
as used in discussions of Bell's theorem and to the notion of \emph{entanglement}
or quantum non-separability.

Secondly, we will see that the EPR paradox can be demonstrated in
a loophole-free way with current technology, thereby providing conclusive
evidence of the failure of its underlying premises, as opposed to
the current situation with Bell inequalities, as pointed out in Chapter
\ref{cha:Introduction} on page \pageref{loopholes}.

Thirdly, we will relate the EPR paradox to the concept of \emph{steering}
originally defined by Schrödinger \cite{Schroedinger1935} in a reply
to EPR%
\footnote{In this same seminal article he coined the term \emph{entanglement
}and introduced his infamous cat paradox\emph{.}%
} and recently formalised by Wiseman and co-authors\textbf{ }\cite{Wiseman2007}\textbf{,
}confirming the latter's claim%
\footnote{While their claim is correct, their proof was incomplete. They consider
a particular instance of the EPR-Reid criteria (a more general instance
of which we will analyse in Section \ref{sec:The-EPR-Reid-criterion})
and show that for a certain class of states that criterion is violated
if and only if the state is steerable. However, the EPR-Reid criteria
are a subset of all possible criteria for the EPR paradox and their
proof only considers a specific class of states. In this thesis we
will see that the EPR paradox and steering are in fact quite generally
equivalent.%
} that any demonstration of the EPR paradox is also a demonstration
of steering. We will further show that the converse is also true:
any demonstration of steering is also a demonstration of the EPR paradox.\textbf{
}With reference to that work we will then see that the EPR paradox
(or steering) constitutes a new class of (non-)locality intermediate
between the classes of quantum (non-)separability and Bell (non-)locality.
This classification could prove important in the context of Quantum
Information Processing, therefore it is desirable to formulate criteria
to determine to which classes a given state (or a set of correlations)
belongs.

The original EPR paradox was based on position and momentum observables.
Bohm \cite{Bohm1951} extended the example of EPR to the case of discrete
observables, particularly to the case of two spin-1/2 particles. That
is the version that was used by Bell in deriving his famous inequalities
and it has played a central role in our understanding of quantum entanglement.
Both the original argument of EPR and Bohm's version, however, rely
on perfect correlations, which are obviously experimentally unattainable.
That situation was corrected by Margaret Reid in 1989 \cite{Reid1989},
when she derived experimental criteria for demonstration of the EPR
paradox which were applicable to realistic situations where noise
and losses are inevitable. The Reid criteria are a standard tool in
Quantum Optics, and have been used for demonstrations of the EPR paradox
with continuous-variables \cite{Ou1992,Zhang2000,Silberhorn2001,Bowen2003,Howell2004}
where quadrature measurements play the role of the position and momentum
observables. 

However, there is to date no such tool for the case of discrete observables.
Experiments by Wu and Shaknov \cite{Wu1950} gave evidence for discrete
EPR correlation, but because detection efficiencies were extremely
low, only a small fraction of emitted pairs were detected, meaning
that {}``no-enhancement'' assumptions \cite{Clauser1978} were incorporated. 

We will derive new criteria that can be applied to discrete observables,
both for the case originally envisaged by Bohm and to other classes
of states, even in presence of inefficiencies in the preparation or
detection procedures. The extent to which these inefficiencies are
allowed will be studied in detail. Using one of these criteria, \textcolor{black}{we
show that the loop-hole free demonstration of the EPR-Bohm paradox
is predicted for considerably lower detection efficiencies} than required
for Bell's theorem. This disparity is even more striking in the case
of \emph{macroscopic} fields, where we propose for feasible efficiencies
to demonstrate a type of EPR-Bohm correlation \textcolor{black}{from
which one can deduce existence of mesoscopic and macroscopic quantum
superpositions.}

While not enough to falsify EPR's local realism, the proposed EPR
paradox experiments do demonstrate a particularly strong form of entanglement.
For this EPR-entanglement, \emph{local} \emph{realism} can \emph{only}
be reconciled with quantum mechanics if one accepts the existence
of an underlying localised \emph{hidden variable} (non-quantum) state.
Put another way, if one can accept \emph{only} \emph{quantum} states,
then the EPR correlation implies nonlocal effects.

\section{The Einstein-Podolsky-Rosen argument\label{sec:The-Einstein-Podolsky-Rosen-argument}}

We will start with a detailed analysis of the original EPR argument,
before finding a suitable mathematical formulation of it. The EPR
paper starts with a distinction between reality and the concepts of
a theory, followed by a critique of the operationalist position, clearly
aimed at the views advocated by Bohr, Heisenberg and the other proponents
of the Copenhagen school.

\begin{quote}
\char`\"{}Any serious consideration of a physical theory must take
into account the distinction between the objective reality, which
is independent of any theory, and the physical concepts with which
the theory operates. These concepts are intended to correspond with
the objective reality, and by means of these concepts we picture this
reality to ourselves.

In attempting to judge the success of a physical theory, we may ask
ourselves two questions: (1) `Is the theory correct?' and (2) `Is
the description given by the theory complete?' It is only in the case
in which positive answers may be given to both of these questions,
that the concepts of the theory may be said to be satisfactory. The
correctness of the theory is judged by the degree of agreement between
the conclusions of the theory and human experience.\char`\"{} \cite{Einstein1935}
\end{quote}
Any theory will have some concepts which will be used to aid in the
description and prediction of the phenomena which are their subject
matter. In quantum theory, Schrödinger introduced the concept of the
wave function and Heisenberg described the same phenomena with the
more abstract matrix mechanics. EPR argue that we must distinguish
those concepts from the reality they attempt to describe. One can
see the physical constructs of the theory as mere calculational tools
if one wishes, but those authors warn that one must be careful to
avoid falling back into a pure operationalist position; the theory
must strive to furnish a complete picture of reality. 

In Chapter \ref{cha:Concepts}, we argued that a minimal realist-relativistic
framework for physical theories, and certainly the position advocated
by Einstein, can be represented by the conjunction of Axioms 1 and
2 of that chapter, namely that the existence of physical events is
independent of observers or reference frames and that those events
can be associated to points in a relativistic space-time. This framework
makes explicit, as EPR desired, that events are among those things
which are part of the \char`\"{}objective reality, which is independent
of any theory\char`\"{}. With that framework in place, we can abstract
out any specific concepts of a theory and represent the most fundamental
aspects of any description of a phenomenon by what was termed an `ontological
model' or simply `model'. EPR's requirement that the theory is correct
is built into Definition 2 of a model by requiring that it correctly
predicts the phenomenon. Their requirement that it be complete will
be addressed within the ontological model shortly. But first let us
look at the rest of EPR's argument.

EPR follow the previous considerations with a \emph{necessary condition
for completeness:}

\begin{quote}
\textbf{EPR's necessary condition for completeness: }\char`\"{}Whatever
the meaning assigned to the term \emph{complete, }the following requirement
for a complete theory seems to be a necessary one: \emph{every element
of the physical reality must have a counterpart in the physical theory}.\char`\"{}
\cite{Einstein1935}
\end{quote}
Soon afterwards they note that this condition only makes sense if
one is able to decide what are the elements of the physical reality.
Contrary to a common belief, they did not then attempt to \emph{define}
element of physical reality.\emph{ }Instead, they provide a \emph{sufficient
condition of reality}:

\begin{quote}
\textbf{EPR's sufficient condition for reality: }\char`\"{}The elements
of the physical reality cannot be determined by \emph{a priori }philosophical
considerations, but must be found by an appeal to results of experiments
and measurements. A comprehensive definition of reality is, however,
unnecessary for our purpose. We shall be satisfied with the following
criterion, which we regard as reasonable. \emph{If, without in any
way disturbing a system, we can predict with certainty (i.e., with
probability equal to unity) the value of a physical quantity, then
there exists an element of physical reality corresponding to this
physical quantity}.\char`\"{} \cite{Einstein1935}
\end{quote}
Later in the same paragraph it is made explicit that this criterion
is \char`\"{}regarded not as a necessary, but merely as a sufficient,
condition of reality\char`\"{}. This is followed by a discussion that,
in quantum mechanics, if a system is in an eigenstate of an operator
$A$ with eigenvalue $a$, by this criterion, there must be an element
of physical reality corresponding to the physical quantity $A$. \char`\"{}On
the other hand\char`\"{}, they continue, if the state of the system
is a superposition of eigenstates of $A$, \char`\"{}we can no longer
speak of the physical quantity $A$ having a particular value\char`\"{}.
After a few more considerations, they state that \char`\"{}the usual
conclusion from this in quantum mechanics is that \emph{when the momentum
of a particle is known, its coordinate has no physical reality}\char`\"{}\emph{.
}We are left therefore, according to EPR, with two alternatives:

\begin{quote}
\textbf{EPR's dilemma: }\char`\"{}From this follows that either (1)
\emph{the quantum-mechanical description of reality given by the wave
function is not complete or }(2) \emph{when the operators corresponding
to two physical quantities do not commute the two quantities cannot
have simultaneous reality.}\char`\"{} \cite{Einstein1935}
\end{quote}
They justify this by reasoning that \char`\"{}if both of them had
simultaneous reality --- and thus definite values --- these values
would enter into the complete description, according to the condition
for completeness\char`\"{}. And in the crucial step of the reasoning:
\char`\"{}If then the wave function provided such a complete description
of reality it would contain these values; \emph{these would then be
predictable\label{quo:EPR-predictable}} {[}my emphasis]. This not
being the case, we are left with the alternatives stated\char`\"{}.
Brassard and Méthot \cite{Brassard2006} have pointed out that strictly
speaking EPR should conclude that (1) \emph{or }(2), instead of \emph{either}
(1) or (2), since they could not exclude the possibility that (1)
and (2) could be both correct. However, this does not affect EPR's
conclusion. It was enough for them to show that (1) and (2) could
not both be wrong, and therefore if one can find a reason for (2)
to be false, (1) must be true%
\footnote{Brassard and Méthot's conclusion that the EPR argument is logically
unsound is not based on this mistake, which they acknowledge as irrelevant.
Their conclusion is based on a misinterpretation of EPR's paper. They
read the quote \char`\"{}In quantum mechanics it is usually assumed
that the wave function \emph{does} contain a complete description
of the physical reality {[}...]. We shall show however, that this
assumption, together with the criterion of reality given above, leads
to a contradiction\char`\"{}, as stating that $\neg(1)\wedge(2)\rightarrow false$.
If that was the correct formalisation of the argument I would agree
with their conclusion. However, by \char`\"{}criterion of reality
given above\char`\"{} EPR mean their sufficient condition of reality,
not statement $(2)$.%
}.

The next section in EPR's paper intends to find a reason for (2) to
be false, that is, to find a circumstance in which one can say that
there are simultaneous elements of reality associated to two non-commuting
operators. They consider a composite system composed of two spatially
separated subsystems $S_{A}$ and $S_{B}$ which are prepared, by
way of a suitable initial interaction, in an entangled state of the
type\begin{equation}
|\Psi\rangle=\sum_{n}c_{n}|\psi_{n}\rangle_{A}\otimes|u_{n}\rangle_{B},\label{eq:entangled1}\end{equation}
where the $|\psi_{n}\rangle_{A}$ denote a basis of eigenstates of
an operator, say $\hat{O}_{1}$, of subsystem $S_{A}$ and $|u_{n}\rangle_{B}$
denote some (normalised but not necessarily orthogonal) states of
$S_{B}$. If one measures the quantity $\hat{O}_{1}$ at $S_{A}$,
and obtains an outcome corresponding to eigenstate $|\psi_{k}\rangle_{A}$
the global state is reduced to $|\psi_{k}\rangle_{A}\otimes|u_{k}\rangle_{B}$.
If, on the other hand, one chooses to measure a non-commuting observable
$\hat{O}_{2}$, with eigenstates $|\phi_{s}\rangle_{A}$, one should
instead use the expansion\begin{equation}
|\Psi\rangle=\sum_{s}c'_{s}|\phi_{s}\rangle_{A}\otimes|v_{s}\rangle_{B},\label{eq:entangled2}\end{equation}
where $|v_{s}\rangle_{B}$ represent another set of normalised states
of $S_{B}$. Now if the outcome of this measurement is, say, the one
corresponding to $|\phi_{r}\rangle_{A}$, the global state is thereby
reduced to $|\phi_{r}\rangle_{A}\otimes|v_{r}\rangle_{B}$. Therefore,
\char`\"{}as a consequence of two different measurements performed
upon the first system, the second system may be left in states with
two different wave functions\char`\"{}. This is just what Schrödinger
later termed \emph{steering, }and we'll return to that later. Now
enters the crucial assumption of locality.

\begin{quote}
\textbf{EPR's locality assumption: }\char`\"{}Since at the time of
measurement the two systems no longer interact, no real change can
take place in the second system in consequence of anything that may
be done to the first system.\char`\"{} \cite{Einstein1935}
\end{quote}
\char`\"{}Thus\char`\"{}, conclude EPR, \char`\"{}\emph{it is possible
to assign two different wave functions to the same} \emph{reality}\char`\"{}.
They now consider a specific example where those different wave functions
are eigenstates of two non-commuting operators. If the initial state
was of type\begin{equation}
\Psi(x_{A},x_{B})=\int_{-\infty}^{\infty}e^{ix{}_{A}p/\hbar}e^{-ix{}_{B}p/\hbar}dp,\label{eq:EPRstate}\end{equation}
then if one measures momentum $\hat{p}^{A}$ at $S_{A}$ and finds
outcome $p$, the reduced state of subsystem $S_{B}$ will be the
one associated with outcome $-p$ of $\hat{p}^{B}$. On the other
hand, if one measures position $\hat{x}^{A}$ and finds outcome $x$,
the reduced state of $S_{B}$ will be the one corresponding to outcome
$x$ of $\hat{x}^{B}$. By measuring position or momentum at $S_{A}$,
one can predict with certainty the outcome of the same measurement
on $S_{B}$. But $\hat{p}^{B}$ and $\hat{x}^{B}$ correspond to non-commuting
operators. EPR conclude from this that

\begin{quote}
\char`\"{}In accordance with our criterion of reality, in the first
case we must consider the quantity {[}$\hat{p}^{B}$] as being an
element of reality, in the second case the quantity {[}$\hat{x}^{B}$]
is an element of reality. But, as we have seen, both wave functions
{[}corresponding to $-p$ and $x$] belong to the same reality.\char`\"{}
\cite{Einstein1935}
\end{quote}
In other words, by using the sufficient condition for reality, the
assumption of locality and the predictions for the entangled state
under consideration, EPR conclude that there must be elements of reality
associated to a pair of non-commuting operators. So the (2) horn of
EPR's dilemma proved before is closed, leaving as the only alternative
option (1), namely, that the quantum mechanical description of physical
reality is incomplete.

In hindsight, as we now know that the premise of locality is not entirely
justified, we can read EPR's argument as demonstrating the incompatibility
between the premises of locality, the completeness of Quantum Mechanics
and some of its predictions. However, one could block the conclusion
of the argument by rejecting those statistical predictions required
to formulate the argument. This move is particularly easy to be made
since the necessary predictions are of perfect correlations, unobtainable
in practice due to unavoidable inefficiency in preparation and detection
of real physical systems. This problem was considered by W. H. Furry
already in 1936 \cite{Furry1936} but experimentally useful criteria
for the EPR paradox were only proposed in 1989 by Margaret Reid \cite{Reid1989}.

\section{The EPR-Reid criterion\label{sec:The-EPR-Reid-criterion}}

The essential difference in the derivation of the EPR-Reid criteria
\cite{Reid1989} and the original EPR argument is in a modification
of the sufficient condition for reality. This could be stated as the
following:

\begin{verse}
\textbf{Reid's sufficient condition of reality:} If, without in any
way disturbing a system, we can predict with \emph{some specified
uncertainty} the value of a physical quantity, then there exists a
\emph{probabilistic} element of physical reality which determines
this physical quantity with at most that specific uncertainty.
\end{verse}
\begin{figure}
\begin{centering}
\includegraphics[width=13cm]{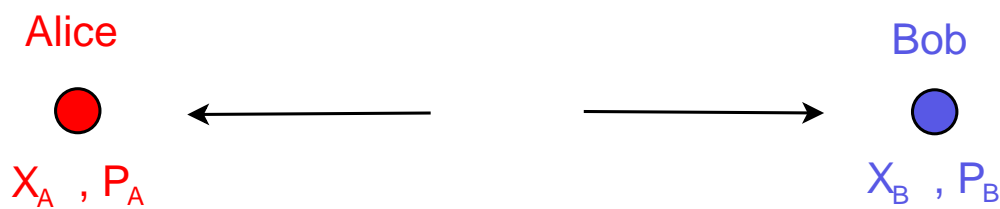}
\par\end{centering}

\caption{\label{fig:EPR} The EPR scenario. Alice and Bob are two spatially
separated observers who can perform one of two (position or momentum)
measurements available to each of them.}

\end{figure}
The scenario considered is the same as the one for the EPR paradox
above, as depicted in Fig. \ref{fig:EPR}, but one does not need a
state which predicts the perfect correlations considered by EPR. Instead,
the two experimenters, Alice and Bob, can measure the conditional
probabilities of Bob finding outcome $x_{B}$ in a measurement of
$\hat{x}_{B}$ given that Alice finds outcome $x_{A}$ in a measurement
of $\hat{x}_{A}$, i.e., $P(x_{B}|x_{A})$. Similarly they can measure
the conditional probabilities $P(p_{B}|p_{A})$ and the unconditional
probabilities $P(x_{A})$, $P(p_{A})$. We denote by $\Delta^{2}(x^{B}|x^{A})$,
$\Delta^{2}(p^{B}|p^{A})$ the variances of the conditional distributions
$P(x_{B}|x_{A})$, $P(p_{B}|p_{A})$, respectively. Reid now defines
the \emph{average inference variances\begin{eqnarray}
\Delta_{inf}^{2}(x^{B}|\hat{x}^{A}) & = & \sum_{x^{A}}P(x^{A})\Delta^{2}(x^{B}|x^{A})\nonumber \\
\Delta_{inf}^{2}(p^{B}|\hat{p}^{A}) & = & \sum_{p^{A}}P(p^{A})\Delta^{2}(p^{B}|p^{A}).\label{eq:aveinfvar}\end{eqnarray}
}The notation $\Delta_{inf}^{2}(x^{B}|\hat{x}^{A})$ is to indicate
the average inference variance of $x^{B}$ given that Alice measures
$\hat{x}^{A}$ and similarly for $p^{B}$. Reid argues, by use of
the sufficient condition of reality above, that since Alice can, by
measuring either position $\hat{x}^{A}$or momentum $\hat{p}^{B}$,
infer with some uncertainty $\Delta_{inf}(x^{B}|\hat{x}^{A})$ or
$\Delta_{inf}(p^{B}|\hat{p}^{A})$ the outcomes of the corresponding
experiments performed by Bob, and since by the locality condition
of EPR her choice cannot affect the elements of reality of Bob, then
there must be simultaneous probabilistic elements of reality which
determine $\hat{x}^{B}$ and $\hat{p}^{B}$ with at most those uncertainties.
Now by Heisenberg's Uncertainty Principle (HUP), quantum mechanics
imposes a limit to the precision with which one can assign values
to observables corresponding to non-commuting operators $\hat{x}$
and $\hat{p}$. In appropriately rescaled units the relevant HUP reads
$\Delta x\Delta p\geq1$. Therefore, if quantum mechanics is complete
as defined by EPR and the locality condition holds, by use of the
adapted sufficient condition of reality, the limit with which one
could determine the average inference variances above is\begin{equation}
\Delta_{inf}(x^{B}|\hat{x}^{A})\Delta_{inf}(p^{B}|\hat{p}^{A})\geq1.\label{eq:EPR-Reid}\end{equation}
This is the \emph{EPR-Reid criterion}. Violation of that criterion
signifies the EPR paradox. It has been used in experimental demonstration
with continuous-variables \cite{Ou1992,Zhang2000,Silberhorn2001,Bowen2003,Howell2004}
where quadrature measurements play the role of the position and momentum
observables.

\section{Formalising EPR}

I will now propose a mathematical formalisation of the premises of
the EPR argument following the formalism of Chapter \ref{cha:Concepts}.

It is evident that EPR had well in mind the basic Axioms of the MRRF
defined on page \pageref{ver:MRRF}. We have already argued in \ref{sec:The-Einstein-Podolsky-Rosen-argument}
that the first of EPR's desiderata, \char`\"{}Is the theory correct?\char`\"{},
is taken care of automatically by the definition of a `model', since
it is required that it correctly predicts the phenomenon under study.
So we can now use the ontological model to formalise the rest of their
premises.

\subsection{The original argument}

Let us first understand what EPR's sufficient condition of reality
amounts to. It is a criterion which, when satisfied, assigns a physical
variable to the set of variables which have an \char`\"{}element of
reality\char`\"{} associated to them. Let us denote this set by $\mathcal{ER}$.
The setup of the ontological model is essentially the same as that
considered by EPR, so by EPR's sufficient condition of reality, if
after the measurement of an observable $a$ in which outcome $A$
is obtained, one deduces that the probability of now obtaining \textbf{$B$
}in a measurement of $b$ is unity, then there's an element of reality
associated to $b$.

\begin{verse}
\textbf{Definition 1 (EPR's sufficient condition of reality) }\emph{If
the outcome of measurement $b$ is predictable given the outcome of
measurement $a$, then there's an element of reality associated to
$b$, i.e.,}\begin{equation}
P(B|A,a,b,c)\in\{0,1\}\rightarrow b\in\mathcal{ER}.\label{eq:EPRreality}\end{equation}

\end{verse}
The necessary condition for completeness is not so transparent. It
is a priori not obvious what EPR had in mind by \char`\"{}a counterpart
in the physical theory\char`\"{}. However, we can extract the meaning
by looking at when they actually use that condition. It is just in
the justification of what I've called EPR's dilemma, already mentioned
in Section \ref{sec:The-Einstein-Podolsky-Rosen-argument} \vpageref{quo:EPR-predictable}:
\char`\"{}If then the wave function provided such a complete description
of reality it would contain these values; \emph{these would then be
predictable}\char`\"{} {[}my emphasis]. In other words, it is necessary,
for EPR, that the variables corresponding to elements of reality be
predictable by the wave function if it provides a complete description
of reality. The wave function is fully specified by the preparation
procedure, so denoting, following EPR, $C(OQT)\equiv$\char`\"{}\emph{the
quantum-mechanical description of reality given by the wave function
is complete}\char`\"{} we obtain

\begin{verse}
\textbf{Definition 2 (EPR's necessary condition for completeness)
}\emph{If the quantum mechanical description of reality given by the
wave function is complete, then if there's an element of reality associated
to $b$, the value $B$ of $b$ given only the preparation procedure
must be predictable, i.e.}\begin{equation}
C(OQT)\rightarrow(b\in\mathcal{ER}\rightarrow P(B|b,c)\in\{0,1\}).\label{eq:EPRcompleteness}\end{equation}

\end{verse}
We can now derive EPR's dilemma. Assume $C(OQT)$ and take two non-commuting
observables \textbf{$b$ }and $b'$. If there's an element of reality
associated with both of them, then by Definition 2 they must both
be predictable. But we know by the HUP that no quantum model involving
non-commuting operators is predictable. Therefore there cannot be
elements of reality associated to both of two non-commuting operators.
Denote this last statement $NC$. Then $C(OQT)\rightarrow NC$, which
is logically equivalent to $\neg C(OQT)\vee NC$. EPR's dilemma can
now be stated as

\begin{verse}
\textbf{Theorem 1 (EPR's dilemma) }\emph{If the quantum mechanical
description of reality given by the wave function is complete, then
there cannot be elements of reality associated to both of two non-commuting
operators.}
\end{verse}
Now enters the example of the entangled state. With that state, one
can find observables $b$, \textbf{$b'$}, $a$, $a'$ such that $P(B|A,a,b,c)\in\{0,1\}$
and that $P(B'|A',a',b',c)\in\{0,1\}$. Let us denote this observation
$ENT$. Given $ENT$ and Definition 1, there must be elements of reality
associated to both $b$ and $b'$, i.e., $b,b'\in\mathcal{ER}$. This
is where the locality assumption enters the reasoning in a fundamental
way, by requiring that the elements of reality at $B$ do not depend
on the choice of experiment at $A$. Definition 1 used that implicitly
by not mentioning that one should actually carry out the measurements
under consideration, only that one \emph{would }be able to predict
the outcome of $b$ (or $b')$ \emph{if} one decided to measure $a$
(or $a'$). Close to the end of their paper, EPR remark that

\begin{quote}
\char`\"{}One could object to this conclusion on the grounds that
our criterion of reality is not sufficiently restrictive. Indeed,
one would not arrive at our conclusion if one insisted that two or
more physical quantities can be regarded as simultaneous elements
of reality \emph{only when they can be simultaneously measured or
predicted. }{[}...] This makes the reality of $P$ and $X$ depend
upon the process of measurement carried out on the first system, which
does not disturb the second system in any way. No reasonable definition
of reality could be expected to permit this.\char`\"{} \cite{Einstein1935}
\end{quote}
Therefore given the observation of $ENT$ and Definition 1, there
must be elements of reality associated to both non-commuting operators,
and the consequent of Theorem 1 is false. Therefore the antecedent
must be false, and EPR conclude, quantum mechanics is incomplete.

\begin{verse}
\textbf{Theorem 2 (the EPR argument) }\emph{The sufficient condition
of reality and some predictions of quantum mechanics imply that the
quantum mechanical description of reality is incomplete.}
\end{verse}

\subsection{General formalisation}

We can understand what kind of locality EPR had in mind if we revisit
their condition for completeness. In Definition 2 we see that if there's
an element of reality associated to $b$, and quantum mechanics is
complete, then the outcome of $b$ given the quantum state must be
predictable. They implicitly mean that if there's an element of reality
associated to a variable, then if one had the knowledge of enough
facts about the system one would be able to predict the outcome of
a measurement of that observable with certainty. In the language of
the ontological model of Chapter \ref{cha:Concepts}, this means that
there must be a deterministic hidden variable model for the outcome
of $b$. With this consideration Definition 1 implies (I remind the
reader that the expressions are to be understood as valid for all
values of the variables, including $\lambda$)\begin{equation}
P(B|A,a,b,c)\in\{0,1\}\rightarrow b\in\mathcal{ER}\rightarrow P(B|b,c,\lambda)\in\{0,1\},\label{eq:EPRreality2}\end{equation}
 Now it is obvious that \begin{equation}
P(B|A,a,b,c)\in\{0,1\}\rightarrow P(B|A,a,b,c,\lambda)\in\{0,1\},\label{eq:HVobvious}\end{equation}
and to obtain \eqref{eq:EPRreality2} from \eqref{eq:HVobvious} we
just need the assumption of local causality as defined in Chapter
\ref{cha:Concepts}, which we reproduce here for completeness:

\begin{verse}
\textbf{Corollary 1:}\emph{ A model is }\textbf{locally causal}\emph{,
i.e., a model satisfies }\textbf{local causality }\emph{(LC) iff}\begin{equation}
P(B|A,a,b,c,\lambda)=P(B|b,c,\lambda),\label{eq:LC-2}\end{equation}
\emph{plus the corresponding equations for $A$.}
\end{verse}
So the locality concept that goes into Definition 1 is that of local
causality. Remember that this implies that the joint probabilities
factorise, i.e., that any model that satisfies local causality must
have the form \begin{equation}
f(A,B|a,b,c)=\sum_{\lambda\in\Lambda}\, P(\lambda|c)\, P(A|a,c,\lambda)P(B|b,c,\lambda).\label{eq:factorize}\end{equation}

What about completeness? If Orthodox Quantum Theory is complete, then
for pure states it is essentially the assumption that there are no
hidden variables. No possible new information could change the probability
assigned to the outcomes of measurements. Since hidden variables could
be unknowable by any observer even in principle, such as is the case
in Bohmian mechanics, a more correct statement would be that no actual
variables exist in the past light cones of the measurements under
study such that, conditionalised on those variables, the probabilities
of the outcomes of said measurements would be further specified. Formally,
that means that if OQT is complete, then the probabilities of Bob's
outcome, say, can only be those allowed by quantum states $P_{Q}(B|b,c,\lambda)=Tr[\hat{\Pi}_{B}\rho_{c,\lambda}^{B}].$
I use the subscript $Q$ to indicate quantum probabilities, where
\emph{$\hat{\Pi}_{B}$} is a projector onto the subspace corresponding
to outcome $B$ of observable $\hat{b}$ while $\rho_{c,\lambda}^{B}$
is a positive unit-trace operator for subsystem $B$ associated with
the preparation\emph{ $c$ }and with the unknown variable $\lambda$\emph{}%
\footnote{Remember that $\lambda$ are not necessarily \char`\"{}hidden\char`\"{}
variables in the sense of being fundamentally unknowable, but rather
are variables which are ignored by the preparation procedure. They
may be actually unknowable, such as the hidden variables in Bohmian
mechanics, but they could be just ignored but knowable variables.
For example, if one prepares a mixed state, there could be in principle
knowable further variables that would specify a pure quantum state,
and those are included in $\lambda.$%
}\emph{. }This is a strong constraint, since by Heisenberg's Uncertainty
Principle, not all probabilities are allowed to be associated simultaneously
to two non-commuting operators $\hat{b}$ and $\hat{b}'.$ By those
considerations, we arrive at the first main result of this Chapter.

\begin{verse}
\textbf{Proposition 1: }\emph{The conjunction of local causality and
completeness implies }\begin{equation}
f(A,B|a,b,c)=\sum_{\lambda\in\Lambda}\, P(\lambda|c)\, P_{Q}(A|a,c,\lambda)P_{Q}(B|b,c,\lambda).\label{eq:LC+COMP}\end{equation}

\end{verse}
This is equivalent to the statement that there exists a separable
(i.e., non-entangled) state which correctly describes the phenomenon.
An entangled, or non-separable \cite{Werner1989} quantum state $\rho$
for two subsystems A and B, is one which cannot be written as a convex
combination of product states, i.e., as\[
\rho=\sum_{i}\eta_{i}\rho_{i}^{A}\otimes\rho_{i}^{B},\]
where the $\eta_{i}$ are probabilities, i.e., they are real non-negative
numbers and $\sum_{i}\eta_{i}=1,$ and $\rho_{i}^{A}$ and $\rho_{i}^{B}$
are quantum states for subsystems A and B respectively. The consequence
has not been widely recognised before and deserves to be emphasised. 

\begin{quote}
\textbf{Any proof that a phenomenon cannot be described by a separable
state is proof that the phenomenon violates the conjunction of local
causality and completeness of operational quantum theory.}
\end{quote}
That is, the conjunction of local causality and completeness of quantum
theory, the basic premises of the EPR argument, can be shown to be
inconsistent with observation \emph{without the need for EPR's sufficient
condition of reality. }The reason is that \eqref{eq:LC+COMP} is the
most general model which is compatible with both of these premises,
allowing for the most general distribution of local elements of reality
which are compatible with the assumption that operational quantum
theory is complete. 

However, the EPR argument was asymmetric in that it considered the
effect of knowledge of the outcomes of measurements on one side of
the apparatus, say Alice's, on the state on the other side, say Bob's.
Schrödinger called this effect \emph{steering}, and that arguably
reflects the spirit of the EPR paradox more closely than the general
nonseparability of \eqref{eq:LC+COMP}. We will therefore use the
term \emph{EPR paradox }to refer to this particular kind of violation
of local causality and completeness, and not the general violation
of \pageref{eq:LC+COMP}. This is the topic of the next subsection.

\begin{figure}
\begin{centering}
\includegraphics[width=13cm]{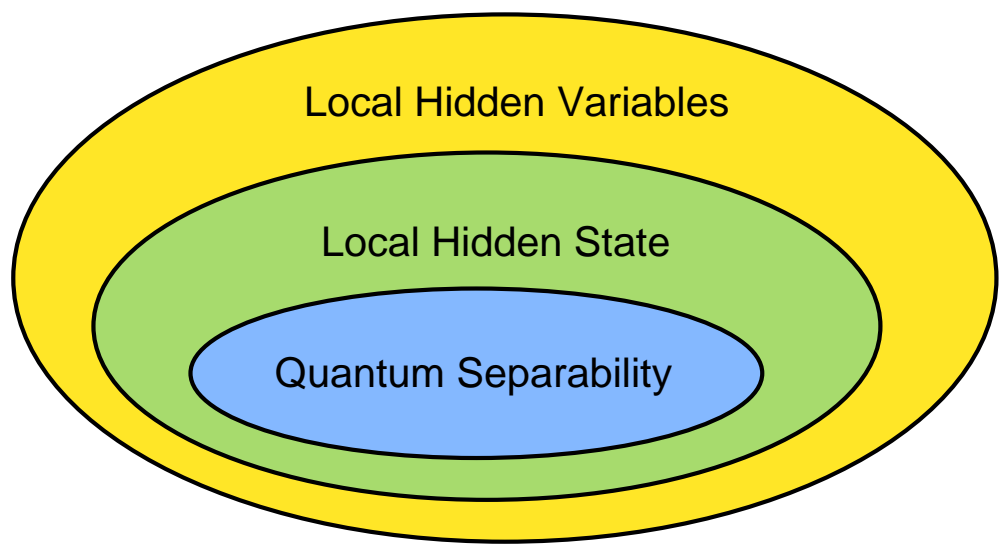}
\par\end{centering}

\caption{\label{fig:hierarchy} The hierarchy of locality models. The diagram
depicts the 'state space', that is, the space of possible quantum
states. If a certain state is separable then all measurements that
can be performed on it have a LHS model and if it has a LHS model
then it will have a LHV model, but the converse implications do not
hold.}

\end{figure}

\subsection{The connection with Steering}

Suppose we assume completeness of OQT for Bob's subsystem, but do
not restrict the hidden variables in Alice's side in any way. That
implies 

\textbf{Proposition 2: }\emph{The conjunction of local causality and
completeness for Bob's subsystem implies }\begin{equation}
f(A,B|a,b,c)=\sum_{\lambda\in\Lambda}\, P(\lambda|c)\, P(A|a,c,\lambda)P_{Q}(B|b,c,\lambda).\label{eq:LC+COMP_Bob}\end{equation}
This was termed a \emph{Local Hidden State (LHS) model }by Wiseman
and co-authors \cite{Wiseman2007}, and was proposed as a formalisation
of the concept of steering, which can best be described in the quantum
information fashion, that is, as a \emph{task}.

The task is the following. Alice wants to convince Bob that she can
act on his state, or \emph{steer} it, at a distance. If Bob does not
trust Alice and believes he has a local quantum state, he will believe
that the measurement statistics of their experimental outcomes will
be given by \eqref{eq:LC+COMP_Bob}, since that is the most general
way in which his local quantum state can be classically correlated
with Alice's measurement outcomes. Therefore if by looking at Alice's
and his own joint measurement probabilities, Bob cannot find a model
of form \eqref{eq:LC+COMP_Bob}, he'll be convinced that Alice's choice
of experiment can somehow steer his own state.

In that letter, Wiseman \emph{et al. }answered a few questions about
steering. They have demonstrated that, as applied to states, the concept
of Bell non-locality, or the inexistence of a Local Hidden Variable
(LHV) model, is strictly stronger than steerability and steerability
is strictly stronger than non-separability, that is, that if a quantum
state demonstrates Bell non-locality then it necessarily demonstrates
steering but not vice-versa, and if it demonstrates steering it necessarily
demonstrates non-separability, but not vice-versa. This hierarchy
is schematised in Fig. \ref{fig:hierarchy}.

Here we will be interested in a different question. Our purpose is
to experimentally demonstrate steering, which by the preceding arguments
represent a demonstration of the EPR paradox. When do the experimental
outcomes violate \eqref{eq:LC+COMP_Bob}? We want to find clear experimental
signatures, in the form of inequalities, which, when violated, imply
steering or the EPR paradox. We'll first show that the LHS model directly
imply the EPR-Reid criterion for continuous-variables, and then we'll
derive new criteria for the EPR-Bohm paradox.

\subsection{The EPR-Reid criterion re-derived}

It is easy to show that for the original EPR state, the measurement
outcomes cannot be described in the form \eqref{eq:LC+COMP_Bob},
To see that, note that for all $\lambda$, Alice can choose to measure
$a$ or $a'$ and thereby predict with certainty either $b$ or $b'.$
But the value of $\lambda$ is independent of Alice's choice. So for
all $\lambda$, the outcomes of $b$ and $b'$ must be determined,
that is, it must be simultaneously the case that $P_{Q}(B|b,c,\lambda)\in\{0,1\}$
and $P_{Q}(B|b',c,\lambda)\in\{0,1\}$. But no quantum state allows
that probability assignment, and therefore the LHS model cannot describe
the joint statistics.

But in the laboratory there are no such perfect correlations. What
about the more general case of imperfect detection and preparation
efficiencies? And what about other states which do not predict such
correlations even with perfect efficiencies?

Recall that we defined, for the situation considered in Section \ref{sec:The-EPR-Reid-criterion},
average inference variances as $\Delta_{inf}^{2}(x^{B}|\hat{x}^{A})=\sum_{x^{A}}P(x^{A})\Delta^{2}(x^{B}|x^{A}),\:\Delta_{inf}^{2}(p^{B}|\hat{p}^{A})=\sum_{p^{A}}P(p^{A})\Delta^{2}(p^{B}|p^{A}).$
Consider the LHS model of \eqref{eq:LC+COMP_Bob}, which applied to
the situation at hand reads (omitting henceforth unnecessary notation)\begin{equation}
P(x^{A},x^{B})=\sum_{\lambda\in\Lambda}\, P(\lambda)\, P(x^{A}|\lambda)P_{Q}(x^{B}|\lambda).\label{eq:LHS P(xA,xB)}\end{equation}
The conditional probability distribution of $x^{B}$ given an outcome
$x^{A}$ of $\hat{x}^{A}$ is then\begin{equation}
P(x^{B}|x^{A})=\sum_{\lambda\in\Lambda}\frac{P(\lambda)\, P(x^{A}|\lambda)}{P(x^{A})}P_{Q}(x^{B}|\lambda).\label{eq:LHS P(xB|xA)}\end{equation}
It is a general result that if a probability distribution can be written
in the form of a convex combination of further normalised probability
distributions, \begin{equation}
P(z)=\sum_{i}P(i)P(z|i),\label{eq:aveconvex}\end{equation}
then the variance of the resulting combination is larger than or equal
to the combination of the variances, i.e., \begin{equation}
\Delta^{2}z\geq\sum_{i}P(i)\Delta^{2}(z|i).\label{eq:varconvex}\end{equation}
Eq. \eqref{eq:LC+COMP_Bob} has the form of \eqref{eq:aveconvex},
so the variance $\Delta^{2}(x^{B}|x^{A})$ of the conditional distribution
\eqref{eq:LHS P(xB|xA)} must satisfy the inequality \begin{equation}
\Delta^{2}(x^{B}|x^{A})\geq\sum_{\lambda}\frac{P(\lambda)\, P(x^{A}|\lambda)}{P(x^{A})}\Delta_{Q}^{2}(x^{B}|\lambda),\label{eq:EPR-Reid1}\end{equation}
where $\Delta_{Q}^{2}(x^{B}|\lambda)$ represents the variance of
$P_{Q}(x^{B}|\lambda).$ Substituting \eqref{eq:EPR-Reid1} into the
definition of $\Delta_{inf}^{2}(x^{B}|\hat{x}^{A})$ we obtain \begin{eqnarray}
\Delta_{inf}^{2}(x^{B}|\hat{x}^{A}) & \geq & \sum_{\lambda,x^{A}}P(x^{A})\frac{P(\lambda)\, P(x^{A}|\lambda)}{P(x^{A})}\Delta_{Q}^{2}(x^{B}|\lambda)\nonumber \\
 & = & \sum_{\lambda}P(\lambda)\Delta_{Q}^{2}(x^{B}|\lambda).\label{eq:infineq1}\end{eqnarray}
A similar procedure produces the equivalent inequality for $\Delta_{inf}^{2}(p^{B}|\hat{p}^{A})$.
We now define two vectors \begin{eqnarray}
u & = & (\sqrt{P(\lambda_{1})}\Delta_{Q}(x^{B}|\lambda_{1}),\,\sqrt{P(\lambda_{2})}\Delta_{Q}(x^{B}|\lambda_{2}),...)\nonumber \\
v & = & (\sqrt{P(\lambda_{1})}\Delta_{Q}(p^{B}|\lambda_{1}),\,\sqrt{P(\lambda_{2})}\Delta_{Q}(p^{B}|\lambda_{2}),...),\label{eq:twovectors}\end{eqnarray}
and note that $\sum_{\lambda}P(\lambda)\Delta_{Q}^{2}(x^{B}|\lambda)=|u|^{2}$
and similarly for $p^{B}$ and $v$. By use of the Cauchy-Schwarz
inequality, i.e., $|u||v|\geq u\cdot v$ we obtain\begin{eqnarray}
\Delta_{inf}^{2}(x^{B}|\hat{x}^{A})\Delta_{inf}^{2}(p^{B}|\hat{p}^{A}) & \geq & \sum_{\lambda}P(\lambda)\Delta_{Q}^{2}(x^{B}|\lambda)\sum_{\lambda'}P(\lambda')\Delta_{Q}^{2}(p^{B}|\lambda')\nonumber \\
 & = & |u|^{2}|v|^{2}\geq(u\cdot v)^{2}\nonumber \\
 & = & \left\{ \sum_{\lambda}P(\lambda)\Delta_{Q}(x^{B}|\lambda)\Delta_{Q}(p^{B}|\lambda)\right\} ^{2}.\label{eq:EPR-Reid2}\end{eqnarray}
Now remember that $\Delta_{Q}(x^{B}|\lambda)$ and $\Delta_{Q}(p^{B}|\lambda)$
represent standard deviations over the \emph{same} quantum state $\rho_{\lambda}^{B}.$
Therefore they must obey the HUP\begin{equation}
\Delta_{Q}(x^{B}|\lambda)\Delta_{Q}(p^{B}|\lambda)\geq1,\label{eq:EPR-ReidHUP}\end{equation}
and by substituting \eqref{eq:EPR-ReidHUP} in \eqref{eq:EPR-Reid2}
we finally arrive at the EPR-Reid criterion\begin{equation}
\Delta_{inf}(x^{B}|\hat{x}^{A})\Delta_{inf}(p^{B}|\hat{p}^{A})\geq1.\label{eq:EPR-Reidrederived}\end{equation}
So we see that we can derive the EPR-Reid criterion directly from
the premises of local causality and completeness, which imply the
LHS model. No additional condition of reality is necessary.

\section{Criteria for the EPR-Bohm paradox}

In Bohm's EPR paradox, spin measurements $J_{\theta}^{A}$ and $J_{\phi}^{B}$
are performed, simultaneously, on two spatially separated subsystems,
$A$ and $B$. For the quantum states\begin{equation}
|\psi_{j}\rangle=\frac{1}{\sqrt{2j+1}}\sum_{m=-j}^{j}(-1)^{j-m}|j,m\rangle_{A}|j,-m\rangle_{B}\label{eq:bohmstate}\end{equation}
there is a maximum correlation between the results if the same spin
component is measured at each location. \textcolor{black}{Here $|j,m\rangle_{A/B}$
are the eigenstates of $J^{2}$ and $J_{z}$ respectively, for $A/B$.
Opposite outcomes are predicted for $J_{\theta}^{A}$ and0 $J_{\theta}^{B}$,
so that the outcome of measurement of any one of $J_{x}^{B}$, $J_{y}^{B}$,
$J_{z}^{B}$ can be predicted, with absolute certainty, by measurement
of one of the $J_{x}^{A}$, $J_{y}^{A}$, $J_{z}^{A}$.} Bohm's original
variant \cite{Bohm1951} of the EPR argument considered only the Bell-state
$|\psi_{1/2}\rangle.$

\begin{figure}
\begin{centering}
\includegraphics[width=13cm]{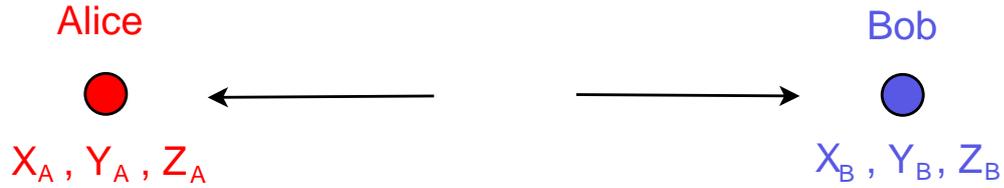}
\par\end{centering}

\caption{\label{fig:EPRBohm} The EPR-Bohm scenario. Alice and Bob are two
spatially separated observers who can perform one of three spin measurements
available to each of them. }

\end{figure}

The EPR-Bohm argument follows analogously to the EPR argument analysed
before in this Chapter, except that instead of being based in the
uncertainty principle for position-momentum, it is based on that for
the spin observables, as depicted in Fig. \ref{fig:EPRBohm}.

One of the uncertainty relations that impose lower bounds on the uncertainties
for spin observables is of the form \begin{equation}
\Delta j_{x}^{B}\Delta j_{y}^{B}\geq|\langle j_{z}^{B}\rangle|/2.\label{eq:hupspin}\end{equation}
 The lower case $j_{\theta}$ indicate the outcomes of measurements
$J_{\theta}$.

To derive experimental criteria for the EPR-Bohm paradox, we define
the \emph{measurable} average inference variance in the \emph{prediction}
of $J_{\phi}^{B}$, based on measurement $J_{\theta}^{A}$ at $A$
\cite{Reid1989}.\begin{equation}
{\normalcolor \Delta_{est}^{2}j_{\phi}^{B}=\langle(j_{\phi}^{B}-j_{\phi|\theta,est}^{B})^{2}\rangle\geq\sum_{j_{\theta}^{A}}P(j_{\theta}^{A})\Delta^{2}(j_{\phi}^{B}|j_{\theta}^{A})=\Delta_{inf}^{2}(j_{\theta}^{B}|J_{\phi}^{A})}\label{eq:est}\end{equation}
Here $j_{\phi|\theta,est}^{B}$ is an inferred estimate for $j_{\phi}^{B}$,
given an outcome $j_{\theta}^{A}$ for{\large{} }$J_{\theta}^{A}$,
and the average is over all outcomes $j_{\phi}^{B}$, $j_{\theta}^{A}$.
The inequality follows because, for a given $j_{\theta}^{A}$, the
estimate that minimises $\langle(j_{\phi}^{B}-j_{\phi|\theta,est}^{B})^{2}\rangle$
is the mean of \textcolor{black}{$P(j_{\phi}^{B}|j_{\theta}^{A})$.}
This minimum is optimal, but not always accessible, in EPR experiments.
From now on we will derive everything for the $\Delta_{inf}^{2}(j_{\theta}^{B}|J_{\phi}^{A})$,
but one should keep in mind that one can measure the less optimal
but possibly more accessible $\Delta_{est}^{2}j_{\phi}^{B}$, modifying
the inequalities in an obvious way.

We will use a technique similar to that used to re-derive the EPR-Reid
criterion in the previous section. We first assume that the measurement
statistics can be described by a LHS model, which implies \begin{equation}
\Delta_{inf}^{2}(j_{\theta}^{B}|J_{\phi}^{A})\geq\sum_{\lambda}P(\lambda)\Delta_{Q}^{2}(j_{\theta}^{B}|\lambda),\label{eq:infineq1spin}\end{equation}
for all $\theta$, $\phi$, by an entirely analogous reasoning as
in the derivation of \eqref{eq:infineq1}.

We now again define two vectors \begin{eqnarray}
u & = & (\sqrt{P(\lambda_{1})}\Delta_{Q}(j_{x}^{B}|\lambda_{1}),\,\sqrt{P(\lambda_{2})}\Delta_{Q}(j_{x}^{B}|\lambda_{2}),...)\nonumber \\
v & = & (\sqrt{P(\lambda_{1})}\Delta_{Q}(j_{y}^{B}|\lambda_{1}),\,\sqrt{P(\lambda_{2})}\Delta_{Q}(j_{y}^{B}|\lambda_{2}),...),\label{eq:twovectors2}\end{eqnarray}
and by using the Cauchy-Schwarz inequality\begin{eqnarray}
\Delta_{inf}^{2}(j_{x}^{B}|J_{x}^{A})\Delta_{inf}^{2}(j_{y}^{B}|J_{y}^{A}) & \geq & \sum_{\lambda}P(\lambda)\Delta_{Q}^{2}(j_{x}^{B}|\lambda)\sum_{\lambda'}P(\lambda')\Delta_{Q}^{2}(j_{y}^{B}|\lambda')\label{eq:EPR-Bohm1}\\
 & \geq & \left\{ \sum_{\lambda}P(\lambda)\Delta_{Q}(j_{x}^{B}|\lambda)\Delta_{Q}(j_{y}^{B}|\lambda)\right\} ^{2}.\nonumber \end{eqnarray}
But in this case the relevant uncertainty relation is\begin{equation}
\Delta_{Q}(j_{x}^{B}|\lambda)\Delta_{Q}(j_{y}^{B}|\lambda)\geq\frac{1}{2}|\langle j_{z}^{B}|\lambda\rangle|,\label{eq:EPR-BohmHUP}\end{equation}
where $\langle j_{z}^{B}|\lambda\rangle$ represents the average of
$J_{z}^{B}$ over the quantum state $\rho_{\lambda}^{B}$. By substituting
\eqref{eq:EPR-BohmHUP} in \eqref{eq:EPR-Bohm1} \begin{equation}
\Delta_{inf}(j_{x}^{B}|J_{x}^{A})\Delta_{inf}(j_{y}^{B}|J_{y}^{A})\geq\frac{1}{2}\sum_{\lambda}P(\lambda)|\langle j_{z}^{B}|\lambda\rangle|.\label{eq:EPR-Bohm2}\end{equation}
Now note that further specifying the hidden variables cannot decrease
the average of the modulus of the mean, so that \begin{equation}
\sum_{\lambda}P(\lambda)|\langle j_{z}^{B}|\lambda\rangle|\geq\sum_{j_{z}^{A}}P(j_{z}^{A})|\langle j_{z}^{B}|j_{z}^{A}\rangle|\geq|\langle j_{z}^{B}\rangle|\label{eq:furtherhiddenvar}\end{equation}
 and finally we arrive at the following theorem.

\begin{verse}
\textbf{Theorem 3:}\emph{ The following inequality if violated for
spatially separated systems A and B would demonstrate Bohm's EPR paradox}\begin{equation}
\Delta_{inf}(j_{x}^{B}|J_{x}^{A})\Delta_{inf}(j_{y}^{B}|J_{y}^{A})\geq\sum_{j_{z}^{A}}P(j_{z}^{A})|\langle j_{z}^{B}|j_{z}^{A}\rangle|.\label{eq:EPR-Bohmineq1}\end{equation}

\end{verse}
We reiterate that the choice of measuring $J_{x}^{A}$, for example,
to infer $j_{x}^{B}$ was purely conventional. It will be the best
choice in a state of type \eqref{eq:bohmstate}, but nothing special
hangs on which particular quantum observable actually corresponds
to what we called $J_{x}^{A}$, $J_{y}^{A}$ $J_{z}^{A}$. In a practical
situation one should choose whichever observables optimise the violation
of \eqref{eq:EPR-Bohmineq1}. The observables at $B$, however, must
actually correspond to three mutually orthogonal directions, as the
uncertainty principle used is only valid in that situation.

We point out that, from (\ref{eq:furtherhiddenvar}), $\Delta_{inf}(j_{x}^{B}|J_{x}^{A})\Delta_{inf}(j_{y}^{B}|J_{y}^{A})<|\langle j_{z}^{B}\rangle|/2$
also signifies the EPR paradox, as shown by Bowen et al \cite{Bowen2002b},
who demonstrated this inequality experimentally for Stokes operators
\cite{Korolkova2007} that represent the polarisation of optical fields.
For the states (\ref{eq:bohmstate}), $\langle j_{Z}^{B}\rangle=0$
and this latter form is not useful. 

\begin{figure}
\begin{centering}
\includegraphics[width=13cm]{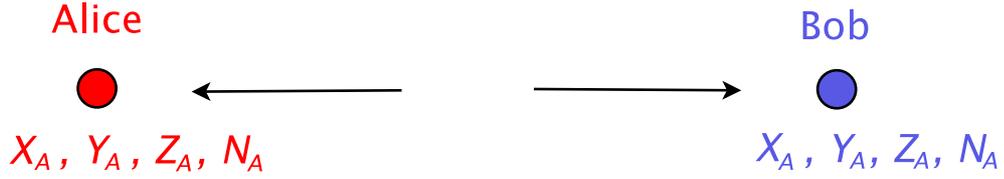}
\par\end{centering}

\caption{\label{fig:EPR-Bohm_extended}The extended EPR-Bohm scenario. Alice
and Bob are two spatially separated observers who can perform one
of four measurements (one of three \char`\"{}spin\char`\"{} components
or a total number measurement) available to each of them.}

\end{figure}

The states (\ref{eq:bohmstate}) violate (\ref{eq:EPR-Bohmineq1})
and can be investigated experimentally using parametric down conversion
\cite{Lamas-Linares2001}. We re-express (\ref{eq:bohmstate}) using
Schwinger's formalism \textcolor{black}{\begin{eqnarray}
{\normalcolor {\color{red}{\normalcolor |\psi_{j}\rangle}}} & = & \frac{1}{N!\sqrt{N+1}}(a_{+}^{\dagger}b_{-}^{\dagger}-a_{-}^{\dagger}b_{+}^{\dagger})^{N}|0\rangle,\label{eq:drum}\end{eqnarray}
where $a_{\pm}$ ar}e boson operators for orthogonally-polarised field
modes of $A$, a similar set is defined for $B$, $j=N/2$ (where
here $j$ represents the eigenvalues of $J^{2}$ as in state \eqref{eq:bohmstate})
and $|0\rangle$ is the multi-mode vacuum. \textcolor{black}{We define
the Schwinger spin operators \begin{eqnarray}
J_{x}^{A} & = & \frac{1}{2}\left(\hat{a}_{-}\hat{a}_{+}^{\dagger}+\hat{a}_{-}^{\dagger}\hat{a}_{+}\right)\nonumber \\
J_{y}^{A} & = & \frac{1}{2i}\left(\hat{a}_{-}\hat{a}_{+}^{\dagger}-\hat{a}_{-}^{\dagger}\hat{a}_{+}\right)\nonumber \\
J_{z}^{A} & = & \frac{1}{2}\left(\hat{a}_{+}^{\dagger}\hat{a}_{+}-\hat{a}_{-}^{\dagger}\hat{a}_{-}\right)\nonumber \\
N^{A} & = & \left(\hat{a}_{+}^{\dagger}\hat{a}_{+}+\hat{a}_{-}^{\dagger}\hat{a}_{-}\right).\label{eq:Schwinger}\end{eqnarray}
The situation of the EPR-Bohm setup is therefore extended with number
measurements, as in Fig. \ref{fig:EPR-Bohm_extended}. }

\textcolor{black}{Fields $a_{\pm}$ and $b_{\pm}$ are spatially separated,
and impinge on polarising beam splitters to enable a measure of the
photon number differences, $J_{z}^{A}$ and $J_{z}^{B}$, the spin
measurements $J_{x/y}^{A}$, $J_{x/y}^{B}$ being made with phase
shifts that immediately precede the polariser. }

A case of experimental interest is $j=1/2$, the Bell state, but with
detection efficiency $\eta$, so that there is a non-zero probability
($1-\eta$) of photons \emph{not} being detected. The outcome of no
detection corresponds to $0$ for both $\hat{a}_{+}^{\dagger}\hat{a}_{+}$
and $\hat{a}_{-}^{\dagger}\hat{a}_{-}$ and hence $0$ for $J_{\theta}^{A}$
(similarly for $J_{\phi}^{B}$). Calculation reveals \textcolor{black}{\begin{equation}
\Delta_{inf}^{2}(j_{x}^{B}|j_{x}^{A})=\Delta_{inf}^{2}(j_{y}^{B}|j_{y}^{A})=\eta(1-\eta^{2})/4\label{eq:EPR-Bohmcalc1}\end{equation}
and \begin{equation}
\sum_{j_{z}^{A}}P(j_{z}^{A})|\langle j_{z}^{B}|j_{z}^{A}\rangle|=\eta^{2}/2,\label{eq:EPR-Bohmcalc2}\end{equation}
so the paradoxical correlations are obtained where $\eta>0.62$.}
This is consistent with the Werner state \cite{Werner1989} \begin{equation}
\rho_{w}=(1-p_{s})\frac{\mathbb{I}}{4}+p_{s}|\psi_{1/2}\left\rangle \right\langle \psi_{1/2}|\label{eq:Werner}\end{equation}
 ($\frac{\mathbb{I}}{4}$ is the maximally mixed density operator)
which yields the EPR paradox when $p_{s}>0.62$. Entanglement occurs
when, and only when, $p_{s}>0.33$, so the EPR paradox is more difficult
to confirm \cite{Bowen2003,Wiseman2007}.

The states (\ref{eq:drum}) can be generated using two parametric
amplifiers \cite{Clauser1978} as modelled by the interaction Hamiltonian
\textcolor{black}{\begin{equation}
H=i\hbar\kappa(a_{+}^{\dagger}b_{-}^{\dagger}-a_{-}^{\dagger}b_{+}^{\dagger})-i\hbar\kappa(a_{+}b_{-}-a_{-}b_{+})\label{eq:parampham}\end{equation}
Assuming the initial state to be a vacuum, the solution after a time
$t$ is a superposition of the (\ref{eq:drum}). The predictions of
a particular $|\psi_{j}\rangle$ could be tested by restricting to
the ensemble with a fixed $N^{B}$.} Most interesting is the limit
of large $\langle N^{B}\rangle$. We therefore propose to detect the
Bohm-EPR correlation using the full solution of $H$. For macroscopic
systems, measurement of all the conditional probabilities can be difficult.
We present an alternative Bohm-EPR criterion.

\begin{verse}
\textbf{Theorem} \textbf{4:} \emph{Violation of the following inequality
for spatially separated $A$ and $B$ reveals an EPR paradox}

\begin{equation}
\Delta_{inf}^{2}(j_{x}^{B}|J_{x}^{A})+\Delta_{inf}^{2}(j_{y}^{B}|J_{y}^{A})+\Delta_{inf}^{2}(j_{z}^{B}|J_{z}^{B})\geq\frac{\langle N^{B}\rangle}{2}\label{eq:bohmlocalsumcrit}\end{equation}

\end{verse}
The proof is analogous to that of \eqref{eq:EPR-Bohmineq1}, except
that we follow \cite{Toth2004,Hofmann2003a} to write a quantum uncertainty
relation \begin{equation}
\Delta^{2}j_{x}^{B}+\Delta^{2}j_{y}^{B}+\Delta^{2}j_{z}^{B}\geq\Delta^{2}N^{B}/4+\langle N^{B}\rangle/2.\label{eq:HUPHofmann}\end{equation}

\begin{figure}
\begin{centering}
\includegraphics[width=13cm]{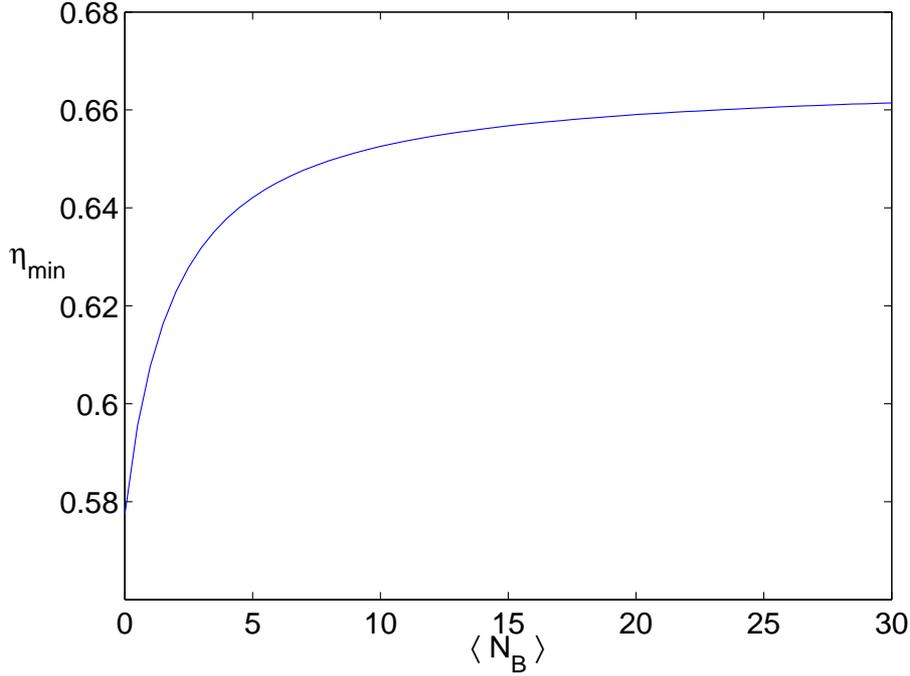}
\par\end{centering}

\caption{\textcolor{black}{Minimum detection efficiency for violation of (\ref{eq:bohmlocalsumcrit})
by (\ref{eq:parampham}) versus mean photon number.}}
\label{fig:EPR-Bohm1}
\end{figure}

Inequality (\ref{eq:bohmlocalsumcrit}) is tested for (\ref{eq:parampham}),
\textcolor{black}{with detection efficiency $\eta$. The $\Delta_{est}^{2}j_{\theta}^{B}$
of \eqref{eq:est} is defined for the linear estimate \cite{Reid1989}
$j_{\theta,est}^{B}=gj_{\theta}^{A}$, where $g=-\langle j_{\theta}^{B}j_{\theta}^{A}\rangle/\langle j_{\theta}^{A}j_{\theta}^{A}\rangle$
to minimise $\Delta_{est}^{2}j_{\theta}^{B}$, giving solutions \begin{equation}
\Delta_{est}^{2}j_{\theta}^{B}=\langle(j_{\theta}^{B})^{2}\rangle-\langle j_{\theta}^{B}j_{\theta}^{A}\rangle^{2}/\langle(j_{\theta}^{A})^{2}\rangle=\eta sinh^{2}r(1-\eta^{2}+2\eta(1-\eta)sinh^{2}r)/2(1+\eta sinh^{2}r)\label{eq:EPR-BOhmcalc3}\end{equation}
 and \begin{equation}
\langle N^{B}\rangle=2\eta sinh^{2}r,\label{eq:EPR-Bohmcalc4}\end{equation}
 where $r=|\kappa|t$. Fig. \ref{fig:EPR-Bohm1} plots the minimum
efficiency $\eta$ required for violation of (\ref{eq:bohmlocalsumcrit}),
to indicate a test of macroscopic EPR for large $\langle N^{B}\rangle$
and $\eta>0.66$.}

I should clarify that inequalities \eqref{eq:EPR-Bohmineq1} and \eqref{eq:bohmlocalsumcrit}
of Theorems 3 and 4 are both completely general and valid for arbitrary
detector efficiencies, and the explicit introduction of the efficiencies
$\eta$ in the above discussions are to model what would happen in
an experimental situation. All that one needs for their experimental
evaluation are the inferred variances (or their estimates), defined
in \eqref{eq:est}, and either the probability distributions $P(j_{z}^{A})$
and conditional averages $\langle j_{z}^{B}|j_{z}^{A}\rangle$ for
\eqref{eq:EPR-Bohmineq1} or the average $\langle N^{B}\rangle$ and
variance $\Delta^{2}N^{B}$ for \eqref{eq:bohmlocalsumcrit}. For
the latter, a further clarification is needed that the inequality
is derived for arbitrary number distributions.

\begin{figure}
\begin{centering}
\includegraphics[width=13cm]{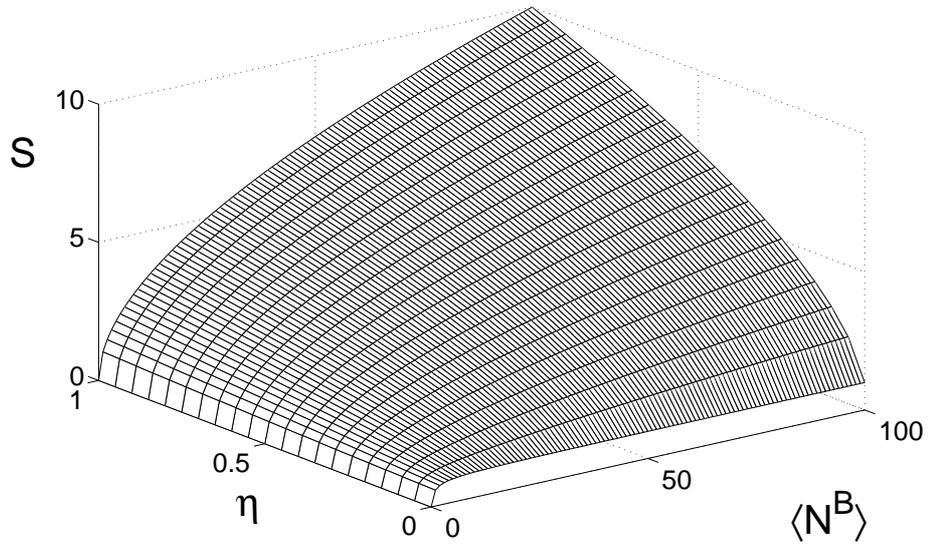}
\par\end{centering}

\caption{\textcolor{black}{Minimum size $S$ of superpositions that may be
inferred from violation of \eqref{eq:macro} with the Hamiltonian
of \eqref{eq:parampham}versus mean photon number.}}
\label{fig:EPR-Bohm2}
\end{figure}

The existence of macroscopic quantum superpositions \cite{Cavalcanti2006}
are implied by violation of the EPR-Bohm inequality where $\langle N^{B}\rangle$
is large%
\footnote{Following \cite{Cavalcanti2006} and the formalism to be presented
in Chapter \ref{cha:Macro-Super}, we suppose $\rho$ to be expressible
as a mixture of \emph{microscopic superpositions} of $J_{x}^{B}$
and $J_{y}^{B}$ \emph{only}, so $\rho=\sum_{i}P_{i}\rho_{i}=\sum_{i}P_{i}|\psi_{i,S}\rangle\langle\psi_{i,S}|$,
where each $|\psi_{i,S}\rangle$ is a superposition of eigenstates
$|j_{x}^{B}\rangle$ of $J_{z}^{B}$ separated by no more than $S$.
For such $\rho_{i}$ there is the constraint $\Delta^{2}j_{x}^{B}\leq S^{2}/4$;
similarly $\Delta^{2}j_{y}^{B}\leq S^{2}/4$. For each $\rho_{i}$
(being a quantum state) $\Delta_{inf}^{2}j_{z}^{B}+\Delta^{2}j_{y}^{B}+\Delta^{2}j_{x}^{B}\geq\langle N^{B}\rangle/2$
(see next paragraph), so $\Delta_{inf}^{2}j_{z}^{B}\geq\langle N^{B}\rangle/2-S^{2}/2$.
This inequality must also hold for the mixture $\rho$.

To prove the uncertainty relation above, define the reduced state
$\rho_{j_{z}^{A}}^{B}$ of $B$ given result $j_{z}^{A}$ for $J_{z}^{A}$;
its variance for $J_{z}^{B}$ is $\Delta^{2}(J_{z}^{B}|j_{z}^{A})$.
Now $\rho^{B}=Tr_{A}\rho=\sum_{j_{z}^{A}}P(j_{z}^{A})\rho_{j_{z}^{A}}^{B}$
so $\Delta^{2}J_{x/y}^{B}\geq\sum_{j_{z}^{A}}P(j_{z}^{A})\Delta^{2}(J_{x/y}^{B}|j_{z}^{A})$,
and since $\rho_{j_{z}^{A}}^{B}$ is a quantum state satisfying $\Delta^{2}J_{x}^{B}+\Delta^{2}J_{y}^{B}+\Delta^{2}J_{z}^{B}\geq\langle N^{B}\rangle/2$,
we get $\Delta_{inf}^{2}J_{z}^{B}+\Delta^{2}J_{x}^{B}+\Delta^{2}J_{y}^{B}\geq\sum_{j_{z}^{A}}P(j_{z}^{A})[\sum_{I=x,y,z}\Delta^{2}(J_{I}^{B}|j_{z}^{A})]\geq\langle N^{B}\rangle/2$.%
}. Measurement of a tiny $\Delta_{est}^{2}j_{z}^{B}$ implies the existence
of large superpositions of the eigenstates $|j_{x}^{B}\rangle$ of
$J_{x}^{B}$ (or $J_{y}^{B}$)\textcolor{black}{,} superpositions
with a range in $j_{x}^{B}$ of at least \begin{equation}
S=\sqrt{\langle N^{B}\rangle-2\Delta_{est}^{2}j_{z}^{B}},\label{eq:macro}\end{equation}
\textcolor{black}{which becomes $\approx\sqrt{\eta\langle N^{B}\rangle}$
for (\ref{eq:parampham}) with $\langle N^{B}\rangle$ large (Fig.
\ref{fig:EPR-Bohm2}). I won't attempt to explain this point in much
detail as it will be the subject of Chapter }\ref{cha:Macro-Super}\textcolor{black}{.}

\chapter{\textcolor{black}{\label{cha:CV-Bell}Continuous-Variables Bell inequalities}}

As we have seen in Chapter \ref{cha:EPR-paradox}, Einstein, Podolsky
and Rosen (EPR), in their famous 1935 paper \cite{Einstein1935},
demonstrated the incompatibility between the premises of \emph{``}local
realism\emph{\char`\"{}}%
\footnote{We have already seen in Chapters \ref{cha:Concepts} and \ref{cha:EPR-paradox}
what this term means and what assumptions particularly are needed
for the EPR argument. Having that in mind, I will keep using the vague
but popular terminology \char`\"{}local realism\char`\"{} for simplicity. %
} and the completeness of quantum mechanics. The original EPR pape\textcolor{black}{r
used continuous position and momentum variables, and relied on their
commutation relations, via the corresponding uncertainty principle.
Bohm \cite{Bohm1951} introduced, in 1951, his version of the EPR
paradox with spin observables. This was the version that was used
by Bell \cite{Bell1964} to prove his famous theorem showing that
quantum mechanics predicts results which can rule out the whole class
of local hidden variable (LHV) theories. It is hard to overemphasise
the importance of this result, which has even been called {}``the
most profound discovery of science'' \cite{Stapp1977}. However, the
original Bell inequality, and all of its generalisations, are directly
applicable only to the case of discrete observables. The main purpose
of this chapter is to close the circle and derive a class of Bell-type
inequalities applicable to continuous-variables (CV) correlations,
together with multipartite generalisations.}

\section{Motivation}

The last decades have witnessed the birth of whole new areas of enquiry,
which aim to harness these non-classical correlations predicted by
Quantum Mechanics towards information-processing applications. Bell
inequality violations have been shown to be relevant for quantum teleportation
\cite{Popescu1994a,Horodecki1996}, quantum key distribution \cite{Ekert1991,Scarani2001,Acin2004,Barrett2005a,Acin2006},
and reduction of communication complexity \cite{Brukner2004}. However
there are still many unanswered questions. Bell inequality violation
was once thought to be equivalent to entanglement, until Werner \cite{Werner1989}
showed that there are entangled states that do not violate a large
class of Bell inequalities. Gisin \cite{Gisin1991} proved that all
bipartite pure entangled states violate the CHSH \cite{Clauser1969}
inequality, a result which was generalised to the multipartite case
by Popescu and Rohrlich \cite{Popescu1992}. Later, Horodecki \emph{et
al.} \cite{Horodecki1995} devised an analytic criterion to determine
whether a mixed state of two qubits have a LHV description. Apart
from this simple case, there is no known general method to decide
whether a quantum state violates some Bell inequality.

For $n$ parties, $m$ measurements per party and $o$ outcomes, it
is well-known that the set of correlations allowed by LHV theories
can be represented as a \emph{convex polytope}, a multi-dimensional
geometrical structure formed by all convex combinations (linear combinations
where the coefficients are probabilities, i.e., they are non-negative
and sum to one) of a finite number of vertices. The vertices of this
polytope are the classical pure states --- the states with well-defined
values for all variables \cite{Pitowsky1989,Peres1999,Gisin2007}.
The tight Bell inequalities are associated with the linear facets
of the polytope. It is a computationally hard problem to list all
Bell inequalities for given $(n,m,o)$, and full numerical characterisations
have been accomplished only for small values of those parameters.

Nevertheless, some special cases can be considered. One can derive
classes of Bell inequalities which are recursively defined in terms
of those parameters. There are classes for $(n,2,2)$ \cite{Mermin1990,Ardehali1992,Belinskii1993,Werner2001},
$(n,3,2)$ \cite{Zukowski2006}, $(2,2,o)$ \cite{Mermin1980,Collins2002a}
and even $(n,2,o)$ \cite{Cabello2002d,Son2006a}, $(n,m,2)$ \cite{Zukowski1993,Laskowski2004,Nagata2006}
and $(2,m,o)$ \cite{Collins2004}. However, no class of Bell inequalities
has previously been derived without \emph{any} reference to the number
$o$ of outcomes or to their bound. Any real experiment will always
yield a finite number of outcomes but are there constraints imposed
by LHV theories that are independent of any \emph{particular} discretisation,
and can be explicitly written even in the limit $o\rightarrow\infty$?
This question goes beyond a challenge proposed in a recent paper by
Gisin \cite{Gisin2007}, where he proposes a list of open questions
regarding Bell inequalities, which was (in other words) to find inequalities
valid for arbitrary but \emph{fixed} $o$.\textcolor{black}{ }Our
answer is yes; and the derivation is much more straightforward than
in the case of the usual Bell-type inequalities which are restricted
to a particular set of outputs.

\textcolor{black}{We derive a class of inequalities for local realism
that directly} uses correlations of measurements, with no restriction
to spin measurements or discrete binning. They are not only valid
for arbitrary but fixed $o$, but they do not mention the number of
outcomes $o$ in their derivation at all. The new inequalities are
remarkably simple. They place no restriction o\textcolor{black}{n
the number of possible outcomes, and the contrast between the classical
and quantum bounds involves commutation relations in a central way.
They must be satisfied b}y any observations in a LHV theory, whether
having discrete, continuous or unbounded outcomes. We can immediately
re-derive previously known Bell-type inequalities, obtaining at the
same time their quantum-mechanical bounds by considering the non-commutativity
of the observables involved. We also display quantum states that directly
violate the new inequalities for continuous, unbounded measurements,
\textcolor{black}{even in the macroscopic, large} $n$ limit \cite{Mermin1990,Drummond1983,Peres1999,Reid2001}.
We show that \textcolor{black}{the new Bell violations survive the
effects of finite} generation and detection efficiency. This is very
surprising, in view of the many examples in which decoherence rapidly
destroys macroscopic superpositions \cite{Zurek2003}.

\textcolor{black}{Apart from this intrinsic interest, these inequalities
are relevant to an important scientific problem. No experiment has
yet produced a Bell inequality violation without introducing either
locality or detection loopholes. }As emphasised by Gisin in his recent
article \cite{Gisin2007}, \char`\"{}quantum nonlocality is so fundamental
for our world view that it deserves to be tested in the most convincing
way\char`\"{}. \textcolor{black}{One path towards this goal is to
use continuous-variables and efficient homodyne detection, which allows
much higher detection efficiency than is feasible with discrete spin
or photo-detection measurements. A number of loop-hole free proposals
exist in the literature, but they all use Bell \cite{Leonhardt1995,Gilchrist1998,Auberson2002,Wenger2003,Garcia-Patron2004}
or Hardy \cite{Yurke1999} inequalities with a dichotomic binning
of the results (which usually lead to small violations), or else a
parity or pseudo-spin approach \cite{Banaszek1999,ChenZB2002a,Son2006b}
which cannot be realized with efficient homodyne detection.}

\section{The variance inequality}

We will focus on the \emph{correlation functions of observables} for
$n$ sites or observers, each equipped with $m$ possible apparatus
settings to make their causally separated measurements. We consider
any real, complex or vector function $F\left(\mathbf{X},\mathbf{Y},\mathbf{Z},\ldots\right)$
of local observations $X_{i},Y_{i},Z_{i}$ at each site $i$, which
in an LHV theory are all functions of hidden variables $\lambda$.
In a real experiment the different terms in $F$ may not all be measurable
at once, because they may involve different choices of incompatible
observables. The assumption of \emph{locality} enters the reasoning
by requiring that the local choice of observable does not affect the
correlations between variables at different sites, and therefore that
the averages are taken over the same hidden variable ensemble $P\left(\lambda\right)$
for all terms. We introduce averages over the LHV ensemble (there's
no loss of generality in considering deterministic LHVs, since, according
to Fine's Theorem \textcolor{black}{2.8, a phenomenon violates local
causality if and only if it violates local determinism} \cite{Fine1982}),\begin{equation}
\left\langle F\right\rangle =\int P\left(\lambda\right)F\left(\mathbf{X}\left(\lambda\right),\mathbf{Y}\left(\lambda\right),\mathbf{Z}\left(\lambda\right),\ldots\right)d\lambda.\label{eq:LHV average}\end{equation}

Our LHV inequality uses the simple result that any function of random
variables has a non-negative variance, \begin{equation}
\left|\left\langle F\right\rangle \right|^{2}\leq\langle\left|F\right|^{2}\rangle.\label{eq:variance}\end{equation}
 We can also give a bound \begin{equation}
\langle\left|F\right|^{2}\rangle\leq\langle\left|F\right|^{2}\rangle_{sup},\label{eq:bound}\end{equation}
where the subscript denotes the supremum (least upper bound), in which
products of incompatible observables are replaced by their maximum
achievable values. This is necessary since if we are not able to measure
both $X_{i}$ and $Y_{i}$ simultaneously, a \textcolor{black}{general
LHV model could predict any achievable correlation \cite{Seevinck2007}.}

The same variance inequality applies to the corresponding Hermitian
operator $\hat{F}$ in quantum mechanics. While the observables at
different sites commute --- they can be simultaneously measured ---
those at the same site do not, so operator ordering must be included.
This enables us to see how quantum theory can violate the variance
bound for an LHV.

\section{An example}

As an example, we will apply this variance inequality to a well-known
case. Consider two dichotomic observables $X_{i},Y_{i}$ per site
$i$, the outcomes of which are $\pm1$. We define $F_{1}\equiv X_{1}$,
$F_{1}'\equiv Y_{1}$ , and then inductively construct \cite{Gisin1998}:

\begin{equation}
F_{n}\equiv\frac{1}{2}(F_{n-1}+F_{n-1}')X_{n}+\frac{1}{2}(F_{n-1}-F_{n-1}')Y_{n},\label{eq:Fn}\end{equation}
where $F_{n}'$ can be obtained from $F_{n}$ by the exchange $X_{i}\longleftrightarrow Y_{i}$.
In calculating $F_{n}^{2}$ we'll keep track of the local commutators
just to make the contrast with quantum mechanics clearer. For real
random variables it is obvious that the commutators are zero; at a
first read one can just regard that as the act of an overzealous classical
theorist. For real variables $X,\, Y$, the commutator is defined
in the same way as for the corresponding operators, i.e., $[X,Y]\equiv XY-YX$.
The anti-commutator is defined by $[X,Y]_{+}\equiv XY+YX$. Then \begin{multline}
F_{n}^{2}=\frac{1}{4}\left\{ (F_{n-1}^{2}+F_{n-1}'^{2})(X_{n}^{2}+Y_{n}^{2})\right.\\
+[F_{n-1},F_{n-1}']_{+}(X_{n}^{2}-Y_{n}^{2})+(F_{n-1}^{2}-F_{n-1}'^{2})[X_{n},Y_{n}]_{+}\\
\left.-[F_{n-1},F_{n-1}'][X_{n},Y_{n}]\right\} .\label{eq:Fn2}\end{multline}

Since $\hat{X}_{n}^{2}=\hat{Y}_{n}^{2}=1$, we can show that \textcolor{black}{$F_{n}^{2}=F_{n}'^{2}$
and \begin{equation}
F_{n}^{2}=F_{n-1}^{2}-\frac{1}{4}\left[F_{n-1},F_{n-1}'\right]\left[X_{n},Y_{n}\right].\label{eq:FN2 - 2}\end{equation}
}In a LHV theory, the term which involves commutators will be zero
since $[X(\lambda),Y(\lambda)]=X(\lambda)Y(\lambda)-Y(\lambda)X(\lambda)=0$.
H\textcolor{black}{ence by induction $F_{n}^{2}=F_{1}^{2}=1$ and
the variance inequality (\ref{eq:variance}) becomes:} \begin{equation}
-1\leq\left\langle F_{n}\right\rangle \leq1.\label{eq:MABK}\end{equation}
 This is the Mermin-Ardehali-Belinskii-Klyshko (MABK) \cite{Mermin1990,Ardehali1992,Belinskii1993}
Bell inequality, which reduces to the well-known Bell-CHSH \cite{Clauser1969}
inequality for $n=2$.

\subsection{Quantum bound}

We can now calculate the quantum mechanical bound by writing the variance
inequality \eqref{eq:variance} and substituting the functions in
\eqref{eq:FN2 - 2} by their corresponding operators\begin{eqnarray}
\bigl\langle\hat{F}_{n}\bigr\rangle_{Q}^{2} & \leq & \bigl\langle\hat{F}_{n}^{2}\bigr\rangle_{Q}=\bigl\langle\hat{F}_{n-1}^{2}-\frac{1}{4}[\hat{F}_{n-1},\hat{F}_{n-1}'][\hat{X}_{n},\hat{Y}_{n}]\bigr\rangle_{Q}\nonumber \\
 & \leq & \bigl\Vert\hat{F}_{n-1}^{2}\bigr\Vert+\frac{1}{4}\bigl\Vert[\hat{F}_{n-1},\hat{F}_{n-1}']\bigr\Vert\bigl\Vert[\hat{X}_{n},\hat{Y}_{n}]\bigr\Vert,\label{eq:QM Bound 1}\end{eqnarray}
where the norm $\left\Vert A\right\Vert $ denotes the modulus of
the maximum value of $\langle\hat{A}\rangle_{Q}$ over all quantum
states. The norm of the second commutator has the bound $\Vert[\hat{X}_{n},\hat{Y}_{n}]\Vert\leq2$.
It's easy to show that $[\hat{F}_{n},\hat{F}_{n}']=\hat{F}_{n-1}^{2}[\hat{X}_{n},\hat{Y}_{n}]+[\hat{F}_{n-1},\hat{F}_{n-1}']$
and therefore $\Vert[\hat{F}_{n},\hat{F}_{n}']\Vert\leq2\Vert\hat{F}_{n-1}^{2}\Vert+\Vert[\hat{F}_{n-1},\hat{F}_{n-1}']\Vert$
. Solving the recursion relation by noting that $\Vert\hat{F_{1}^{2}}\Vert=\frac{1}{2}\Vert[\hat{X}_{1},\hat{Y}_{1}]\Vert=1$
we finally arrive at the bound \begin{equation}
\langle\hat{F_{n}}\rangle_{Q}^{2}\leq2^{n-1}.\label{eq:QMbound}\end{equation}
 This can be attained with the generalised GHZ states \cite{Gisin1998},
which therefore violate (\ref{eq:variance}).

\section{Continuous-variables inequalities}

Inspired by those results, we now demonstrate an LHV inequality that
is directly applicable to unbounded continuous variables, in particular
field quadrature operators. The choice of the function $F_{n}$ in
(\ref{eq:Fn}) is not optimal though, since the variance in general
involves incompatible operator products that have no upper bound.

\textcolor{black}{To overcome this problem, consider a complex function
$C_{n}$ of the local real observables $\{X_{k},\, Y_{k}\}$ defined
as:\begin{equation}
C_{n}=\tilde{X_{n}}+i\tilde{Y_{n}}=\prod_{k=1}^{n}(X_{k}+iY_{k})\,,\label{eq:CnXnYn}\end{equation}
 so that the modulus square only involves compatible operator products,
i.e. \begin{equation}
\left|C_{n}\right|^{2}=\prod_{k=1}^{n}(X_{k}^{2}+Y_{k}^{2}).\label{eq:mod square}\end{equation}
 Applying the variance inequality to both $\tilde{X_{n}}$ and $\tilde{Y_{n}}$,
we find that: }

\textcolor{black}{\begin{equation}
\langle\tilde{X}_{n}\rangle^{2}+\langle\tilde{Y_{n}}\rangle^{2}\leq\langle\prod_{k=1}^{n}(X_{k}^{2}+Y_{k}^{2})\rangle\label{eq:New Ineq final}\end{equation}
This is the main result of this chapter. Given the assumption of local
hidden variables, this inequality must be satisfied for any set of
observables $X_{k}$, $Y_{k}$, regardless of their spectrum. }

\subsection{Quantum violations}

The fact that we have neglected the commutators in deriving \eqref{eq:New Ineq final}
hints that quantum mechanics might predict a violation. We define
quadrature operators\begin{eqnarray}
\hat{X}_{k} & = & \hat{a}_{k}e^{-i\theta_{k}}+\hat{a}_{k}^{\dagger}e^{i\theta_{k}}\nonumber \\
\hat{Y}_{k} & = & \hat{a}_{k}e^{-i(\theta_{k}+s_{k}\pi/2)}+\hat{a}_{k}^{\dagger}e^{i(\theta_{k}+s_{k}\pi/2)},\label{eq:XkYk}\end{eqnarray}
where $\hat{a}_{k},\hat{a}_{k}^{\dagger}$ are the boson annihilation
and creation operators at site $k$ and $s_{k}\in\{-1,1\}$. 

We now define the \textcolor{black}{operator \begin{equation}
\hat{Z}_{k}\equiv\hat{X}_{k}+i\hat{Y}_{k}\label{eq:Zk}\end{equation}
 and note that it follows that $\hat{C_{n}}=\prod_{k=1}^{n}\hat{Z}_{k}$.}
The definition of $\hat{Y}_{k}$ allows for the choice of the relative
phase with respect to $\hat{X}_{k}$ to be $\pm\pi/2$. Depending
on $s_{k}$, for each $k$ \textcolor{black}{either $\hat{Z}_{k}=2\hat{a}_{k}e^{-i\theta_{k}}$
or $\hat{Z}_{k}=2\hat{a}_{k}^{\dagger}e^{i\theta_{k}}$. Denoting
$\hat{A}_{k}(1)=\hat{a}_{k}$ and $\hat{A}_{k}(-1)=\hat{a}_{k}^{\dagger}$,
the term in the LHS of (\ref{eq:New Ineq final}) in quantum mechanics
is then \begin{equation}
|\langle\prod_{k}\hat{Z}_{k}\rangle_{Q}|^{2}=|2^{n}e^{i\sum_{k}s_{k}\theta_{k}}\langle\prod_{k}\hat{A}_{k}(s_{k})\rangle_{Q}|^{2}.\label{eq:LHS}\end{equation}
The RHS becomes \begin{equation}
\langle\prod_{k=1}^{n}(X_{k}^{2}+Y_{k}^{2})\rangle_{Q}=\langle\prod_{k=1}^{n}(4\hat{a}_{k}^{\dagger}\hat{a}_{k}+2)\rangle_{Q}\label{eq:RHS}\end{equation}
 regardless of the phase choices. To violate (\ref{eq:New Ineq final})
we must therefore find a state that satisfies}

\textcolor{black}{\begin{equation}
\Bigl|\Bigl\langle\prod_{k}\hat{A}_{k}(s_{k})\Bigr\rangle_{Q}\Bigr|^{2}>\Bigl\langle\prod_{k}\bigl(\hat{a}_{k}^{\dagger}\hat{a}_{k}+\frac{1}{2}\bigr)\Bigr\rangle_{Q},\label{eq:New Ineq a}\end{equation}
}which is surprisingly insensitive to relative phases between the
quadrature measurements at different sites.

This violation of a continuous variable Bell inequality can be realised
within quantum mechanics. Consider an even \textcolor{black}{number
of sites, choosing $s_{k}=1$ for the first half of them and $s_{k}=-1$
for the remaining. To maximise the LHS we need a superposition of
terms which are coupled by that product of annihilation/creation operators.
One choice is a state of type}

\textcolor{black}{\begin{equation}
\left|\Psi_{S}\right\rangle =c_{0}\left|0,\dots,0,1,\dots,1\right\rangle +c_{1}\left|1,\dots,1,0,\dots,0\right\rangle ,\label{eq:state}\end{equation}
where in the first term the first $n/2$ modes are occupied by zero
photons and the remaining by $1$; conversely for the second term.
With that choice of state the LHS of \eqref{eq:New Ineq a} becomes
$|c_{0}|²|c_{1}|²$, which is maximised by $|c_{0}|^{2}=|c_{1}|^{2}=\frac{1}{2}$.
The RHS is $(\frac{3}{2})^{\frac{n}{2}}(\frac{1}{2})^{\frac{n}{2}}$
independently of the amplitudes $c_{0},\, c_{1}$. Dividing the LHS
by the RHS, inequality (\ref{eq:New Ineq a}) becomes $\frac{1}{4}\left(\frac{4}{3}\right)^{\frac{n}{2}}\leq1$,
which is vio}lated for $n\geq10$, and the violation grows exponentially
with the number of sites.

\subsection{Feasibility}

While setting up the homodyne detectors necessary for this observation
is challenging, the complexity of this task scales linearly with the
number of modes. A more stringent constraint is most likely in the
state preparation, but we can relate state (\ref{eq:state}) to a
class of states of great experimental interest. They can be achieved
from a generalised GHZ state of $n/2$ photons, \begin{equation}
|GHZ(n)\rangle=\frac{1}{\sqrt{2}}(|H\rangle^{\otimes\frac{n}{2}}+|V\rangle^{\otimes\frac{n}{2}}),\label{eq:GHZn}\end{equation}
where $|H\rangle$ and $|V\rangle$ respectively represent single-particle
states of horizontal and vertical polarisation --- by splitting each
mode with a polarising beam splitter. Therefore violation of (\ref{eq:New Ineq final})
can be observed in the ideal case with a 5-qubit photon polarisation
GHZ state and homodyne detection. 

An interesting question is the effect of decoherence, both from state
preparation error \cite{Jang2006} and detector inefficiency. The
usual Bell-CHSH violations have an efficiency threshold \cite{Garg1987}
of $83\%$. This has not yet been achieved for single-photon counting.
Homodyne detection is remarkably efficient by comparison (\cite{Marquardt2007}
report up to 94.4\%, for example). However, the effect of detector
efficiency is easily included by assuming that each detected photon
mode is preceded by a beam splitter with intensity transmission $\eta<1$.
This changes both the LHS and RHS, so that the inequality becomes
\textcolor{black}{$\frac{4\eta^{2}}{2\eta+1}\leq4^{2/n}$, giving
a threshold efficiency requirement of $\eta>\eta_{min}$, where \begin{equation}
\eta_{min}=(1+\sqrt{1+4^{1-2/n}})/4^{1-2/n}.\label{eq:etamin}\end{equation}
}

This \emph{reduces} at large $n$ to an asymptotic value of $\eta_{\infty}=0.80902$.
Unexpectedly, the Bell violation (which signifies a quantum superposition)
is less sensitive to detector inefficiency in the macroscopic, large
$n$ limit. The minimum detector efficiency $\eta_{n}$ at finite
$n$ is plotted in Fig. 1, together with the minimum state preparation
fidelity $\epsilon_{min}$ in the case of ideal detectors, where we
model the density matrix as \begin{equation}
\hat{\rho}=\epsilon|\Psi_{S}\rangle\langle\Psi_{S}|+(1-\epsilon)\hat{I}.\label{eq:werner}\end{equation}

\begin{figure}
\begin{centering}
\includegraphics[width=11cm]{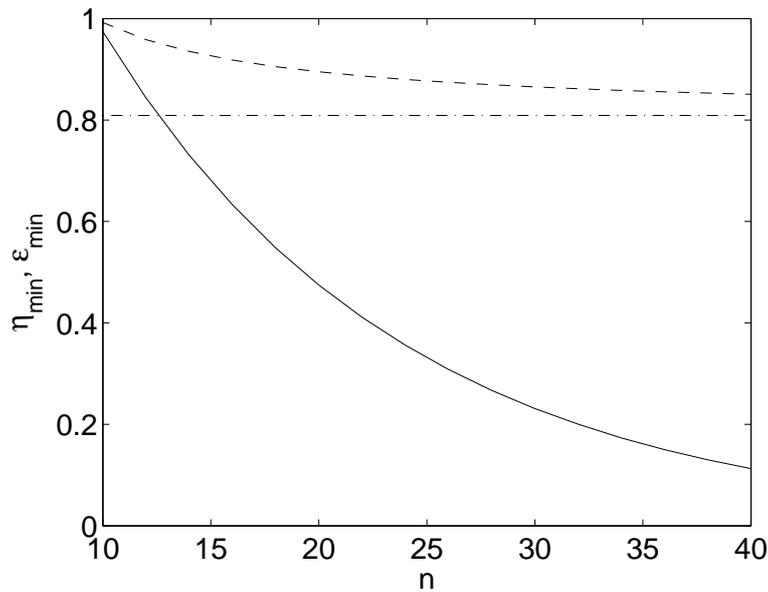}
\par\end{centering}

\caption{Minimum state preparation fidelity $\epsilon_{min}$ for ideal detectors
(solid line), and minimum detection efficiency $\eta_{min}$ for ideal
state preparation (dashed line) required for violation of (\ref{eq:New Ineq final})
as a function of the number of modes. The asymptotic value of $\eta_{min}$
is indicated by the dash-dotted line.}

\end{figure}

\section{No-go proof for first-moment correlation C.V. Bell inequalities}

We will finally prove that there are no LHV inequalities possible
if one considers only the first-moment correlations between continuous
variables in different sites. We will show this explicitly for the
simplest case and indicate how to generalise to arbitrary numbers
of parties and settings. Consider first $n=2$ parties, Alice and
Bob, each of which can choose between $m=2$ observables: $X_{a},Y_{a}$
for Alice and $X_{b},Y_{b}$ for Bob. Each measurement yields an outcome
in the real numbers. The first-moment correlation functions for each
of the $4$ possible configurations are just the averages $\langle X_{a}X_{b}\rangle$,
$\langle X_{a}Y_{b}\rangle$, $\langle Y_{a}X_{b}\rangle$, $\langle Y_{a}Y_{b}\rangle$.
Given those $4$ experimental outcomes, can we find a local hidden
variable model which reproduces them? 

\textcolor{black}{We construct an explicit example. Consider a hidden-variable
state $S$ where the hidden variables are the measured values $\mathbf{X},\,\mathbf{Y}$,
in an equal mixture of four classical pure} \textcolor{black}{states
$S_{k}=\left(X_{a},Y_{a},X_{b},Y_{b}\right)_{k}$ defined by}

\textcolor{black}{\begin{equation}
\begin{split}S_{1} & =2\,(1,0,\langle X_{a}X_{b}\rangle,0)\\
S_{2} & =2\,(1,0,0,\langle X_{a}Y_{b}\rangle)\\
S_{3} & =2\,(0,1,\langle Y_{a}X_{b}\rangle,0)\\
S_{4} & =2\,(0,1,0,\langle Y_{a}Y_{b}\rangle).\end{split}
\label{eq:stateSdef}\end{equation}
}

\textcolor{black}{Each of the states $S_{k}$ assigns a nonzero value
to only one of the $4$ correlation functions. Since the probability
of each of the states in the equal mixture is $1/4$, we have for
example $\langle X_{a}X_{b}\rangle_{S}=\frac{1}{4}\sum_{i}\langle X_{a}X_{b}\rangle_{S_{i}}=\langle X_{a}X_{b}\rangle$.}

\textcolor{black}{Satisfying the two-site correlations using the state
$S$ defined by \eqref{eq:stateSdef} leaves us with uncontrolled
values for the single-site correlations, for instance $\langle X_{b}\rangle_{S}=\frac{1}{2}(\langle X_{a}X_{b}\rangle+\langle Y_{a}X_{b}\rangle)$.
One might object to the fact that this is not equal to $\langle X_{b}\rangle$
in general. However, we may correct these lower order correlations
by adding four more states ($S_{5}$ to $S_{8}$) and changing the
prefactors multiplying $S_{1}$ to $S_{4}$ to compensate for their
reduced weight in the equal mixture. Crucially, adding these extra
states to $S$ in this manner does not modify the values of correlations
such as $\langle X_{a}X_{b}\rangle$. As an example, we exhibit the
state \begin{equation}
S_{5}=8\,(0,0,\langle X_{b}\rangle-(\langle X_{a}X_{b}\rangle+\langle Y_{a}X_{b}\rangle)/\sqrt{8},0),\label{eq:S5}\end{equation}
which corrects the single expectation value $\langle X_{b}\rangle_{S}$
to $\langle X_{b}\rangle$. }

\textcolor{black}{The proof generalises easily to arbitrary $n$ and
$m$. In that case, there are $m^{n}$ possible combinations of measurements
which yield $n$-site correlations. Denoting the $j^{\mathrm{th}}$
observable at site $i$ by $X_{i}^{j}$, each combination is specified
by a sequence of indices $(j_{1},j_{2},\dotsc,j_{n})$. For each combination
of measurements, we define a hidden variable state which assigns nonzero
values only to the variables which appear in the associated correlation
function $\langle\prod_{i=1}^{n}X_{i}^{j_{i}}\rangle$. In analogy
to the example above, we can always choose the values of the hidden
variables associated to $X_{i}^{j_{i}}$ such that their product is
equal to $m^{n}\langle\prod_{i=1}^{n}X_{i}^{j_{i}}\rangle$. Since
all other $m^{n}-1$ states defined in this way will give a value
of zero to this particular correlation function, and given that the
probability associated with each of those states is $1/m^{n}$, we
reproduce all correlations as desired. As indicated in the example,
additional first moment correlations involving} \textcolor{black}{\emph{less}}
\textcolor{black}{than $n$ sites can be included in the LHV model
by adding additional states to $S$ in a way which doesn't affect
the $n$-site correlations. Thus, any possible observation of first
moment correlations may be explained using a LHV model, and hence
these correlations alone cannot violate any Bell inequality. In other
words, the minimum requirement for a correlation Bell inequality with
continuous, unbounded variables, is to use not just the first but
also the second moments at each site.}

\section{Concluding remarks}

In conclusion, in this chapter we have derived a new class of Bell-type
inequalities valid for continuous and unbounded experimental outcomes.
We have shown that the same procedure allows one to derive the MABK
class of Bell inequalities and their corresponding quantum bounds.
That derivation makes it explicit that non-zero commutators --- associated
with the incompatibility of the local observables --- are the essential
ingredient responsible for the discrepancy between quantum mechanics
and local hidden variable theories. The new Bell-type inequality derived
here can be directly applied to continuous variables without the need
for a specific binning of the measurement outcomes. Surprisingly,
quantum mechanics predicts exponentially increasing violations of
the inequality for macroscopically large numbers of sites, even including
realistic decoherence effects like inefficient state preparation,
and a detector loss at \emph{every} site.

\chapter{\label{cha:Macro-Super}Generalised Macroscopic Superpositions}

\section{Introduction}

Since Schrödinger's seminal essay of 1935 \cite{Schroedinger1935},
in which he introduced his famous cat paradox, there has been a great
deal of interest and debate on the subject of the existence of a superposition
of two macroscopically distinguishable states. This issue is closely
related to the so-called \emph{measurement problem} \cite{Zurek1991}.
Some attempts to solve this problem, such as that of Ghirardi, Rimini,
Weber and Pearle \cite{Ghirardi1986,Ghirardi1990}, introduce modified
dynamics that cause a collapse of the wave function, effectively limiting
the size of allowed superpositions.

It thus becomes relevant to determine whether a superposition of states
with a certain level of distinguishability can exist experimentally
\cite{Marshall2003}. Evidence \cite{Brezger2002,Brune1996,Friedman2000,Raimond2001,Monroe1996,Auffeves2003,DeMartini2005,Ourjoumtsev2006}
for quantum superpositions of two distinguishable states has been
put forward for a range of different physical systems including SQUIDs,
trapped ions, optical photons and photons in microwave high-Q cavities.
Signatures for the size of superpositions have been discussed by Leggett
\cite{Leggett2002} and, more recently, by Korsbakken et al \cite{Korsbakken2007}.
Theoretical work suggests that the generation of a superposition of
two truly macroscopically distinct states will be greatly hindered
by decoherence \cite{Caldeira1985,Zurek2003}. 

In a recent paper \cite{Cavalcanti2006}, we suggested to broaden
the concept of detection of macroscopic superpositions, by focusing
on signatures that confirm, for some experimental instance, a failure
of microscopic/mesoscopic superpositions to predict the measured statistics.
This approach is applicable to a broader range of experimental situations
based on macroscopic systems, where there would be a macroscopic range
of outcomes for some observable, but not necessarily just two that
are macroscopically distinct. Recent work by Marquardt et al \cite{Marquardt2007}
reports experimental application of this approach. 

The paradigmatic example \cite{Leggett1984,Leggett1985,Brune1996,Monroe1996,Friedman2000,Raimond2001,Brezger2002,Auffeves2003,DeMartini2005,Ourjoumtsev2006}
of a macroscopic superposition involves two states $\psi_{+}$ and
$\psi_{-}$, macroscopically distinct in the sense that the respective
outcomes of a measurement $\hat{x}$ fall into regions of outcome
domain, denoted $+$ and $-$, that are macroscopically different.
We argue in \cite{Cavalcanti2006} that a superposition of type \begin{equation}
\psi_{+}+\psi_{0}+\psi_{-},\label{eq:mmsupgen}\end{equation}
that involves a range of states but with only some pairs (in this
case $\psi_{+}$and $\psi_{-}$) macroscopically distinct must also
be considered a type of macroscopic superposition (we call these \emph{generalised
macroscopic superpositions}), in the sense that it displays a nonzero
off-diagonal density matrix element $\langle\psi_{+}|\rho|\psi_{-}\rangle$,
connecting two macroscopically distinct states, and hence cannot be
constructed from microscopic superpositions of the basis states of
$\hat{x}$. Such superpositions \cite{Mermin1980,Drummond1983,Peres1992a,Reid2002}
are predicted to be generated in certain key macroscopic experiments,
that have confirmed continuous-variable \cite{Ou1992,Zhang2000,Silberhorn2001,Bowen2002b,Schori2002,Bowen2003,Josse2003,Josse2004,Laurat2005,Corney2006,Suzuki2006}
squeezing and entanglement, spin squeezing and entanglement of atomic
ensembles \cite{Julsgaard2001,Julsgaard2004}, and entanglement and
violations of Bell inequalities for discrete measurements on multi-photon
systems \cite{DeMartini1998,Leibfried2004,Roos2004,Lamas-Linares2001}.

We derive criteria for the detection of the generalised macroscopic
(or $S$-scopic) superpositions using continuous variable measurements.
These criteria confirm that a macroscopic system cannot be described
as any mixture of only microscopic (or $s$-scopic, where $s<S$)
quantum superpositions of eigenstates of $\hat{x}$. We show how to
apply the criteria to detect generalised $S$-scopic superpositions
in squeezed and entangled states that are of experimental interest. 

The generalised macroscopic superpositions still hold interest from
the point of view of Schrödinger's discussion \cite{Schroedinger1935}
of the apparent incompatibility of quantum mechanics with macroscopic
realism. This is so because such superpositions cannot be represented
as a mixture of states which give outcomes for $\hat{x}$ that always
correspond to one or other (or neither) of the macroscopically distinct
regions $+$ and $-$. The quantum mechanical paradoxes associated
with the generalised macroscopic superposition (\ref{eq:mmsupgen})
have been discussed in previous works \cite{Cavalcanti2006,Mermin1980,Drummond1983,Reid2005,Cavalcanti2007b}.

The criteria derived in this chapter take the form of inequalities.
Their derivation utilises the uncertainty principle and the assumption
of certain types of mixtures. In this respect they are similar to
criteria for inseparability that have been derived by Duan \emph{et
al.} and Simon \cite{Duan2000,Simon2000} and Hofmann and Takeuchi
\cite{Hofmann2003a}. Rather than testing for failure of separable
states, however, they test for failure of a phase space {}``macroscopic
separability'', where it is assumed that a system is always in a
mixture (never a superposition) of macroscopically separated states.

We will note that one can be more general in the derivation of the
inequalities, adopting the approach of Leggett and Garg \cite{Leggett1985}
to define a macroscopic reality without reference to any quantum concepts.
One may consider a whole class of theories, which we refer to as the
\emph{minimum uncertainty theories} (MUT) and to which quantum mechanics
belongs, for which the uncertainty relations hold and the inequalities
therefore follow, based on this macroscopic reality. The experimental
confirmation of violation of these inequalities will then lead to
demonstration of a new type of Einstein-Podolsky-Rosen argument (or
{}``paradox'') \cite{Einstein1935}, in which the inconsistency
of a type of macroscopic ($S$-scopic) reality with the completeness
of quantum mechanics is revealed \cite{Cavalcanti2006,Reid2005}.
A direct analogy exists with the original EPR argument, which is a
demonstration of the incompatibility of local realism with the completeness
of quantum mechanics \cite{Reid2003,Wiseman2006,Reid2007tb}. In our
case, the inconsistency of the completeness of quantum mechanics is
shown to be with a macroscopic reality rather than the local reality
of the original EPR argument.

\section{Generalised $S$-scopic Coherence \label{sec:Generalized-S-scopic-Coherence}}

We introduce in this Section the concept of a generalised $S$-scopic
coherence \cite{Cavalcanti2006}, which we define in terms of failure
of certain types of mixtures. In the next Section, we link this concept
to that of the generalised $S$-scopic superpositions (\ref{eq:mmsupgen}).

We consider a system which is in a statistical mixture of two component
states. For example, if one attributes probabilities $\wp_{1}$ and
$\wp_{2}$ to underlying quantum states $\rho_{1}$ and $\rho_{2}$,
respectively (where $\rho_{i}$ denotes a quantum density operator),
then the state of the system will be described as a mixture, which
in quantum mechanics is represented as \begin{equation}
\rho=\wp_{1}\rho_{1}+\wp_{2}\rho_{2}.\label{eq:QM mixture}\end{equation}
This can be interpreted as \char`\"{}the state is \emph{either} $\rho_{1}$
with probability $\wp_{1}$, \emph{or} $\rho_{2}$ with probability
$\wp_{2}$\char`\"{}. The probability for an outcome $x$ of any measurable
physical quantity $\hat{x}$ can be written, for a mixture of the
type (\ref{eq:QM mixture}), as\begin{equation}
P(x)=\wp_{1}P_{1}(x)+\wp_{2}P_{2}(x),\label{eq:probmacro}\end{equation}
where $P_{i}(x)$ ($i=1,2$) is the probability distribution of $x$
in the state $\rho_{i}$. 

More generally, in any physical theory, the specification of a state
$\rho$ (where here $\rho$ is just a symbol to denote the state,
but not necessarily a density matrix) fully specifies the probabilities
of outcomes of all experiments that can be performed on the system.
If we then have with probability $\wp_{1}$ a state $\rho_{1}$ which
predicts for each observable $\hat{x}$ a probability distribution
$P_{1}(x)$ and with probability $\wp_{2}$ a second state which predicts
$P_{2}(x)$, then the probability distribution for any observable
$\hat{x}$ given such mixture is of the form \eqref{eq:probmacro}.

The concept of coherence can now be introduced. 

\begin{verse}
\textbf{Definition 1:} \emph{The state of a physical system displays}
\textbf{coherence}\emph{ between two outcomes $x_{1}$ and $x_{2}$
of an observable $\hat{x}$ iff the state $\rho$ of the system cannot
be considered a statistical mixture of some underlying states $\rho_{1}$
and $\rho_{2}$, where $\rho_{1}$ assigns probability zero for $x_{2}$
and $\rho_{2}$ assigns probability zero for $x_{1}$.}
\end{verse}
This definition is independent of quantum mechanics. Within quantum
mechanics it implies that the quantum density matrix representing
the system cannot be decomposed in the form (\ref{eq:QM mixture}).
Thus, for example, $\rho=\frac{1}{\sqrt{2}}(|\psi_{+}\rangle\langle\psi_{+}|+|\psi_{-}\rangle\langle\psi_{-}|),$
where $|\psi_{\pm}\rangle=[|x_{1}\rangle\pm|x_{2}\rangle]/\sqrt{2},$
does not display coherence between $x_{1}$ and $x_{2}$ because it
can be rewritten to satisfy (\ref{eq:QM mixture}). The definition
will allow a state to be said to have coherence between $x_{1}$ and
$x_{2}$ if and only if there is no possible ensemble decomposition
of that state which allows an interpretation as a mixture (\ref{eq:QM mixture}),
so that the system cannot be regarded as being in one or other of
the states that can generate at most one of $x_{1}$ or $x_{2}$. 

We next define the concept of \emph{generalised $S$-scopic coherence}. 

\begin{figure}
\begin{centering}
\includegraphics[width=11cm]{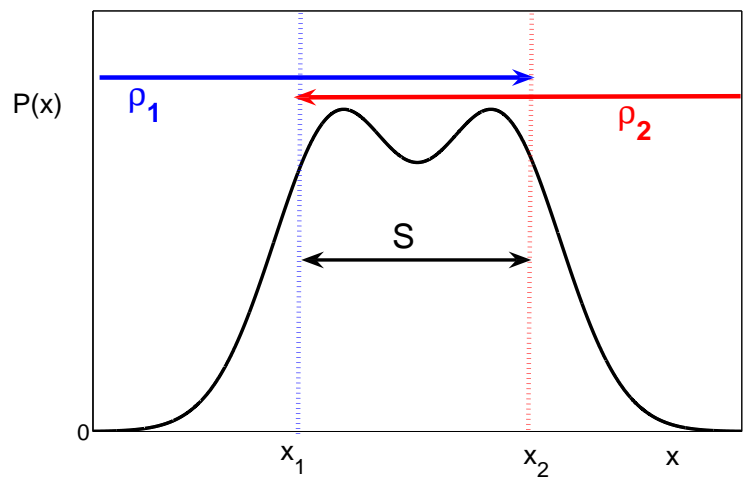}
\par\end{centering}

\caption{\label{fig:macsup1}Probability distribution for outcomes $x$ of
measurement $\hat{x}$. If $x_{1}$ and $x_{2}$ are macroscopically
separated, then we might expect the system to be described as the
mixture (\ref{eq:QM mixture}), where $\rho_{1}$ encompasses outcomes
$x<x_{2}$, and $\rho_{2}$ encompasses outcomes $x>x_{1}$. This
means an absence of generalised macroscopic coherence, as defined
in Section \ref{sec:Generalized-S-scopic-Coherence}.}

\end{figure}

\begin{verse}
\textbf{Definition 2:} \emph{We say that the state displays }\textbf{generalised
$S$-scopic coherence}\emph{ iff there exist $x_{1}$ and $x_{2}$
with $x_{2}-x_{1}\geq S$ (we take $x_{2}>x_{1}$), such that $\rho$
displays coherence between outcomes $x\leq x_{1}$ and $x\geq x_{2}$.
This coherence will be said to be }\textbf{macroscopic}\emph{ when
$S$ is macroscopic.} 
\end{verse}
If there is \emph{no} generalised\emph{ }$S$-scopic coherence, then
the system can be described as a mixture (\ref{eq:QM mixture}) where
now states $\rho_{1}$ and $\rho_{2}$ assign nonzero probability
only for $x<x_{2}$ and $x>x_{1}$ respectively. This situation is
depicted in Fig. \ref{fig:macsup1}.

An important clarification is needed at this point. It is clearly
a vague matter to determine when $S$ is macroscopic. What is important
is that we are able to push the boundaries of experimental demonstrations
of $S$-scopic coherence to larger values of $S$. We will keep the
simpler terminology, but the reader might want to understand \emph{macroscopic}
as \emph{$S$-scopic} throughout the text\emph{.}

Generalised macroscopic coherence amounts to a loss of what we will
call a \emph{generalised macroscopic reality}. The simpler form of
macroscopic reality that involves only two states macroscopically
distinct has been discussed extensively by Leggett \cite{Leggett1984,Leggett1985}.
This simpler case would be applicable to the situation of Fig. 1 if
there were zero probability for result in the intermediate region
$x_{1}<x<x_{2}$. Macroscopic reality in this simpler situation means
that the system must be in one or other of two macroscopically distinct
states, $\rho_{1}$ and $\rho_{2}$, that predict outcomes in regions
$x\leq x_{1}$ and $x\geq x_{2}$, respectively. The term {}``macroscopic
reality'' is used \cite{Leggett1985} because the definition precludes
that the system can be in a superposition of two macroscopically distinct
states prior to measurement. \emph{Generalised macroscopic reality}
applies to the broader situation, where probabilities for outcomes
$x_{1}<x<x_{2}$ are not zero, and means that where we have two macroscopically
separated outcomes $x_{1}$ and $x_{2}$, the system can be interpreted
as being in one or other of two states $\rho_{1}$ and $\rho_{2}$,
that can predict \emph{at most} one of $x_{1}$ or $x_{2}$. Again,
the term macroscopic reality is used, because this definition precludes
that the system is a superposition of two macroscopically separated
states that give outcomes $x_{1}$ and $x_{2}$ respectively. 

We note that Leggett and Garg \cite{Leggett1985} define a macroscopic
reality in which they do not restrict to quantum states $\rho_{1}$
and $\rho_{2}$, but allow for a more general class of theories where
$\rho_{1}$ and $\rho_{2}$ can be hidden variable states of the type
considered by Bell \cite{Bell1964}. Such states are not restricted
by the uncertainty relation that would apply to each quantum state,
and hence the assumption of macroscopic reality as applied to these
theories would not lead to the inequalities we derive in this chapter.
This point will be discussed in Section \ref{sec:minimum-uncertainty},
but the reader should note the definition of $S$-scopic coherence
within quantum mechanics means that $\rho_{1}$ and $\rho_{2}$ are
quantum states.

\section{Generalised macroscopic and $S$-scopic quantum Superpositions\label{sec:Generalised-macroscopic-and} }

We now link the definition of generalised macroscopic coherence to
the definition of generalised macroscopic superposition states \cite{Cavalcanti2006}.
Generally we can express $\rho$ as a mixture of pure states $|\psi_{i}\rangle$.
Thus\begin{equation}
\rho=\sum_{i}\wp_{i}|\psi_{i}\rangle\langle\psi_{i}|,\label{eq:mixture}\end{equation}
where we can expand each $|\psi_{i}\rangle$ in terms of a basis set
such as the eigenstates $|x\rangle$ of $\hat{x}$: thus $|\psi_{i}\rangle=\sum_{x}c_{x}|x\rangle$,
the $c_{x}$ being probability amplitudes.

\textbf{Theorem A:} The existence of coherence between outcomes $x_{1}$
and $x_{2}$ of an observable $\hat{x}$ is equivalent, within quantum
mechanics, to the existence of a nonzero off-diagonal element in the
density matrix, i.e, $\left\langle x_{1}\right|\rho\left|x_{2}\right\rangle \neq0$.

\textbf{Proof:} The proof is given in Appendix A. $\blacksquare$

\textbf{Theorem B:} In quantum mechanics, there exists coherence between
outcomes $x_{1}$ and $x_{2}$ of an observable $\hat{x}$ iff in
\emph{any} decomposition (\ref{eq:mixture}) of the density matrix,
there is a nonzero contribution from a superposition state of the
type \begin{equation}
|\psi_{S}\rangle=c_{x_{1}}\left|x_{1}\right\rangle +c_{x_{2}}\left|x_{2}\right\rangle +\sum_{x\neq x_{1},x_{2}}c_{x}\left|x\right\rangle \label{eq:supx}\end{equation}
with $c_{x_{1}}$,$c_{x_{2}}\neq0$. 

\textbf{Proof}: If each $|\psi_{i}\rangle$ cannot be written in the
specific form (\ref{eq:supx}), then each $|\psi_{i}\rangle\langle\psi_{i}|$
is either of form $\rho_{1}$ or $\rho_{2}$, so that we can write
$\rho$ as the mixture (\ref{eq:QM mixture}). Hence the existence
of coherence, which implies $\rho$ cannot be written as (\ref{eq:QM mixture}),
implies the superposition must always exist in (\ref{eq:mixture}).
The converse is also true: if the superposition exists in any decomposition,
then there exists an irreducible term in the decomposition that assigns
nonzero probabilities to both $x_{1}$ and $x_{2}$, and therefore
the density matrix cannot be written as (\ref{eq:QM mixture}). $\blacksquare$

We say that a \emph{generalised $S$-scopic superposition} of states
$|x_{1}\rangle$ and $|x_{2}\rangle$ exists when any decomposition
(\ref{eq:mixture}) must contain a nonzero probability for a superposition
(\ref{eq:supx}), where $x_{1}$ and $x_{2}$ are separated by at
least $S$. Throughout this chapter, we define the \emph{size} $S$
of the generalised superposition \begin{equation}
|\psi\rangle=\sum_{k}c_{k}|x_{k}\rangle\label{eq:gmssize}\end{equation}
(where $|x_{k}\rangle$ are eigenstates of $\hat{x}$ and each $c_{k}\neq0$)
to be the range of its prediction for $\hat{x}$, this range being
the maximum value of $|x_{k}-x_{j}|$ where $|x_{k}\rangle$ and $|x_{j}\rangle$
are any two components of the superposition (\ref{eq:gmssize}) (so
$c_{k}$,$c_{j}\neq0$).

From the above discussions it follows that within quantum mechanics,
the existence of generalised $S$-scopic coherence between $x_{1}$
and $x_{2}$ (here $|x_{2}-x_{1}|=S$) implies the existence of a
generalised $S$-scopic superposition of type (\ref{eq:supx}), which
can be written as \begin{equation}
|\psi\rangle=c_{-}\psi_{-}+c_{0}\psi_{0}+c_{+}\psi_{+},\label{eq: QMGMS}\end{equation}
where the quantum state $\psi_{-}$ assigns some nonzero probability
only to outcomes smaller than or equal to $x_{1}$, the quantum state
$\psi_{+}$ assigns some nonzero probability only to outcomes larger
than or equal to $x_{2}$, and the state $\psi_{0}$ assigns nonzero
probabilities only to intermediate values satisfying $x_{1}<x<x_{2}$.
Where $S$ is macroscopic, expression (\ref{eq: QMGMS}) depicts a
\emph{generalised macroscopic superposition} state. In this case then,
only the states $\psi_{-}$ and $\psi_{+}$ are necessarily macroscopically
distinct. We regain the traditional extreme macroscopic quantum state
$c_{-}\psi_{-}+c_{+}\psi_{+}$ when $c_{0}=0$.

\section{Minimum Uncertainty Theories \label{sec:minimum-uncertainty}}

We now follow a procedure similar to that used to derive criteria
useful for the confirmation of inseparability \cite{Duan2000}. The
underlying states $\rho_{1}$ and $\rho_{2}$ comprising the mixture
(\ref{eq:QM mixture}) are themselves quantum states, and so each
will satisfy the quantum uncertainty relations with respect to complementary
observables. This and the assumption of (\ref{eq:QM mixture}) will
imply a set of constraints. The violation of any one of these is enough
to confirm the observation of a generalised macroscopic coherence-
that is, of a generalised macroscopic superposition of type (\ref{eq: QMGMS}).

While our specific aim is to develop criteria for quantum macroscopic
superpositions, we present the derivations in as general a form as
possible to make the point that experimental violation of the inequalities
would imply not only a generalised macroscopic coherence in quantum
theory, but a failure of the assumption (\ref{eq:probmacro}) in all
theories which place the system in a probabilistic mixture of two
states, which we designate by $\rho_{1}$ and $\rho_{2}$, and for
which the appropriate uncertainty relation holds for each of the states.
In this sense, our approach is similar to that of Bell \cite{Bell1964},
except that the assumption used here of minimum uncertainties for
outcomes of measurements would be regarded as more restrictive than
the general hidden variable theories considered by Bell.

We make this point more specific by defining a whole class of theories,
which we refer to as the Minimum Uncertainty Theories (MUT), that
embody the assumption that any state $\rho$ within the theory will
predict the same uncertainty relation for the variances of two incompatible
observables $\hat{x}$ and $\hat{p}$ as is predicted by quantum mechanics.
This is a priori not an unreasonable thing to postulate for a theory
that may differ from quantum mechanics in the macroscopic regime but
agree with all the observations in the well-studied microscopic regime.
Here we will focus on pairs of observables, like position and momentum,
for which the uncertainty bound is a real number, which with the use
of scaling and choice of units will be set to $1$, so we can write
the an uncertainty relation assumed by all MUT's as \begin{equation}
\Delta^{2}x\Delta^{2}p\geq1,\label{eq:MUT UR}\end{equation}
where $\Delta^{2}x$ and $\Delta^{2}p$ are the variances of $x$
and $p$ respectively. This is Heisenberg's uncertainty relation,
and quantum mechanics is clearly a member of MUT. Other quantum uncertainty
relations that will be specifically used in this chapter include\begin{equation}
\Delta^{2}x+\Delta^{2}p\geq2,\label{eq:sumhup}\end{equation}
which follows from (\ref{eq:MUT UR}) and has been useful in derivation
of inseparability criteria \cite{Duan2000}.

\section{Signatures for generalised S-scopic superpositions: Binned domain\label{sec:Signatures-for-generalised}}

In this Section we will derive inequalities that follow if there are
no $s$-scopic superpositions (where $s>S$), so that violation of
these inequalities implies existence of an $S$-scopic superposition
(or coherence), as defined in Sections \ref{sec:Generalized-S-scopic-Coherence}
and \ref{sec:Generalised-macroscopic-and}. The approach is similar
to that often used to detect entangled states. Separability implies
inequalities such as those derived by Duan \emph{et al. }and Simon
\cite{Duan2000,Simon2000}, and their violation thus implies existence
of entanglement. This approach has been used to experimentally confirm
entanglement, as described in reference \cite{Bowen2003}, among others.
An experimental description of the approach we use here has also been
outlined by Marquardt et al. \cite{Marquardt2007}.

We consider two types of criteria for the detection of a generalised
macroscopic superposition (or coherence). The first, of the type considered
in \cite{Cavalcanti2006}, will be considered in this section and
uses \emph{binned outcomes} to demonstrate a generalised $S$-scopic
superposition of states $\psi_{+}$ and $\psi_{-}$ that predict outcomes
in \emph{specified} regions denoted $+1$ and $-1$ respectively (Fig.
2), where these regions are separated by a minimum distance \emph{$S$.}
We expand on some earlier results of \cite{Cavalcanti2006} for completeness
and also introduce new criteria of this type.

\subsection{Single system}

Consider a system $A$ and a macroscopic measurement $\hat{x}$ on
$A$, the outcomes of which are spread over a macroscopic range. We
partition the domain of outcomes $x$ for this measurement into three
regions, labelled $l=-1,0,1$ for the regions $x\leq-S/2$, $-S/2<x<S/2$,
$x\geq S/2$, respectively. The probabilities for outcomes to fall
in those regions are denoted $\wp_{-},$ $\wp_{0}$ and $\wp_{+}$,
respectively (Fig. 2).

\begin{figure}
\begin{centering}
\includegraphics[width=11cm]{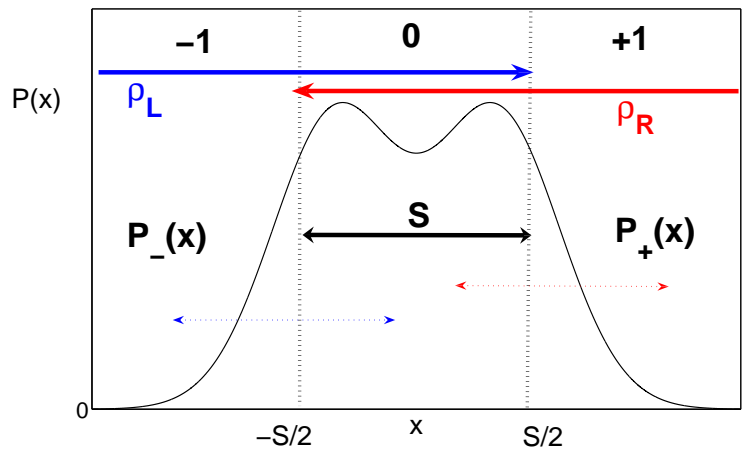}
\par\end{centering}

\caption{Probability distribution for a measurement $\hat{x}$ . We bin results
to give three distinct regions of outcome: $0$, $-1,$$+1$.}

\end{figure}

If there is \emph{no} generalised S-scopic coherence then there is
no coherence between outcomes in $l=1$ and $l=-1$, and the state
of system $A$ can be written as \begin{equation}
\rho_{mix}=\wp_{L}\rho_{L}+\wp_{R}\rho_{R},\label{eq:Binned mixture}\end{equation}
where $\rho_{L}$ predicts outcomes in the region $x<S/2$, $\rho_{R}$
predicts outcomes in the region $x>-S/2$, and $\wp_{L}$ and $\wp_{R}$
are their respective probabilities. The assumption of this mixture
(\ref{eq:Binned mixture}) implies\begin{equation}
P(y)=\wp_{L}P_{L}(y)+\wp_{R}P_{R}(y).\label{eq:probconst}\end{equation}
Here $y$ is the outcome of some measurement that can be performed
on the system, and $P_{R/L}(y)$ is the probability for a result $y$
when the system is specified as being in state $\rho_{R/L}$. Where
the measurement performed is $\hat{x}$, so $y=x$, there is the constraint
on (\ref{eq:probconst}) so that $P_{R}(x)=0$ for $x\leq-S/2$ and
$P_{L}(x)=0$ for $x\geq S/2$.

Now consider an observable $\hat{p}$ (with outcomes $p$) incompatible
with $\hat{x}$, such that the variances are constrained by the uncertainty
relation $\Delta^{2}x\Delta^{2}p\geq1$. Our goal is to derive inequalities
from just two assumptions: firstly, that $\hat{x}$ and $\hat{p}$
are incompatible observables of quantum mechanics (or of a Minimum
Uncertainty Theory), so the uncertainty relation holds for both $\rho_{R/L}$;
and, secondly, that there is no generalised S-scopic coherence.

Violation of these inequalities will imply that one of these assumptions
is false. Within quantum mechanics, for which the first assumption
is necessarily true, that would imply the existence of a generalised
macroscopic superposition of type (\ref{eq: QMGMS}) with outcomes
$x_{1}$ and $x_{2}$ separated by at least $S$.

If the quantum state is of form (\ref{eq:Binned mixture}) or if the
theory satisfies (\ref{eq:probconst}), then \begin{equation}
\Delta^{2}p\geq\wp_{L}\Delta_{L}^{2}p+\wp_{R}\Delta_{R}^{2}p,\label{eq:mixvarp}\end{equation}
where $\Delta²p$, $\Delta²_{L}p$ and $\Delta²_{R}p$ are the variances
of $p$ in the states $\rho_{mix}$, $\rho_{L}$ and $\rho_{R}$,
respectively. This follows simply from the fact the variance of a
mixture cannot be less than the average variance of its component
states. Specifically, if a probability distribution for a variable
$z$ is of the form $P(z)=\sum_{i=1}^{N}\wp_{i}P(z)$, then $\Delta^{2}z=\sum_{i=1}^{N}\wp_{i}\Delta_{i}^{2}z+\frac{1}{2}\sum_{i\neq i'}\wp_{i}\wp_{i'}(\langle z\rangle_{i}-\langle z\rangle_{i'})^{2}$. 

We can now, using (\ref{eq:mixvarp}) and the Cauchy-Schwarz inequality,
derive a bound for a particular function of variances that will apply
if the system is describable as (\ref{eq:Binned mixture})\begin{eqnarray}
(\wp_{L}\Delta_{L}^{2}x+\wp_{R}\Delta_{R}^{2}x)\Delta^{2}p & \geq & [\sum_{i=L,R}\wp_{i}\Delta_{i}^{2}x][\sum_{i=L,R}\wp_{i}\Delta_{i}^{2}p]\nonumber \\
 & \geq & [\sum_{i=L,R}\wp_{i}\Delta_{i}x\Delta_{i}p]^{2}\label{eq:cs}\\
 & \geq & 1.\nonumber \end{eqnarray}

The left hand side is not directly measurable, since it involves variances
of $\hat{x}$ in two states which have overlapping ranges of outcomes.
We must derive an upper bound for $\Delta_{L/R}^{2}x$ in terms of
measurable quantities. For this we partition the probability distribution
$P_{R}(x)$ according to the outcome domains $l=0,1$, into normalised
probability distributions $P_{R0}(x)\equiv P_{R}(x|x<S/2)$, and $P_{+}(x)\equiv P_{R}(x|x\geq S/2)$:\begin{equation}
P_{R}(x)=\wp_{R0}P_{R0}(x)+\wp_{R+}P_{+}(x).\label{eq:domain partition}\end{equation}
Here $\wp_{R+}=\int_{S/2}^{\infty}P_{R}(x)dx=\wp_{+}$ and $\wp_{R0}=\int_{0}^{S/2}P_{R}(x)dx$.
It follows that $\Delta_{R}^{2}x=\wp_{R0}\Delta_{R0}^{2}x+\wp_{R+}\Delta_{+}^{2}x+\wp_{R0}\wp_{R+}(\mu_{+}-\mu_{R0})^{2}$,
where $\mu_{+}$($\Delta_{+}^{2}x$) and $\mu_{R0}$ ($\Delta_{R0}^{2}x$)
are the averages (variances) of $P_{+}(x)$ and $P_{R0}(x)$, respectively.
Using the bounds $\wp_{R0}\leq\wp_{0}/(\wp_{0}+\wp_{+})$, $\Delta_{R0}^{2}x\leq S^{2}/4$,
$\wp_{R+}\leq1$ and $0\leq\mu_{+}-\mu_{R0}\leq\mu_{+}+S/2$, we derive
\begin{equation}
\Delta_{R}^{2}x\leq\Delta_{+}^{2}x+\frac{\wp_{0}}{\wp_{0}+\wp_{+}}[(S/2)^{2}+(\mu_{+}+S/2)^{2}]\label{eq:domainbound}\end{equation}
and, by similar reasoning,\begin{equation}
\Delta_{L}^{2}x\leq\Delta_{-}^{2}x+\frac{\wp_{0}}{\wp_{0}+\wp_{-}}[(S/2)^{2}+(\mu_{-}-S/2)^{2}].\label{eq:domainbound-}\end{equation}
 Here $\mu_{\pm}$ and $\Delta_{\pm}^{2}x$ are the mean and variance
of the measurable $P_{\pm}(x)$, which\foreignlanguage{english}{,}
since the only contributions to the regions + and - are from $P_{R}(x)$
and $P_{L}(x)$ respectively, are defined as the normalised $+$ and
$-$ parts of $P(x)$, so that $P_{+}(x)\equiv P(x|x\geq S/2)$ and
$P_{-}(x)\equiv P(x|x\leq-S/2)$. We substitute (\ref{eq:domainbound})
in (\ref{eq:cs}), and use $\wp_{0}+\wp_{+}\geq\wp_{R}$ and $\wp_{0}+\wp_{-}\geq\wp_{L}$
to derive the final result which is expressed in the following theorem.

\textbf{Theorem 1}: The assumption of no generalised $S$-scopic coherence
between outcomes in regions $+1$ and $-1$ of Fig. 2 (or, equivalently,
of no generalised $S$-scopic superpositions involving two states
$\psi_{-}$ and $\psi_{+}$ predicting outcomes for $\hat{x}$ in
the respective regions $+1$ and $-1$) will imply the uncertainty
relations\begin{equation}
(\Delta_{ave}^{2}x+\wp_{0}\delta)\Delta^{2}p\geq1\label{eq:critsuper}\end{equation}
and\begin{equation}
\Delta_{ave}^{2}x+\Delta^{2}p\geq2-\wp_{0}\delta,\label{eq:sumcrit}\end{equation}
where we define $\Delta_{ave}^{2}x=\wp_{+}\Delta_{+}^{2}x+\wp_{-}\Delta_{-}^{2}x$
and $\delta\equiv\{(\mu_{+}+S/2)^{2}+(\mu_{-}-S/2)^{2}+S²/2\}+\Delta_{+}^{2}x+\Delta_{-}^{2}x$.
Thus, the violation of either one of these inequalities implies the
existence of a generalised $S$-scopic quantum superposition, and
in this case the superposition involves states $\psi_{+}$ and $\psi_{-}$
predicting outcomes for $\hat{x}$ in regions $+1$ and $-1$, of
Fig. 2, respectively.).

As illustrated in Fig.2, the $\Delta_{\pm}^{2}x$ and $\mu_{\pm}$
are the variance and mean of $P_{\pm}(x)$, the normalised distribution
over the domain $l=\pm1$. $\wp_{\pm}$ is the total probability for
a result $x$ in the domain $l=\pm1$, while $\wp_{0}=1-(\wp_{+}+\wp_{-})$.
The measurement of the probability distributions for $\hat{x}$ and
$\hat{p}$ are all that is required to determine whether violation
of the inequality (\ref{eq:critsuper}) or (\ref{eq:sumcrit}) occurs.
Where $\hat{x}$ and $\hat{p}$ correspond to optical field quadratures,
such distributions have been measured, for example, by Smithey et
al \cite{Smithey1993}.

\textbf{Proof}: The assumption of no such generalised $S$-scopic
superposition implies (\ref{eq:Binned mixture}). We have proved that
(\ref{eq:critsuper}) follows. To prove (\ref{eq:sumcrit}), we start
from (\ref{eq:Binned mixture}) and the uncertainty relation (\ref{eq:sumhup}),
and derive a bound that will apply if the system is describable as
(\ref{eq:Binned mixture}): $(\wp_{L}\Delta_{L}^{2}x+\wp_{R}\Delta_{R}^{2}x)+\Delta^{2}p\geq[\sum_{i=L,R}\wp_{i}\Delta_{i}^{2}x]+[\sum_{i=L,R}\wp_{i}\Delta_{i}^{2}p]\geq[\sum_{i=L,R}\wp_{i}[\Delta_{i}^{2}x+\Delta_{i}^{2}p]\geq2$.
Using (\ref{eq:domainbound}), (\ref{eq:domainbound-}) and $\wp_{0}+\wp_{+}\geq\wp_{R}$
and $\wp_{0}+\wp_{-}\geq\wp_{L}$ we get the final result. $\blacksquare$

\subsection{Bipartite systems}

One can derive similar criteria where we have a system comprised of
two subsystems $A$ and $B$. In this case, a reduced variance may
be found in a combination of observables from both subsystems. A common
example is where there is a correlation between the two positions
$X^{A}$ and $X^{B}$ of subsystems $A$ and $B$ respectively, and
also between the two momenta $P^{A}$ and $P^{B}$. Such correlation
was discussed by Einstein, Podolsky and Rosen \cite{Einstein1935}
and is called EPR correlation. If a sufficiently strong correlation
exists, it is possible that both the position difference $X^{A}-X^{B}$
and the momenta sum $P^{A}+P^{B}$ will have zero variance.

Where we have two subsystems that may demonstrate EPR correlation,
we may construct a number of useful complementary measurements that
may reveal generalised macroscopic superpositions. The simplest situation
is where we again consider superpositions with respect to the observable
$X^{A}$ of system $A$. Complementary observables include observables
of the type \begin{equation}
\tilde{P}=P^{A}-gP^{B},\label{eq:ptilde}\end{equation}
 where $g$ is an arbitrary constant and $P^{B}$ is an observable
of system $B$. We denote the outcomes of measurements $X^{A},$ $P^{A},$
$P^{B},$ $\tilde{P}$ by the lower case symbols $x^{A},$ $p^{A},$
$p^{B},$ $\tilde{p}$ respectively. The Heisenberg uncertainty relation
is\begin{equation}
\Delta^{2}x^{A}\Delta_{inf,L}^{2}p^{A}=\Delta^{2}x^{A}\Delta^{2}\tilde{p}\geq1.\label{eq:uncerinf}\end{equation}
We have introduced $\Delta_{inf,L}^{2}p^{A}=\Delta^{2}\tilde{p}$
so that a connection is made with notation used previously in the
context of demonstration of the EPR paradox \cite{Reid1989,Reid2007tb}.
More generally \cite{Reid2003,Reid2007tb}, we define an inference
variance \begin{equation}
\Delta_{inf}^{2}p^{A}=\sum_{p^{B}}P(p^{B})\Delta^{2}(p^{A}|p^{B}),\label{eq:condvar}\end{equation}
which is the average conditional variance for $P^{A}$ at $A$ given
a measurement of $P^{B}$ at $B$. The $\Delta^{2}(p^{A}|p^{B})$
are the variances of the conditional probability distributions $P(p^{A}|p^{B}).$
We note that $\Delta_{inf,L}^{2}p^{A}$ is the linear regression estimate
of $\Delta_{inf}^{2}p^{A}$, but that we have $\Delta_{inf}^{2}p^{A}=\Delta_{inf,L}^{2}p^{A}$
for the case of Gaussian states \cite{Reid2007tb}. The uncertainty
relation \begin{equation}
\Delta^{2}x^{A}\Delta_{inf}^{2}p^{A}\geq1\label{eq:condinfur}\end{equation}
and also $\Delta^{2}p^{A}\Delta_{inf}^{2}x^{A}\geq1$, holds true
for all quantum states \cite{Cavalcanti2007b}, so that we can interchange
$\Delta_{inf}^{2}p^{A}$ with $\Delta_{inf,L}^{2}p^{A}$ in the proofs
and theorems below.

\textbf{Theorem 2}: Where we have a system comprised of subsystems
$A$ and $B$, the absence of generalised $S$-scopic superpositions
with respect to the measurement $X^{A}$ implies\begin{equation}
(\Delta_{ave}^{2}x^{A}+\wp_{0}\delta)\Delta_{inf}^{2}p^{A}\geq1.\label{eqn:eprcat}\end{equation}
$\Delta_{ave}^{2}x^{A}$, $\wp_{0}$ and $\delta$ are defined as
for Theorem 1 for the distribution $P(x^{A})$. $\Delta_{inf}^{2}p^{A}$
is defined by (\ref{eq:condvar}) and involves measurements performed
on both systems $A$ and $B$. The inequality (\ref{eqn:eprcat})
also holds replacing $\Delta_{inf}^{2}p^{A}$ with $\Delta_{inf,L}^{2}p^{A}$
which is defined by (\ref{eq:uncerinf}). Thus violation of (\ref{eqn:eprcat})
implies the existence of the generalised $S$-scopic superposition,
involving states predicting outcomes for $X^{A}$ in regions $+1$
and $-1$.

\textbf{Proof}: The proof follows in identical fashion to that of
Theorem 1, except in this case the $\rho_{L}$ and $\rho_{R}$ of
(\ref{eq:Binned mixture}) are states of the composite system, and
there is no constraint on these except that the domain for outcomes
of $X^{A}$ is restricted as specified in the definition of $\rho_{R/L}$.
The expansion (\ref{eq:mixture}) for the density matrix as a mixture
is $\rho=\sum_{r}\wp_{r}|\psi_{r}\rangle\langle\psi_{r}|$ where now
$\psi_{r}=\sum_{i,j}c_{i,j}|x_{i}\rangle_{A}|x_{j}\rangle_{B},$ $|x_{j}\rangle_{B}$
being eigenstates of an observable of system $B$ that form a basis
set for states of $B$. The generalised superposition (\ref{eq:supx})
thus becomes in this bipartite case\begin{equation}
|\psi_{r}\rangle=c_{1}\left|x_{1}\right\rangle _{A}|u_{1}\rangle_{B}+c_{2}\left|x_{2}\right\rangle _{A}|u_{2}\rangle_{B}+\sum_{i\neq1,2}c_{ij}\left|x_{i}\right\rangle _{A}|x_{j}\rangle_{B},\label{eq:supbi}\end{equation}
where $|u_{1}\rangle$ and $|u_{2}\rangle$ are pure states for system
$B$. If we assume no generalised $S$-scopic superposition, then
$\rho$ can be written without contribution from a state of form (\ref{eq:supbi})
and we can write $\rho$ as (\ref{eq:Binned mixture}). The constraint
(\ref{eq:Binned mixture}) implies $P(\tilde{p})=\sum_{I=R,L}\wp_{I}P_{I}(\tilde{p})$
where $P_{R|L}(\tilde{p})$ is the probability distribution of $\tilde{p}$
for state $\rho_{R/L}$. Thus (\ref{eq:mixvarp}) also holds for $\tilde{p}$
replacing $p$, as do all the results (\ref{eq:domain partition})-(\ref{eq:domainbound-})
involving the variances of $x$. Also, (\ref{eq:mixvarp}) holds for
$\Delta_{inf}^{2}p^{A}$ (see Appendix B). Thus we prove Theorem 2
by following (\ref{eq:mixvarp})-(\ref{eq:critsuper}). $\blacksquare$

In order to violate the inequality (\ref{eqn:eprcat}), we would look
to minimise $\Delta_{inf}^{2}p^{A}$, or $\Delta_{inf,L}^{2}p^{A}=\Delta^{2}\tilde{p}$.
For the optimal EPR states, $P^{A}+P^{B}$ has zero variance, and
one would choose for $\tilde{p}$ the case of $g=-1$, so that $\tilde{p}=p^{A}+p^{B}$,
where $p^{B}$ is the result of measurement of $P^{B}$ at $B$. This
case gives $\Delta_{inf}^{2}p^{A}=0$. More generally for quantum
states that are not the ideal case of EPR, our choice of $\tilde{p}$
becomes so as to optimise the violation of (\ref{eqn:eprcat}) and
will depend on the quantum state considered. This will be explained
further in Section \ref{sec:Predictions-of-particular}.

A second approach is to use as the macroscopic measurement a linear
combination of observables from both systems $A$ and $B$, so for
example we might have $\hat{x}=(X^{A}+X^{B})/\sqrt{2}$ and $\hat{p}=(P^{A}+P^{B})/\sqrt{2}$.
Relevant uncertainty relations include (based on $|[X^{A},P^{A}]|=2$
which gives $\Delta x^{A}\Delta p^{A}\geq1$) \begin{equation}
\Delta(x^{A}+x^{B})\Delta(p^{A}+p^{B})\geq2\label{eq:prodsum}\end{equation}
and \begin{equation}
\Delta^{2}(x^{A}+x^{B})+\Delta^{2}(p^{A}+p^{B})\geq4.\label{eq:sumsumur}\end{equation}
and from these we can derive criteria for generalised S-scopic coherence
and superpositions. 

\textbf{Theorem 3}: The following inequalities if violated will imply
existence of generalised S-scopic superpositions. \begin{equation}
\Bigl(\Delta_{ave}^{2}(\frac{x^{A}+x^{B}}{\sqrt{2}})+\wp_{0}\delta\Bigr)\Delta^{2}(\frac{p^{A}+p^{B}}{\sqrt{2}})\geq1\label{eq:critsuper bipartite}\end{equation}

and\begin{equation}
\Delta_{ave}^{2}(\frac{x^{A}+x^{B}}{\sqrt{2}})+\Delta^{2}(\frac{p^{A}+p^{B}}{\sqrt{2}})\geq2-\wp_{0}\delta.\label{eq:sumcrit bipartite}\end{equation}
We write in terms of the normalised quadratures so that, following
(\ref{eq:prodsum}), $\Delta^{2}(\frac{x^{A}+x^{B}}{\sqrt{2}})<1$
would imply squeezing of the variance below the quantum noise level.
The quantities $\Delta_{ave}^{2}x$, $\wp_{0}$ and $\delta$ are
defined as for Theorem 1, but we note that $P(x)$ in this case is
the distribution for $\hat{x}=(X^{A}+X^{B})/\sqrt{2}$. $S$ now refers
to the size of the superposition of $(X^{A}+X^{B})/\sqrt{2}$. 

\textbf{Proof:} In this case the $\rho_{R/L}$ of (\ref{eq:Binned mixture})
are defined as specified originally in (\ref{eq:Binned mixture})
but where $x$ is now defined as the outcome of the measurement $\hat{x}=(X^{A}+X^{B})/\sqrt{2}$.
The failure of the form (\ref{eq:Binned mixture}) for $\rho$ is
equivalent to the existence of a generalised superposition of type
(\ref{eq:supbi}) where now $|x_{i}\rangle$ refers to eigenstates
of $X^{A}+X^{B}$. Thus the eigenstates $|x_{i}\rangle$ are of the
general form $|x_{i}\rangle=\sum_{x_{j}}c_{j}|x_{j}\rangle_{A}|x_{i}-x_{j}\rangle_{B}$.
The mixture (\ref{eq:Binned mixture}) implies (\ref{eq:mixvarp})
where now $p$ refers to the outcome of $\hat{p}=(P^{A}+P^{B})\sqrt{2}$,
and will imply a similar inequality for $\hat{x}$. Application of
uncertainty relation (\ref{eq:prodsum}) for the products can be used
in (\ref{eq:cs}), and the proof of \eqref{eq:critsuper bipartite}
follows as in (\ref{eq:mixvarp})-(\ref{eq:critsuper}) of Theorem
1. The second result follows by applying the procedure for proof of
(\ref{eq:sumcrit}) but using the sum uncertainty relation (\ref{eq:sumsumur}).
$\blacksquare$

\section{Signatures of non-locatable generalised S-scopic superpositions\label{sec:Signatures-of-non-locatable}}

A second set of criteria will be developed to demonstrate that a generalised
S-scopic superposition exists, so that two states comprising the superposition
predict respective outcomes separated by at least size S, but in this
case there is the disadvantage that no information is obtained regarding
the regions in which these outcomes lie. 

This lack of information is compensated by a far simpler form of the
inequalities and increased sensitivity of the criteria. For pure states,
a measurement of squeezing $\Delta p$ implies a state that when written
in terms of the eigenstates of $x$ is a superposition such that $\Delta x\geq1/\Delta p$.
With increasing squeezing, the extent $S$ of the superposition increases.
To develop a simple relationship between $S$ and $\Delta p$ for
mixtures, we assume that there is no such generalised coherence between
any outcomes of $\hat{x}$ separated by a distance larger than $S$.
This approach gives a simple connection between the minimum size of
a superposition describing the system and the degree of squeezing
that is measured for this system. The drawback is the loss of direct
information about the location (in phase space for example) of the
superposition. We thus refer to these superpositions as \char`\"{}non-locatable\char`\"{}.

\subsection{Single systems}

We consider the outcome domain of a macroscopic observable $\hat{x}$
as illustrated in Fig. 3, and address the question of whether this
distribution could be predicted from microscopic, or $s$-scopic ($s<S$),
superpositions of eigenstates of $\hat{x}$ alone. 

\begin{figure}
\begin{centering}
\includegraphics[width=11cm]{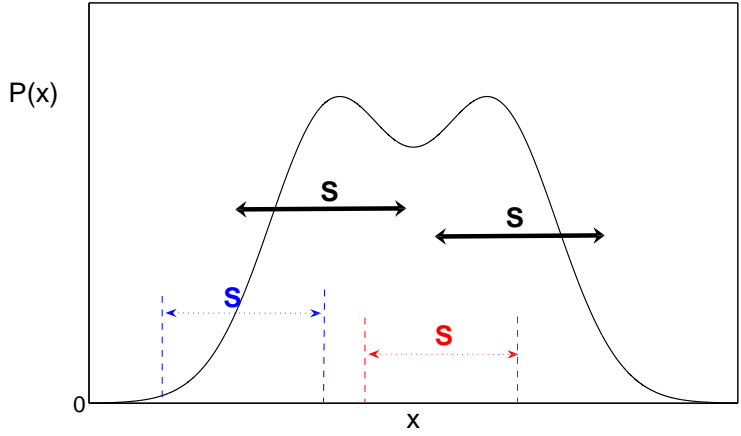}
\par\end{centering}

\caption{We consider an arbitrary probability distribution for a measurement
$\hat{x}$ that gives a macroscopic range of outcomes. }

\end{figure}

The assumption of no generalised $S$-scopic coherence (between any
two outcomes of the domain for $\hat{x}$) or, equivalently, the assumption
of no generalised $S$-scopic superpositions, with respect to eigenstates
of $\hat{x}$, means that the state can be written in the form\begin{equation}
\rho_{S}=\sum_{i}\wp_{i}\rho_{Si},\label{eq:mixmicrosup}\end{equation}
Here each $\rho_{Si}$ is the density operator for a pure quantum
state that is \emph{not} such a generalised $S$-scopic superposition,
so that $\rho_{Si}$ has a range of possible outcomes for $\hat{x}$
separated by less than $S$. Hence $\rho_{Si}=|\psi_{Si}\rangle\langle\psi_{Si}|$
where \begin{equation}
|\psi_{Si}\rangle=\sum_{k}c_{k}|x_{k}\rangle\label{eq:microsup}\end{equation}
but the maximum separation of any two states $|x_{k}\rangle$,$|x_{k'}\rangle$
involved in the superposition (that is with $c_{k},c_{k'}\neq0$ )
is less than $S$, so $|x_{k}-x_{k'}|<S$. 

Assumption (\ref{eq:mixmicrosup}) will imply a constraint on the
measurable statistics, namely that there is a minimum level of uncertainty
in the prediction for the complementary observable $\hat{p}$. The
variances of each $\rho_{Si}$ must be bounded by \begin{equation}
\Delta²_{Si}x<\frac{S²}{4}.\label{eq:dommicrosup}\end{equation}
It is also true that \begin{equation}
\Delta^{2}p\geq\sum_{i}\wp_{i}\Delta²_{Si}p.\label{eq:mixmomentum}\end{equation}
Now the Heisenberg uncertainty relation applies to each $\rho_{Si}$
(the inequality also applies to the MUT's discussed in Section \ref{sec:minimum-uncertainty})
so for the incompatible observables $\hat{x}$ and $\hat{p}$\begin{equation}
\Delta²_{Si}x\Delta²_{Si}p\geq1.\label{eq:hupfor proofgen}\end{equation}
Thus a lower bound on the variance of $p$ follows. \begin{eqnarray}
\Delta^{2}p & \geq & \sum_{i}\wp_{i}\Delta²_{Si}p\label{eq:minfuzp}\\
 & \geq & \sum_{i}\wp_{i}\frac{1}{\Delta_{Si}^{2}x}>\frac{4}{S^{2}}.\nonumber \end{eqnarray}
We thus arrive at the following theorem.

\textbf{Theorem 4:} The assumption of no generalised $S$-scopic coherence
in $\hat{x}$ will imply the following inequality for the variance
of outcomes of the complementary observable $\hat{p}$:\begin{equation}
\Delta p>\frac{2}{S}.\label{eq:nobinning ineq}\end{equation}
The main result of this section follows from Theorem 4 and is that
the observation of a squeezing $\Delta p$ in $\hat{p}$ such that
\begin{equation}
\Delta p\leq2/S\label{eq:critsqsup}\end{equation}
will imply the existence of an $S$-scopic superposition\begin{equation}
c_{x}|x\rangle+c_{x+S}|x+S\rangle+......\label{eq:supS}\end{equation}
namely, of a superposition of eigenstates $|x\rangle$ of $\hat{x}$,
that give predictions for $\hat{x}$ with a range of at least $S$.
The parameter $S$ gives a minimum extent of quantum indeterminacy
with respect to the observable $\hat{x}$. Here $c_{x}$ and $c_{x+S}$
represent non-zero probability amplitudes.

In fact, using our criterion (\ref{eq:critsqsup}) squeezing in $p$
($\Delta p<1$) will rule out any expansion of the system density
operator in terms of superpositions of $|x\rangle$ with $S\leq2$
(Fig. 4). Thus onset of squeezing is evidence of the onset of quantum
superpositions of size $S>2$, the size $S=2$ corresponding to the
vacuum noise level. This noise level may be taken as a level of reference
in determining the relative size of the superposition. The experimental
observation \cite{Suzuki2006} of squeezing levels of $\Delta p\approx0.4$
confirms superpositions of size at least $S=5$.

\subsection{Bipartite systems}

For composite systems comprised of two subsystems $A$ and $B$ upon
which measurements $X^{A}$, $P^{A}$, $X^{B}$, $P^{B}$ can be performed,
the approach of the previous section leads to the following theorem.

\textbf{Theorem 5a}. The assumption of no generalised $S$-scopic
coherence with respect to $X^{A}$ implies \begin{equation}
\Delta_{inf}p^{A}>\frac{2}{S}.\label{eq:nobinning ineq2}\end{equation}
$\Delta_{inf}^{2}p^{A}$ is defined as in (\ref{eq:condvar}). The
result also holds on replacing $\Delta_{inf}^{2}p^{A}$ with $\Delta_{inf,L}^{2}p^{A}$
as defined in (\ref{eq:uncerinf}).

\textbf{Theorem 5b}. The assumption of no generalised $S$-scopic
coherence with respect to $\hat{x}=(X^{A}+X^{B})/\sqrt{2}$ implies\begin{equation}
\Delta(\frac{p^{A}+p^{B}}{\sqrt{2}})>\frac{2}{S}.\label{eq:resultcompsum}\end{equation}

\textbf{Proof:} The proof follows as for Theorem 4, but using the
uncertainty relations (\ref{eq:uncerinf}) and (\ref{eq:prodsum}) in
(\ref{eq:minfuzp}) instead of (\ref{eq:hupfor proofgen}). $\blacksquare$

The observation of squeezing such that \eqref{eq:nobinning ineq2}
is violated, i.e \begin{equation}
\Delta_{inf}p^{A}\leq2/S\label{eq:critsqsupinf}\end{equation}
will imply the existence of an $S$-scopic superposition\begin{equation}
c_{x}|x\rangle_{A}|u_{1}\rangle_{B}+c_{x+S}|x+S\rangle_{A}|u_{2}\rangle_{B}+......\label{eq:supSent}\end{equation}
namely, of a superposition of eigenstates $|x\rangle_{A}$ that give
predictions for $X^{A}$ separated by at least $S$. Similarly, the
observation of two-mode squeezing such that \eqref{eq:resultcompsum}
is violated, i.e. \begin{equation}
\Delta(\frac{p^{A}+p^{B}}{\sqrt{2}})\leq2/S,\label{eq:critsump}\end{equation}
will imply existence of an $S$-scopic superposition of eigenstates
of the normalised position sum $(X^{A}+X^{B})/\sqrt{2}$.

\section{Criteria for generalised $S$-scopic coherent state superpositions\label{sec:Criteria-for-generalised} }

The criteria developed in the previous section may be used to rule
out that a system is describable as a mixture of coherent states,
or certain superpositions of them. If a system can be represented
as a mixture of coherent states $|\alpha\rangle$ the density operator
for the quantum state will be expressible as\begin{equation}
\rho=\int P(\alpha)|\alpha\rangle\langle\alpha|d{}^{2}\alpha\label{eq:GSrep}\end{equation}
which is, since $P(\alpha)$ is positive for a mixture, the Glauber-Sudarshan
P-representation \cite{Glauber1963,Sudarshan1963}. The quadratures
$\hat{x}$ and $\hat{p}$ are defined as $x=a+a^{\dagger}$ and $p=(a-a^{\dagger})/i$,
so that $\Delta x=\Delta p=1$ for this minimum uncertainty state,
where here $a$, $a^{\dagger}$ are the standard boson creation and
annihilation operators, so that $a|\alpha\rangle=\alpha|\alpha\rangle$.
Proving failure of mixtures of these coherent states would be a first
requirement in a search for macroscopic superpositions, since such
mixtures expand the system density operator in terms of states with
equal yet minimum uncertainty in each of $x$ and $p$, that therefore
do not allow significant macroscopic superpositions in either.

The coherent states form a basis for the Hilbert space of such bosonic
fields, and any quantum density operator can thus be expanded as a
mixture of coherent states or their superpositions. It is known \cite{Walls1983}
that systems exhibiting squeezing ($\Delta p<1$) cannot be represented
by a positive Glauber-Sudarshan representation, and hence onset of
squeezing implies the existence of some \emph{superposition} of coherent
states. A next step is to rule out mixtures of \emph{$s_{\alpha}$-scopic
superpositions} of coherent states . To define what we mean by this,
we consider superpositions \begin{equation}
|\psi_{s_{\alpha}}\rangle=\sum_{i}c_{i}|\alpha_{i}\rangle\label{eq:cohssup}\end{equation}
where for any $|\alpha_{i}\rangle$, $|\alpha_{j}\rangle$ such that
$c_{i},c_{j}\neq0$, we have $|\alpha_{i}-\alpha_{j}|\leq s_{\alpha}$
for all $i$, $j$ ($s_{\alpha}$ is a positive number). We note that
for a coherent state $|\alpha\rangle,$ $\langle x\rangle=2\alpha$.
Thus the separation of the states with respect to $\hat{x}$ is defined
as $S_{\alpha}=2s_{\alpha}$. The {}``separation'' of the two coherent
states $|-\alpha\rangle$ and $|\alpha\rangle$ (where $\alpha$ is
real) in terms of $x$ corresponds to $S_{\alpha}=4\alpha=2s_{\alpha}$,
as illustrated in Fig. \ref{fig:figsups5}.

\begin{figure}
\begin{centering}
\includegraphics[width=11cm]{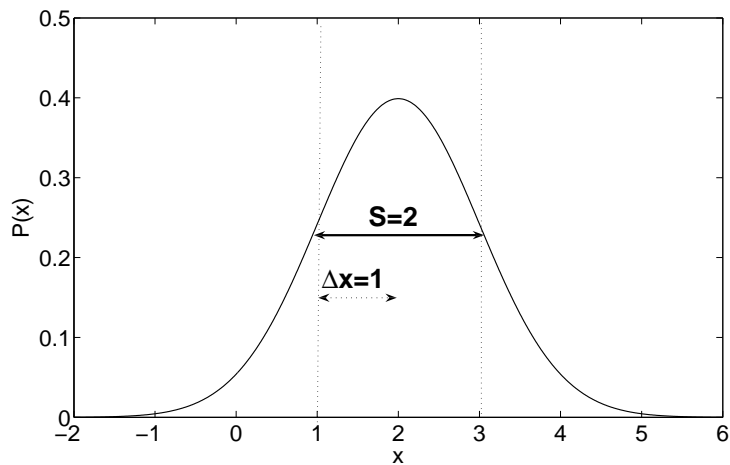}
\par\end{centering}

\caption{P(x) for a coherent state $|\alpha\rangle$: $\Delta x=\Delta p=1$.
\label{fig:figcoh4}}

\end{figure}

We next ask whether the density operator for the system can be described
in terms of the $s_{\alpha}$-scopic coherent superpositions, so that
\begin{equation}
\rho=\sum_{r}\wp_{r}|\psi_{s_{\alpha}}^{r}\rangle\langle\psi_{s_{\alpha}}^{r}|,\label{eq:mixcohS}\end{equation}
where each $|\psi_{s_{\alpha}}^{r}\rangle$ is of the form (\ref{eq:cohssup}).
Each $|\psi_{s_{\alpha}}^{r}\rangle$ predicts a variance in $x$
which has an upper limit given by that of the superposition $(1/\sqrt{2})\{e^{i\pi/4}|-s_{\alpha}/2\rangle+e^{-i\pi/4}|s_{\alpha}/2\rangle\}$.
This state predicts a probability distribution $P(x)=\frac{1}{2}\sum_{\pm}P_{G\pm}(x)$
where \begin{equation}
P_{G\pm}(x)=\frac{1}{\sqrt{2\pi}}\exp[\frac{-(x-\pm s_{\alpha})^{2}}{2}]\label{eq:coherent-state prob}\end{equation}
(Fig. \ref{fig:figsups5}), which corresponds to a variance $\Delta^{2}x=\langle x^{2}\rangle=1+s_{\alpha}^{2}=1+S_{\alpha}^{2}/4$.
This means each $|\psi_{s_{\alpha}}^{r}\rangle$ is constrained to
allow only $\Delta^{2}x\leq1+s_{\alpha}^{2}$, which implies for each
$|\psi_{s,r}\rangle$ a lower bound on the variance $\Delta^{2}p$
so that $\Delta^{2}p\geq1/\Delta^{2}x\geq1/(1+s_{\alpha}^{2})$. Thus
using the result for a mixture (\ref{eq:mixcohS}), we get that if
indeed (\ref{eq:mixcohS}) can describe the system, the variance in
$p$ is constrained to satisfy $\Delta^{2}p\geq1/(1+s_{\alpha}^{2})$.

\begin{figure}
\begin{centering}
\includegraphics[width=11cm]{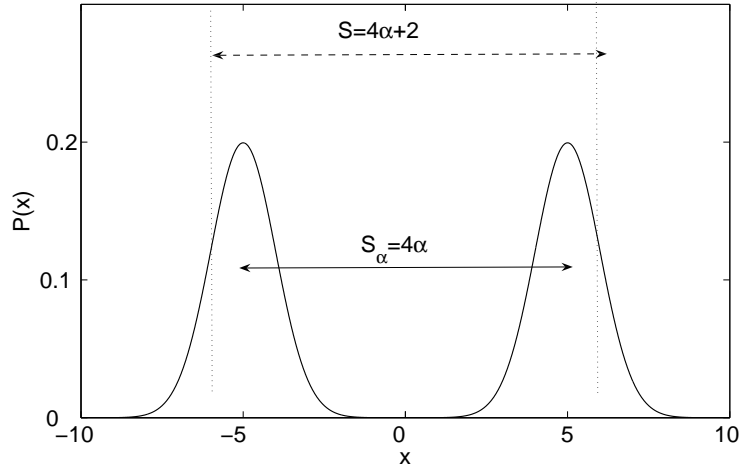}
\par\end{centering}

\caption{(a) P(x) for a superposition of coherent states $(1/\sqrt{2})\{e^{i\pi/4}|-\alpha\rangle+e^{-i\pi/4}|\alpha\rangle\}$
(here the scale is such that $\Delta x=1$ for the coherent state
$|\alpha\rangle$). \label{fig:figsups5}}

\end{figure}

Thus observation of squeezing $\Delta^{2}p<1$, so that the inequality\begin{equation}
\Delta^{2}p<1/(1+s_{\alpha}^{2})\label{eq:supcohresult}\end{equation}
is violated, will allow deduction of superpositions of coherent states
with separation at least $s_{\alpha}$. This separation corresponds
to a separation of $S_{\alpha}=2s_{\alpha}$ in $x$ between the two
corresponding Gaussian distributions (Fig. \ref{fig:figsups5}), on
the scale where $\Delta^{2}x=1$ is the variance predicted by each
coherent state. 

We note that measured values of squeezing $\Delta p\approx0.4$ \cite{Suzuki2006}
would imply $s_{\alpha}\gtrsim2.2$. This confirms the existence of
a superposition of type \begin{equation}
|\psi_{S}\rangle=\sum_{i}c_{i}|\alpha_{i}\rangle=c_{-}|-\alpha_{0}\rangle+...+c_{+}|\alpha_{0}\rangle\label{eq:gencohsupstexp}\end{equation}
where a separation of at least $s_{\alpha}=|\alpha_{i}-\alpha_{j}|=2.2$
occurs between two coherent states comprising the superposition, so
that we may write $\alpha_{0}=1.1$. Note we have defined reference
axes in phase space selected so that the $x$-axis is the line connecting
the two most separated states $|\alpha_{i}\rangle$and $|\alpha_{j}\rangle$so
that $|\alpha_{i}-\alpha_{j}|=2\alpha_{0}$ and the $p$-axis cuts
bisects this line. The (\ref{eq:gencohsupstexp}) can be compared
with experimental reports \cite{Ourjoumtsev2006} of generation of
extreme coherent superpositions of type $(1/\sqrt{2})\{e^{i\pi/4}|-\alpha_{0}\rangle+e^{-i\pi/4}|\alpha_{0}\rangle\}$
where $|\alpha_{0}|^{2}=0.79$, implying $\alpha_{0}=0.89$. The corresponding
generalised $s_{\alpha}-$scopic superposition (\ref{eq:gencohsupstexp})
as confirmed by the squeezing measurement involves at least the two
extreme states with $|\alpha_{0}|^{2}=1.2$, but could include other
coherent states with $|\alpha_{0}|<1.1$.

\section{Predictions of particular quantum states\label{sec:Predictions-of-particular}}

We will now consider experimental tests of the inequalities derived
above. An important point is that the criteria presented are \emph{sufficient}
to prove the existence of generalised macroscopic superpositions,
but there are many macroscopic superpositions which do not satisfy
the above criteria. Nevertheless there are some systems of current
experimental interest which do allow for violation of the inequalities.
We analyse such cases below, noting that the violation would be predicted
without the experimenter needing to make assumptions about the particular
state involved.

\subsection{Coherent states}

The wave function for the coherent state $|\alpha\rangle$ is \begin{equation}
\left\langle x|\alpha\right\rangle =\frac{1}{(2\pi)^{\frac{1}{4}}}\exp\{\frac{-x^{2}}{4}+\alpha x-|\alpha|^{2}\}.\label{eq:cohwfun}\end{equation}
This gives the expansion in the continuous basis set $|x\rangle$,
the eigenstates of $\hat{x}$. Thus for the coherent state\begin{equation}
|\alpha\rangle=\sum_{x}c_{x}|x\rangle=\int\langle x|\alpha\rangle|x\rangle dx.\label{eq:cohexp}\end{equation}
The probability distribution for $\hat{x}$ is the Gaussian (Fig.
\ref{fig:figcoh4})\begin{equation}
P(x)=|\langle x|\alpha\rangle|^{2}=\frac{1}{(2\pi)^{\frac{1}{2}}}\exp\{\frac{-(x-2\alpha)^{2}}{2}\},\label{eq:cohgauspx}\end{equation}
(we take $\alpha$ to be real) centred at $2\alpha$ and with variance
$\Delta^{2}x=1$

\begin{figure}
\begin{centering}
\includegraphics[width=11cm]{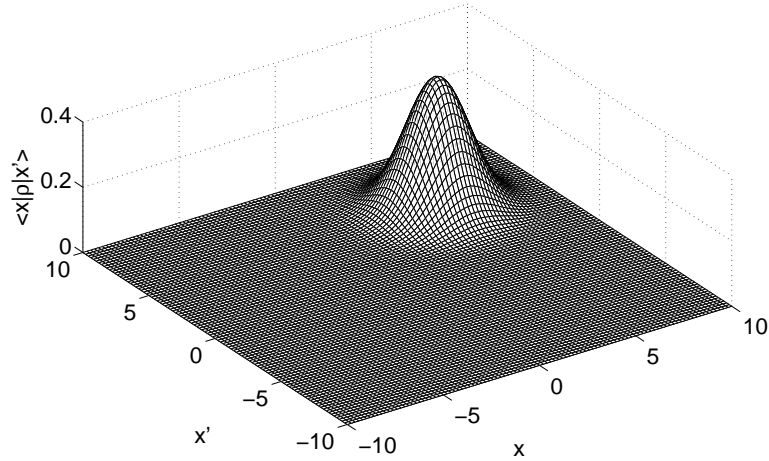}
\par\end{centering}

\caption{Plot of $\langle x|\rho|x'\rangle$ for a coherent state $|\alpha\rangle$.
\label{fig:figcoh3D6} }

\end{figure}

The coherent state possesses nonzero off-diagonal elements $\langle x|\rho|x'\rangle$
where $|x-x'|$ is large and thus strictly speaking can be regarded
as a generalised macroscopic superposition. However, as $x$ and $x'$
deviate from $2\alpha$, the matrix elements decay rapidly, and the
off-diagonal elements decay rapidly with increasing separation. \begin{equation}
\langle x|\rho|x'\rangle=\frac{1}{(2\pi)^{\frac{1}{2}}}\exp\{\frac{-(x-2\alpha)^{2}}{4}+\frac{-(x'-2\alpha)^{2}}{4}\}\label{eq:cohdiaele}\end{equation}
In effect then, the off-diagonal elements become zero for significant
separations $|x-x'|\geq1$ (Fig.\ref{fig:figcoh3D6}). We can expect
that the detection of the macroscopic aspects of this superposition
will be difficult. Since $\Delta p=1$, it follows that we can use
the criterion (\ref{eq:nobinning ineq}) to prove coherence between
outcomes of $x$ separated by at most $S=2$ (Fig. \ref{fig:figcoh4}),
which corresponds to the separation $S=2\Delta x$.

\subsection{Superpositions of coherent states}

The superposition of two coherent states \cite{Walls1985,Yurke1986}\begin{equation}
|\psi\rangle=(1/\sqrt{2})\{e^{i\pi/4}|-\alpha\rangle+e^{-i\pi/4}|\alpha\rangle\},\label{eq:scatys}\end{equation}
where $\alpha$ is real and large is an example of a macroscopic superposition
state. The wave function in the position basis is \[
\langle x|\psi\rangle=\frac{-ie^{i\pi/4}e^{[-x^{2}/4-\alpha_{0}^{2}]}}{\sqrt{2}(2\pi)^{\frac{1}{4}}}\{e^{\alpha x}+ie^{-\alpha x}\}.\]

We consider the two complementary observables $\hat{x}$ and $\hat{p}$,
and note that the probability distribution $P(x)$ for $\hat{x}$
displays two Gaussian peaks centred on $x=\pm2\alpha$ (Fig. \ref{fig:figsups5}):
$P(x)=\frac{1}{2}\sum_{\pm}P_{G\pm}(x)$ where $P_{G\pm}(x)=\exp[{-(x-\pm2\alpha)^{2}/2}]/\sqrt{2\pi}$.
Each Gaussian has variance $\Delta^{2}x=1$.

\begin{figure}
\begin{centering}
\includegraphics[width=13cm]{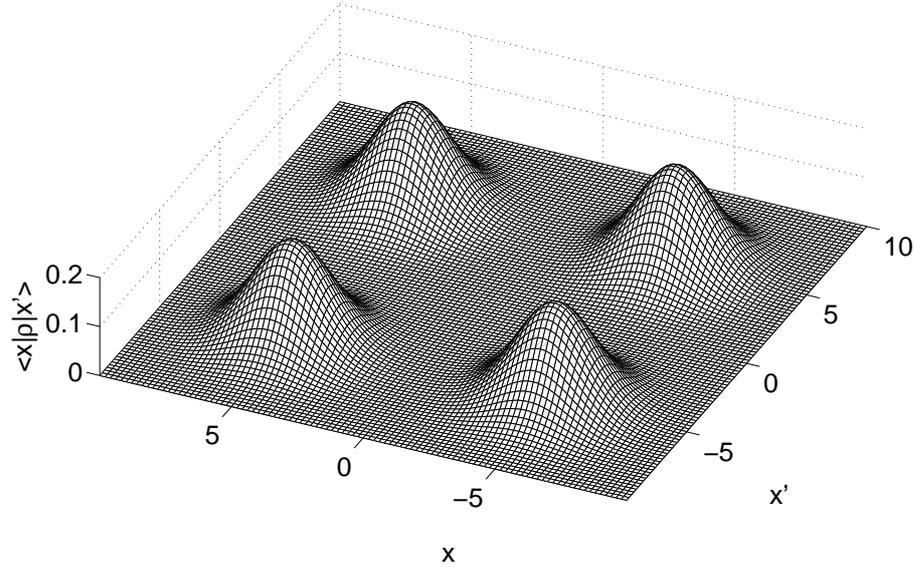}
\par\end{centering}

\caption{Plot of $\langle x|\rho|x'\rangle$ for the superposition state (\ref{eq:scatys}).
\label{fig:pxysup3d7}}

\end{figure}

The macroscopic nature of the superposition is reflected in the significant
magnitude of the off-diagonal elements $\langle x|\rho|x'\rangle$
where $x=\pm2\alpha$ and $x'=\mp2\alpha$, corresponding to $|x-x'|=4\alpha$.
In fact \begin{equation}
|\langle x|\rho|x'\rangle|=\frac{e^{\frac{-(x^{2}+x'^{2})}{4}-2\alpha_{0}^{2}}}{\sqrt{2\pi}}\sqrt{\cosh(2\alpha x)\cosh(2\alpha x')},\label{eq:supcohdiaele}\end{equation}
 as is plotted in Fig. \ref{fig:pxysup3d7} and which for these values
of $x$ and $x'$ becomes $\frac{(1-e^{-8\alpha^{2}})}{2(2\pi)^{\frac{1}{2}}}$.
With significant off-diagonal elements connecting macroscopically
different values of $x$, this superposition is a good example of
a generalised macroscopic superposition (\ref{eq: QMGMS}). 

Nonetheless we show that the simple linear criteria (\ref{eq:nobinning ineq})
and (\ref{eq:critsuper}) derived from (\ref{eq:mixture}) are not
sufficiently sensitive to detect the extent of the macroscopic coherence
of this superposition state (\ref{eq:scatys}), even though the state
(\ref{eq:scatys}) cannot be written in the form (\ref{eq:Binned mixture}).
We point out that it may be possible to derive further nonlinear constraints
from (\ref{eq:Binned mixture}) to arrive at more sensitive criteria.

\begin{figure}
\begin{centering}
\includegraphics[width=11cm]{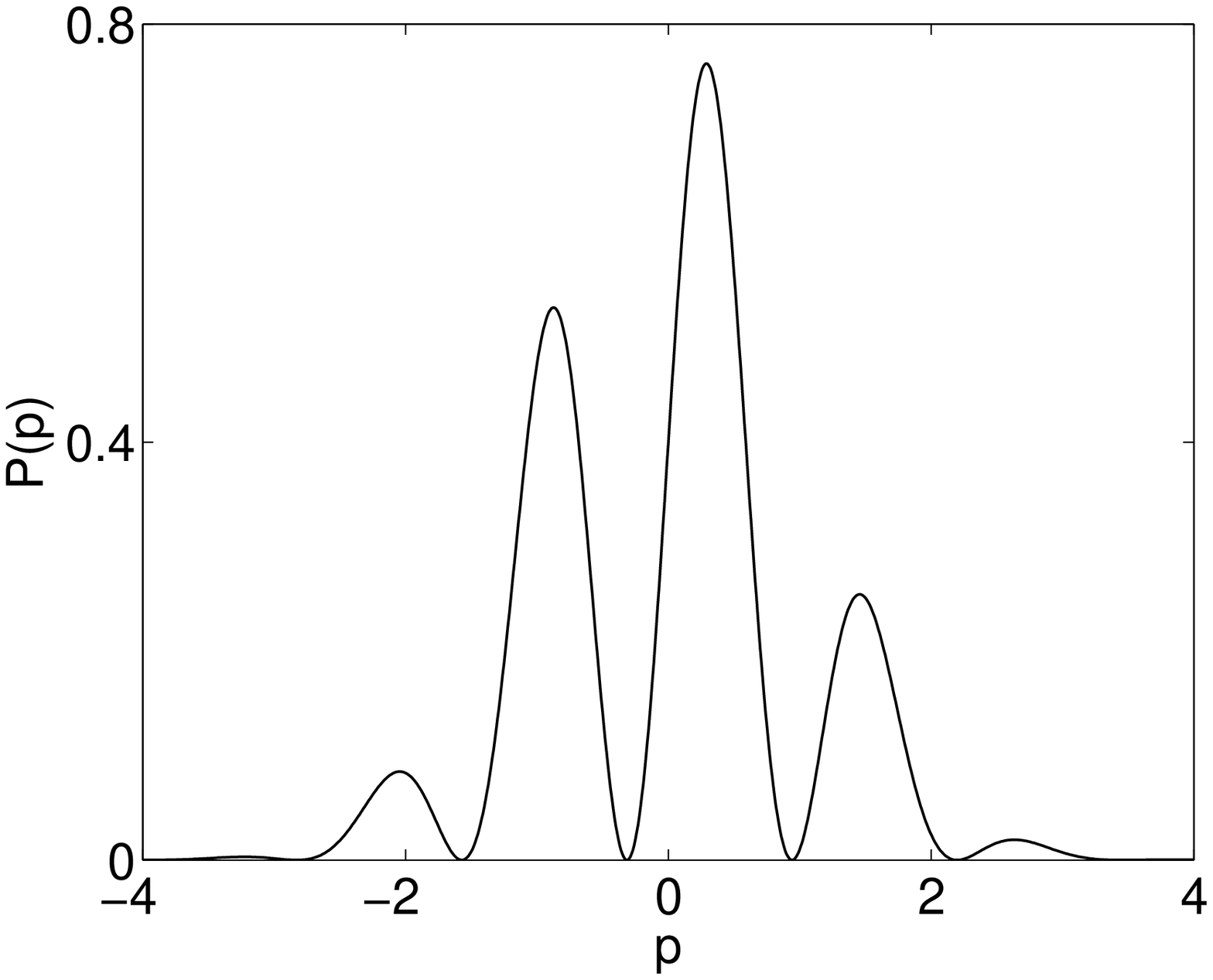}~~\\
 \includegraphics[width=11cm]{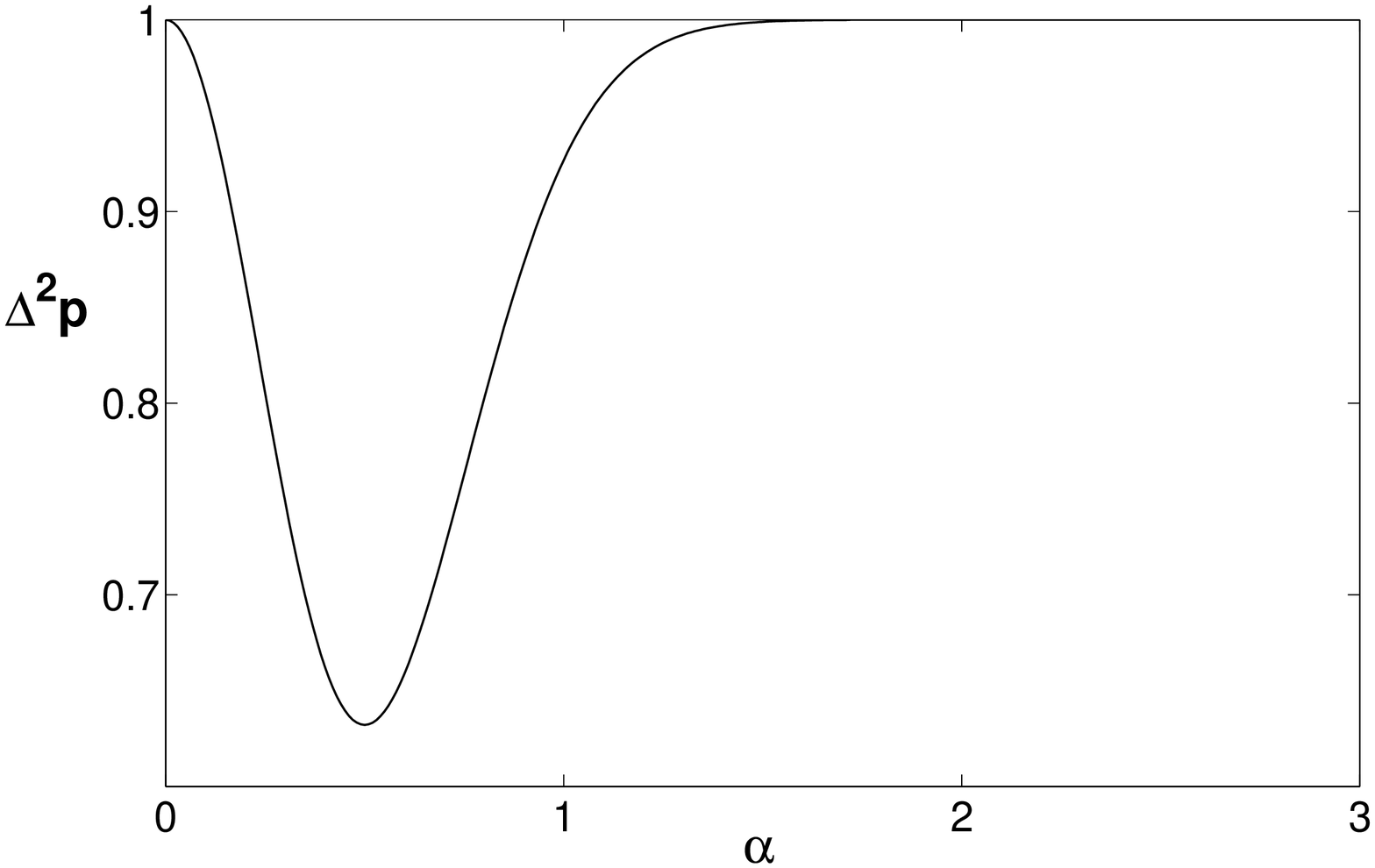}
\par\end{centering}

\caption{(a) P(p) for a superposition (\ref{eq:scatys}) of two coherent states
where $\alpha=2.5$ and (b) the reduced variance $\Delta^{2}p<1$,
versus $\alpha$. \label{fig:fringes8}}

\end{figure}

To investigate what can be inferred from criteria (\ref{eq:nobinning ineq}),
we note that $\hat{x}$ is the macroscopic observable. The complementary
observable $\hat{p}$ has distribution $P(p)=\exp{[-p^{2}/2]}(1+\sin{2\alpha p)}/\sqrt{2\pi}$
which exhibits fringes and has variance $\Delta^{2}p=1-4\alpha^{2}\exp{[-4\alpha^{2}]}$
(Fig. \ref{fig:fringes8}). There is a maximum squeezing of $\Delta^{2}p\approx0.63$
at $\alpha=0.5$. However, the squeezing diminishes as $\alpha$ increases,
so the criterion becomes less effective as the separation of states
of the macroscopic superposition increases. The maximum separation
$S$ that could be conclusively inferred from this criterion is $S\approx2.5$
at $\alpha=0.5$.

As discussed in Section \ref{sec:Criteria-for-generalised}, the detection
of squeezing in $p$ is enough to confirm the system is \emph{not}
that of the mixture \begin{equation}
\rho=1/2(|\alpha\rangle\langle\alpha|+|-\alpha\rangle\langle-\alpha|)\label{eq:mixcoh}\end{equation}
of the two coherent states. In fact, the squeezing rules out that
the system is any mixture of coherent states. We note though that
since the degree of squeezing $\Delta p$ is small, our criteria is
not sensitive enough to rule out superpositions of macroscopically
separated coherent states.

\subsection{Squeezed states}

\begin{figure}
\begin{centering}
\includegraphics[width=12cm]{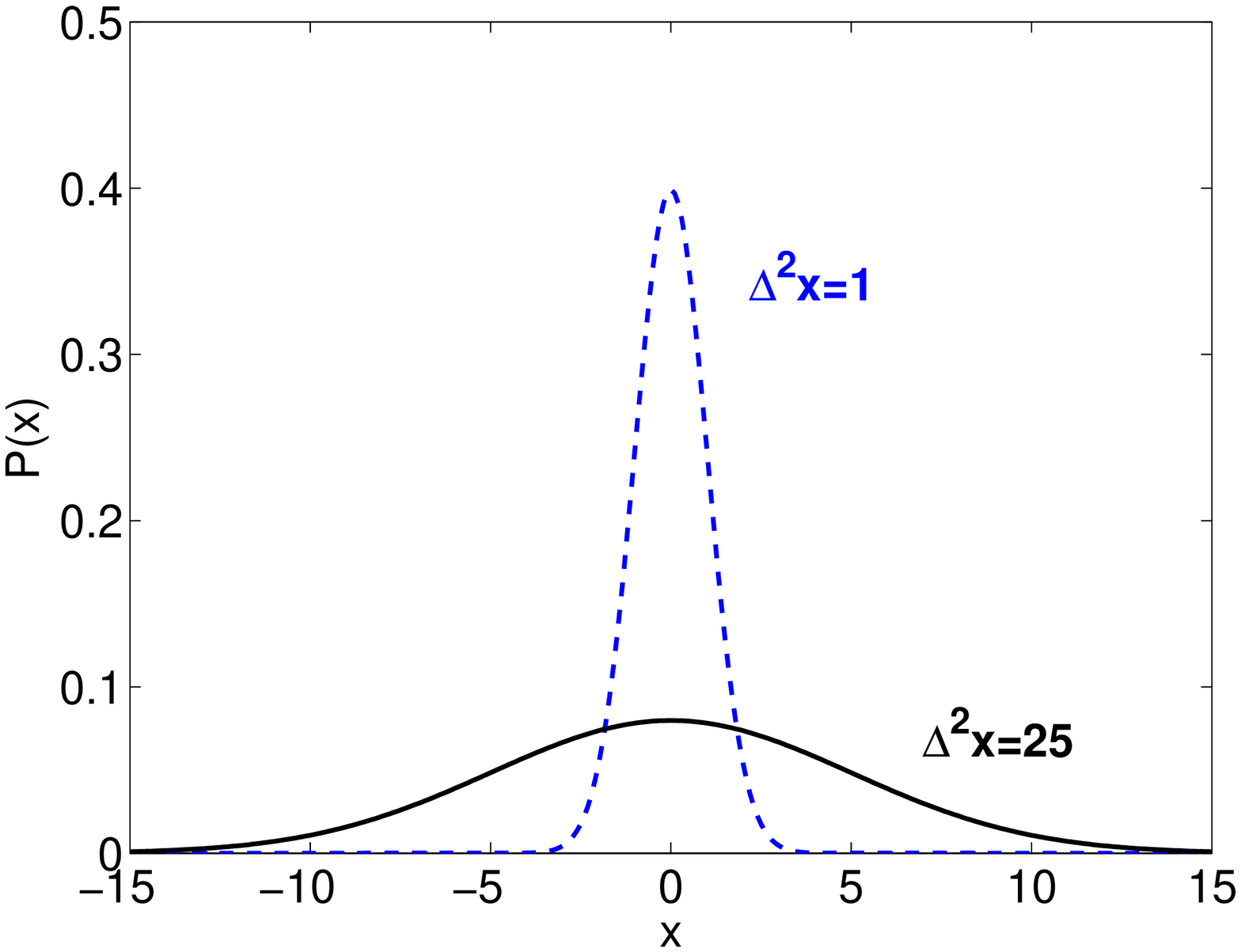}~\\
\includegraphics[width=12cm]{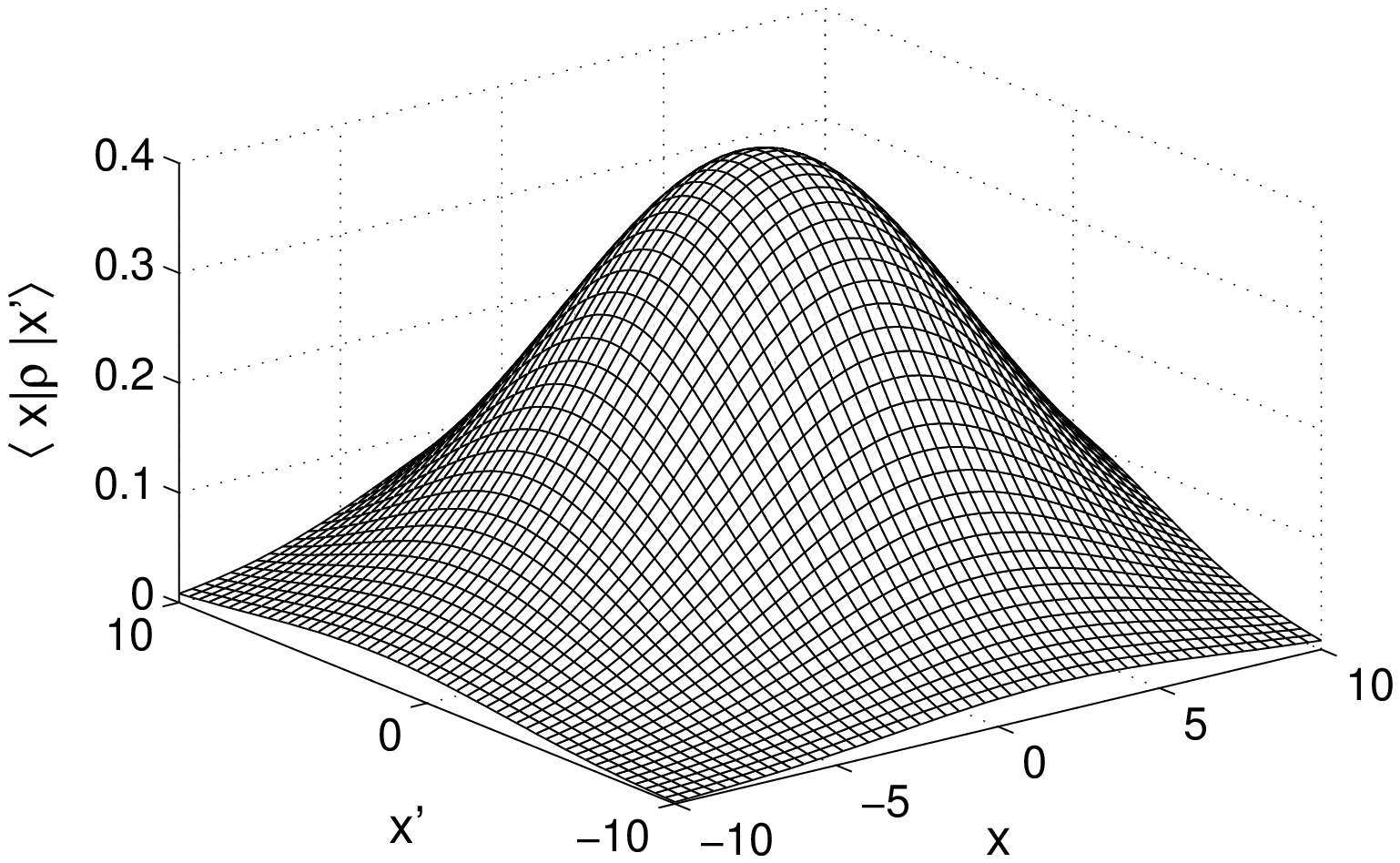}
\par\end{centering}

\caption{(a) Probability distribution for a measurement $X$ for a momentum-squeezed
state. The variance $\Delta^{2}x$ increases with squeezing in $p$,
to give a macroscopic range of outcomes, and for the minimum uncertainty
state (\ref{eq:sqestate}) satisfies $\Delta x\Delta p=1$. (b) The
$\langle x|\rho|x'\rangle$ for a squeezed state (\ref{eq:sqestate})
with $r=13.4$ ($\Delta x=3.67)$ which predicts $\langle a^{\dagger}a\rangle=2.5^{2}$.
\label{fig: gausssquee9}}

\end{figure}

Consider the single-mode momentum squeezed state\cite{Yuen1976} \begin{equation}
|\psi\rangle=e^{r(a²-a^{\dagger2})}\left|0\right\rangle .\label{eq:sqestate}\end{equation}
Here $\left|0\right\rangle $ is the vacuum state. For large values
of $r$ these states are generalised macroscopic superpositions of
the continuous set of eigenstates $|x\rangle$ of $\hat{x}=a+a^{\dagger}$,
with wave function\begin{equation}
\left\langle x|\psi\right\rangle =\frac{1}{(2\pi\sigma)^{\frac{1}{4}}}\exp\{\frac{-x^{2}}{4\sigma}\},\label{eq:sqstatewf}\end{equation}
 and associated Gaussian probability distribution \begin{equation}
P(x)=\frac{1}{(2\pi\sigma)^{\frac{1}{2}}}\exp\{\frac{-x^{2}}{2\sigma}\}.\label{eq:gausp}\end{equation}
The variance is $\sigma=e^{2r}$. As the squeeze parameter $r$ increases,
the probability distribution expands, so that eventually with large
enough $r$, $x$ can be regarded as a macroscopic observable. This
behaviour is shown in Fig. \ref{fig: gausssquee9}. The distribution
for $p$ is also Gaussian but is squeezed, meaning that it has reduced
variance: $\Delta^{2}p<1$. In fact, the (\ref{eq:sqestate}) is a
minimum uncertainty state, with $\Delta^{2}p=1/\sigma=e^{-2r}$. Where
squeezing is significant, the off-diagonal elements $\langle x|\rho|x'\rangle=\langle x|\psi\rangle\langle\psi|x'\rangle$
(where $|x-x'|$ is large) are significant over a large range of $x$
values (Fig. \ref{fig: gausssquee9}). 

\begin{figure}
\begin{centering}
\includegraphics[width=12cm]{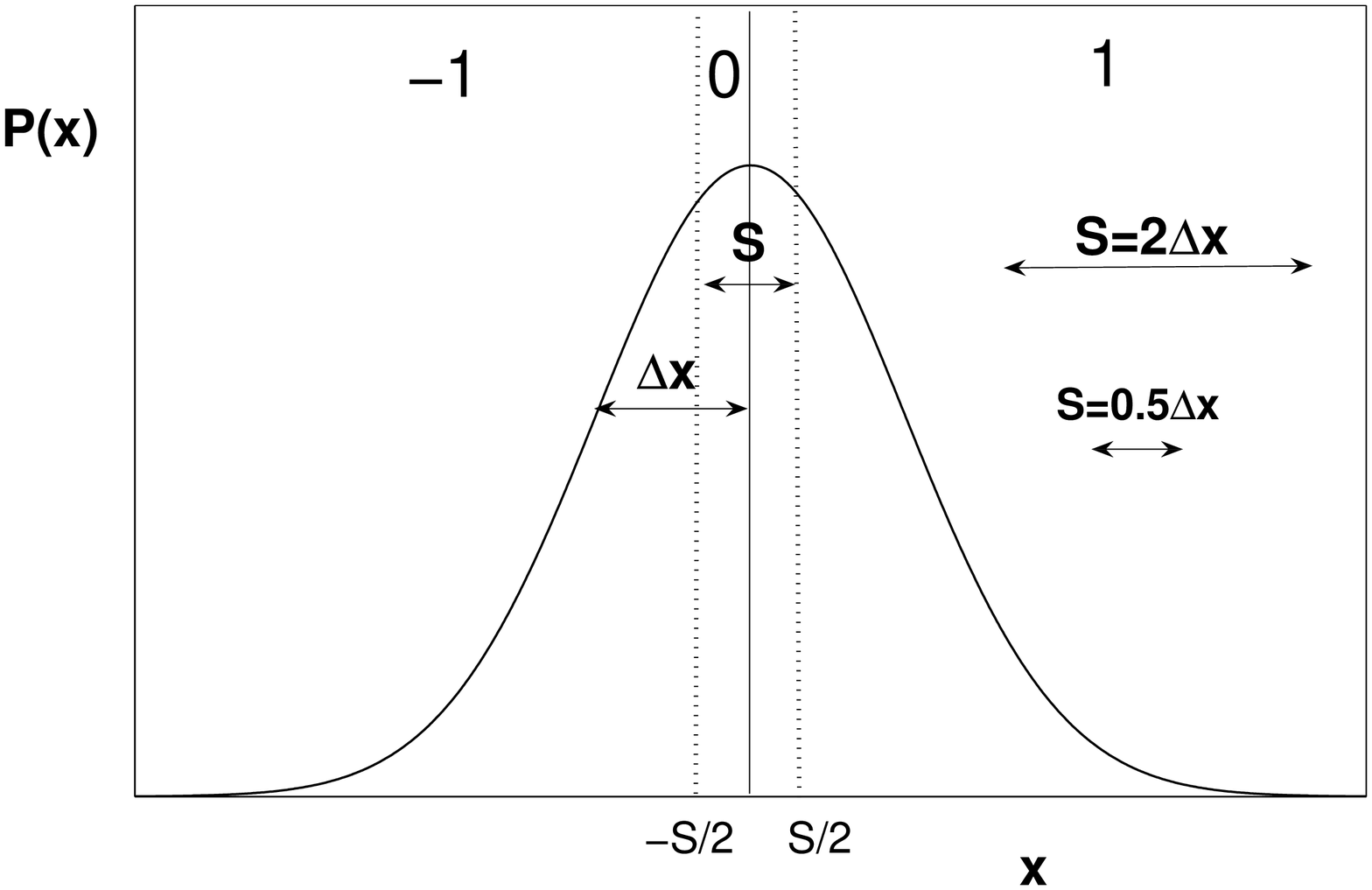}\\
 \includegraphics[width=12cm]{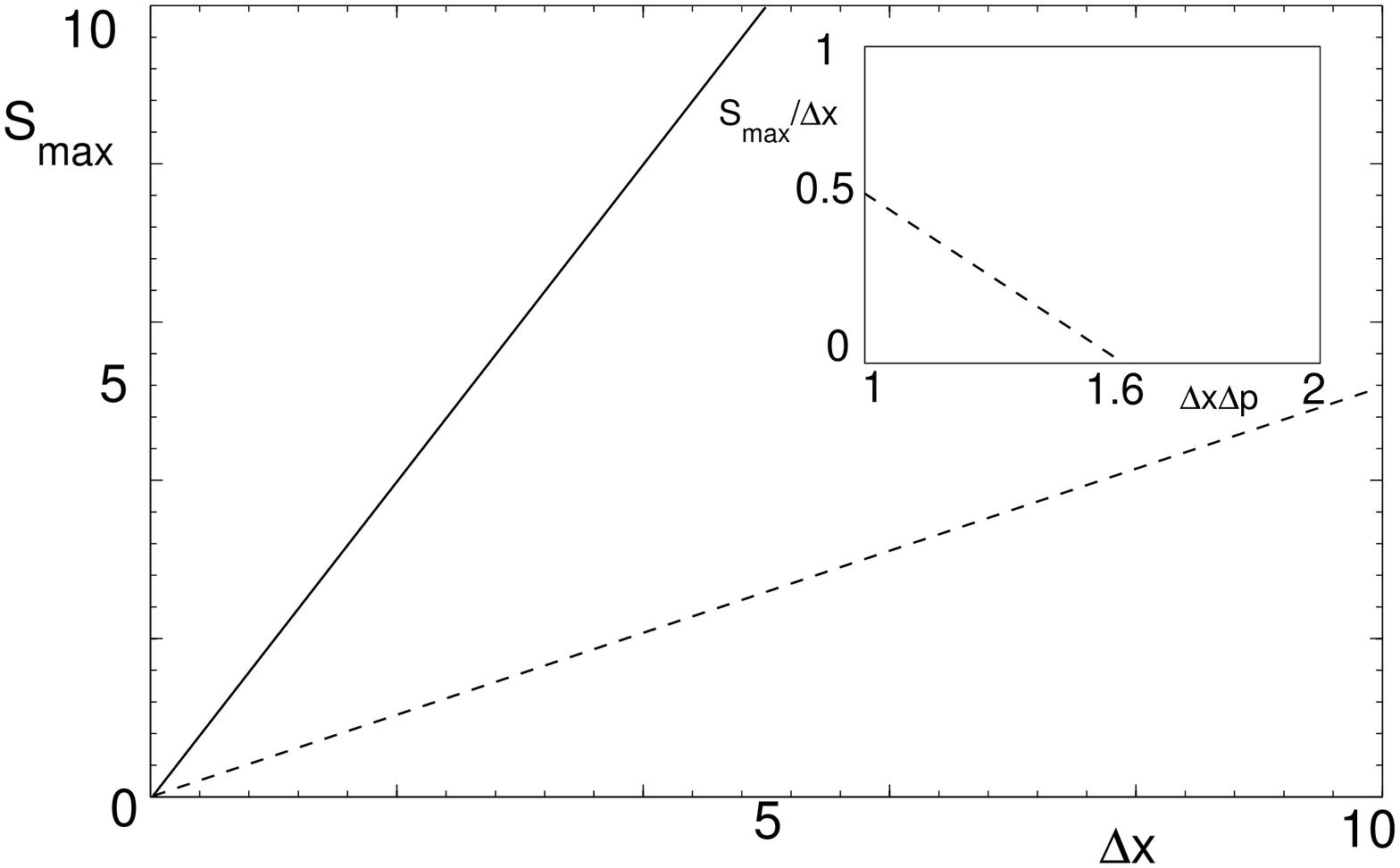}
\par\end{centering}

\caption{Detection of underlying superpositions of size $S$ for the squeezed
minimum uncertainty state (\ref{eq:sqestate}) by violation of (\ref{eq:critsuper})
(dashed line of (b)) and (\ref{eq:nobinning ineq}) (full line of
(b)). $S_{max}$ is the maximum $S$ for which the inequalities are
violated. Inset of (b) shows behaviour of violation of (\ref{eq:critsuper})
for general Gaussian-squeezed states. Inequality (\ref{eq:nobinning ineq})
depends only on $\Delta p$. The size of $S_{max}$ relative to $P(x)$
is illustrated in (a). \label{fig:gauss10}}

\end{figure}

The criterion (\ref{eq:critsuper}) for the binned outcomes is violated
for the ideal squeezed state (\ref{eq:sqestate}) for values of $S$
up to $0.5\sqrt{\sigma}$. The criterion can thus confirm macroscopic
superpositions of states with separation of up to half the standard
deviation of the probability distribution of $x$, even as $\Delta x\rightarrow\infty$.
This behaviour has been reported in \cite{Cavalcanti2006} and is
shown in Fig. \ref{fig:gauss10}.

Squeezed systems that are generated experimentally will not be describable
as the pure squeezed state (\ref{eq:sqestate}). This pure state is
a minimum uncertainty state with $\Delta x\Delta p=1$. Typically
experimental data will generate Gaussian probability distributions
for both $x$ and $p$ and with squeezing $\Delta p<1$ in $p$, but
typically $\Delta x\Delta p>1$. The maximum value of $S$ that can
be proved in this case of the Gaussian states reduces to $0$ as $\Delta x\Delta p$
(or $\Delta x\Delta_{inf}p$) increases to $\sim1.6$. This is shown
in Fig. \ref{fig:gauss10}. Analysis of recent experimental data for
impure states that allows a violation of (\ref{eq:critsuper}) has
been reported by Marquardt et al \cite{Marquardt2007}.

The criterion (\ref{eq:nobinning ineq}), as given by Theorem 4, is
better able to detect the superpositions (Fig. \ref{fig:gauss10}),
particularly where the uncertainty product gives $\Delta x\Delta p>1$,
though in this case the superpositions are non-locatable in phase
space, so that we cannot conclude an outcome domain for the states
involved in the superposition. This criterion depends only on the
squeezing $\Delta p$ in the one quadrature and is not sensitive to
the product $\Delta x\Delta p$. For ideal squeezed states with variance
$\Delta^{2}x=\sigma$, one can prove a superposition of size $S=2\sqrt{\sigma}$,
four times that obtained from (\ref{eq:critsuper}) (Fig. \ref{fig:gauss10}). 

Experimental reports \cite{Suzuki2006} of squeezing of orders $\Delta p\approx0.4$
confirms superpositions of size at least $S=5$, which is 2.5 times
that defined by $S=2$, which corresponds to two standard deviations
of the coherent state, for which $\Delta x=1$ (Fig. \ref{fig:figcoh4}).

\subsection{Two-mode squeezed states}

Next we consider the two- mode squeezed state \cite{Caves1985}

\begin{equation}
e^{r(ab-a^{\dagger}b^{\dagger})}|0\rangle|0\rangle.\label{eq:twomodesqstate}\end{equation}
Here $a,b$ are boson annihilation operators for modes $A$ and $B$
respectively. The wave function $\langle x|\psi\rangle$ and distribution
$P(x)$ are as in (\ref{eq:sqstatewf}) and (\ref{eq:gausp}), but
the variance in $\hat{x}=X^{A}$ is now given by $\sigma=\cosh2r$.
The $\hat{x}=X^{A}$ is thus a macroscopic observable. 

In the two-mode case, the squeezing is in a linear combination $P^{A}+P^{B}$
of the momenta $P^{A}$ and $P^{B}$ at $A$ and $B$, rather than
in the momentum $\hat{p}=P^{A}$ for $A$ itself. The observable that
is complementary to $X^{A}$ is of form $\tilde{P}=P^{A}-gP^{B}$
where $g$ is a constant, which is (\ref{eq:ptilde}) of Section \ref{sec:Signatures-for-generalised}.
We can select to evaluate one of the criteria (\ref{eqn:eprcat}),
(\ref{eq:nobinning ineq2}) or (\ref{eq:resultcompsum}).

Choosing as our macroscopic observable $X^{A}$ and our complementary
one $P^{A}-gP^{B}$, we calculate\begin{equation}
\Delta_{inf}^{2}p^{A}=1/\sigma=1/cosh2r\label{eq:infresult}\end{equation}
for the choice $g=\left\langle P^{A}P^{B}\right\rangle /\left\langle (P^{B})^{2}\right\rangle =-tanhr$
which minimises $\Delta_{inf}^{2}p^{A}$ \cite{Reid1989}. The application
of results to criterion (\ref{eqn:eprcat}) gives the result as in
Fig. \ref{fig:gauss10}, to indicate detection of superpositions of
size $S$ where $S=0.5\sqrt{\sigma}$ for the ideal squeezed state
(\ref{eq:twomodesqstate}), and the result shown in the inset of Fig.
\ref{fig: gausssquee9} if $\Delta x^{A}\Delta_{inf}p^{A}>1$.

The prediction for the criterion of Theorem 3, to detect superpositions
in the position sum $X^{A}+X^{B}$ by measurement of a narrowed variance
in the momenta sum $P^{A}+P^{B}$, is also given by the results of
Fig. \ref{fig:gauss10}. Calculation for the ideal state (\ref{eq:twomodesqstate})
predicts $\Delta^{2}(\frac{p^{A}+p^{B}}{\sqrt{2}})=e^{-2r}$ and $\Delta^{2}(\frac{x^{A}+x^{B}}{\sqrt{2}})=e^{+2r}$
which corresponds to that of the single-mode squeezed state. The prediction
for the maximum value of $S$ of Theorem 3 is therefore given by the
dashed curves of Fig. \ref{fig:gauss10}, and the inset.

A better result is given by (\ref{eq:nobinning ineq2}), if we are
not concerned with the location of the superposition. Where we use
(\ref{eq:nobinning ineq2}), the degree of reduction in $\Delta_{inf}^{2}p^{A}$
determines the size of superposition $S$ that may be inferred. By
Theorem 5, measurement of $\Delta_{inf}p^{A}$ allows inference of
superpositions of eigenstates of $X^{A}$ separated by at least \begin{equation}
S=2/\Delta_{inf}p^{A}.\label{eq:sdpinf}\end{equation}
Realistic states are not likely to be pure squeezed states as given
by (\ref{eq:twomodesqstate}). Nonetheless the degree of squeezing
indicates a size of superposition in $X^{A}$, as given by Theorem
5. Experimental values of $\Delta_{inf}^{2}p^{A}\approx0.76$ have
been reported \cite{Bowen2003}, to give confirmation of superpositions
of size $S\approx2.3$, which is $1.1$ times the level of $S=2$
that corresponds to two standard deviations $\Delta x^{A}=1$ of the
vacuum state (Fig. \ref{fig:figcoh4}). 

More frequently, it is the practice to measure squeezing in the direct
sum $P^{A}+P^{B}$ of momenta. The macroscopic observable is the position
sum $X^{A}+X^{B}$. The reports of measured experimental values indicate
\cite{Laurat2005} $\Delta^{2}(\frac{p^{A}+p^{B}}{\sqrt{2}})\approx0.4$,
which according to Theorem 5 implies superpositions in $(X^{A}+X^{B})/\sqrt{2}$
of size $S\approx3.2$, of order $1.6$ times the standard vacuum
state level . The slightly better experimental result for the superpositions
in the position sum may be understood since it has been shown by Bowen
\emph{et al.} \cite{Bowen2003} that, for the Gaussian squeezed states,
the measurement of $\Delta_{inf}^{2}p^{A}$ is more sensitive to loss
than that of $\Delta^{2}(p^{A}+p^{B})$. The $\Delta_{inf}p^{A}$
is an asymmetric measure that enables demonstration of the EPR paradox
\cite{Reid1989,Reid2003}, a strong form of quantum nonlocality \cite{Reid2007tb,Wiseman2007}.

\section{Concluding remarks}

In this chapter we have derived criteria sufficient to detect generalised
macroscopic (or $S$-scopic) superpositions ($\sum_{k_{1}}^{k_{2}}c_{k}|x_{k}\rangle$)
of eigenstates of an observable $\hat{x}$. For these superpositions,
the important quantity is the value $S$ of the \emph{extent} of the
superposition, which is the range in prediction of the observable
($S$ is the maximum of $|x_{j}-x_{i}|$ where $c_{j},c_{i}\neq0$).
This quantity gives the extent of indeterminacy in the quantum prediction
for $\hat{x}$. In this sense, there is a contrast with the prototype
macroscopic superposition ( of type $c_{2}|x_{2}\rangle+c_{1}|x_{1}\rangle$)
that relates directly to the essay of Schrödinger \cite{Schroedinger1935}.
Such a prototype superposition contains only the two states that have
separation $S$ in their outcomes for $x$. Nonetheless, we have discussed
how the generalised superposition is relevant to testing the ideas
of Schrödinger, in that such macroscopic superpositions are shown
to be inconsistent with the hypothesis of a quantum system being in
at most one of two macroscopically separated states.

We have also defined the concept of a generalised $S$-scopic coherence
and the class of Minimum Uncertainty Theories without direct reference
to quantum mechanics. The former is introduced in Section \ref{sec:Generalized-S-scopic-Coherence}
as the assumption (\ref{eq:probmacro}) and is associated to the failure
of a generalised assumption of macroscopic reality. This assumption is
that the system is in at most one of two macroscopically distinguishable
states, but that these underlying states are not specified to be quantum
states. The assumption of Minimum Uncertainty Theories is that these
component states do at least satisfy the quantum uncertainty relations.
In the derivation of the criteria of this chapter, only two assumptions
are made: that the system does satisfy this generalised macroscopic
($S$-scopic) reality and that the theory is a Minimum Uncertainty
Theory. These assumptions lead to inequalities, which, when violated,
generate evidence that at least one of the assumptions must be incorrect. 

We point out that if, in the event of violation of the inequalities,
we opt to conclude the failure of the Minimum Uncertainty Theory assumption,
then this does not imply quantum mechanics to be incorrect, but rather
that it is incomplete, in the sense that the component states can
themselves not be quantum states. It can be said then that violation
of the inequalities of this chapter implies at least one of the assumptions
of \emph{generalised macroscopic ($S$-scopic) reality} and the\emph{
completeness of quantum mechanics} is incorrect. 

There is a similarity with the Einstein-Podolsky-Rosen argument \cite{Einstein1935}.
In the EPR argument, the assumption of a form of realism (local realism)
is shown to be inconsistent with the completeness of quantum mechanics.
Therefore, as a conclusion of that argument, one is left to conclude
that at least one of \emph{local realism} and the \emph{completeness
of QM} is incorrect \cite{Reid2003,Wiseman2006,Reid2007tb}. EPR opted
for the first and took their argument as a demonstration that quantum
mechanics was incomplete. Only after Bell \cite{Bell1964} was it
shown that this was an incorrect choice. Here, as in the EPR argument,
the assumption of a form of realism (macroscopic ($S$-scopic) realism)
can only be made consistent with the predictions of quantum mechanics
if one allows a kind of theory in which the underlying states are
not restricted by the uncertainty relations \cite{Cavalcanti2006}.

\chapter{\label{cha:Conclusion}Conclusion}

As mentioned in the Introduction, the objective of this thesis was
to contribute to the area or research termed \emph{experimental metaphysics.
}Our modest contribution was in formalising old concepts, proposing
new ones, and finding new results in well-studied areas. We have also
proposed experiments to test each of the major results. It is \emph{experimental}
metaphysics, after all.

In Chapter \ref{cha:Concepts} we set up the appropriate terminology
and the basic concepts. Most of it were simply careful definitions
of standard concepts, but some definitions were new and some results
and consequences may not have been fully appreciated before. 

In Chapter \ref{cha:EPR-paradox} we analysed the original argument
of Einstein, Podolsky and Rosen (EPR) \cite{Einstein1935}, and proposed
a general mathematical form for the assumptions behind that argument,
namely those of \emph{local causality} and \emph{completeness} \emph{of
quantum theory}. That entailed what was termed a Local Hidden State
model by Wiseman \emph{et al. \cite{Wiseman2007}, }which was proposed
as a formalisation of the concept of \emph{steering }first introduced
by Schrödinger \cite{Schroedinger1935} in a reply to the EPR paper.
Violation of any consequences that can be derived from the assumption
of that model therefore implies a demonstration of the EPR paradox.
We have re-derived the well-known EPR-Reid criterion \cite{Reid1989}
for continuous-variables correlations, and derived new ones applicable
to the spin setting considered by Bohm \cite{Bohm1951}.

The spin set-up of the EPR-Bohm paradox was used by Bell \cite{Bell1964}
to derive his now famous theorem demonstrating the incompatibility
of the assumption of local causality and the predictions of quantum
mechanics. The inequalities which bear his name can be derived for
any number of discrete outcomes, but so far there has been no derivation
which can be directly applied to the continuous-variables case of
the original EPR paradox. In Chapter \ref{cha:CV-Bell} we closed
the circle by deriving a class of inequalities which make no explicit
mention about the number of outcomes of the experiments involved,
and can therefore be used in continuous-variables measurements with
no need for binning the continuous results into discrete ones. Apart
from that intrinsic interest, these inequalities could prove important
as a means to perform an unambiguous test of Bell inequalities which
does not suffer from the logical loopholes that plague all experimental
demonstrations so far, since optical homodyne detection can be performed
with high detection efficiency. The technique, which is based on a
simple variance inequality, was also used to re-derive a large class
of well-known Bell-type inequalities and at the same time find their
quantum bound, making explicit from a formal point of view that the
non-commutativity of the local operators is at the heart of the quantum
violations.

Finally, in Chapter \ref{cha:Macro-Super} we addressed the issue
of macroscopic superpositions originally sparked by the infamous \char`\"{}cat
paradox\char`\"{} of Schrödinger \cite{Schroedinger1935}, presented
in the same seminal paper where he coined the terms \emph{entanglement
}and \emph{steering}. We considered macroscopic, mesoscopic and `\emph{S}-scopic'
quantum superpositions of eigenstates of an observable, and developed
some signatures for their existence. We defined the extent, or size
\emph{S} of a (pure-state) superposition, with respect to an observable
\emph{X}, as being the maximum difference in the outcomes of \emph{X}
predicted by that superposition. Such superpositions were referred
to as generalised \emph{S}-scopic superpositions to distinguish them
from the extreme superpositions that superpose only the two states
that have a difference \emph{S} in their prediction for the observable.
We also considered generalised \emph{S}-scopic superpositions of coherent
states. We explored the constraints that are placed on the statistics
if we suppose a system to be described by mixtures of superpositions
that are restricted in size. In this way we arrived at experimental
criteria that are sufficient to deduce the existence of a generalised
\emph{S}-scopic superposition. The signatures developed are useful
where one is able to demonstrate a degree of squeezing.

\bibliographystyle{amsalpha}
\cleardoublepage\addcontentsline{toc}{chapter}{\bibname}\bibliography{Eric.bib}

\appendix

\chapter{Proof of Theorem A of Chapter \ref{cha:Macro-Super}}

We will now prove the statement that coherence between $x_{1}$ and
$x_{2}$ is equivalent to a nonzero off-diagonal element $\langle x_{1}|\rho|x_{2}\rangle$
in the density matrix. As discussed in section \ref{sec:Generalized-S-scopic-Coherence},
within quantum mechanics the statement that there exists coherence
between $x_{1}$ and $x_{2}$ is equivalent to the statement that
there is no decomposition of the density matrix of form (\ref{eq:QM mixture})
where $\rho_{1}$ and $\rho_{2}$ are density matrices such that $\langle x_{1}|\rho_{2}|x_{1}\rangle=\langle x_{2}|\rho_{1}|x_{2}\rangle=0$.
Therefore Theorem A can be reformulated as saying that $\langle x_{1}|\rho|x_{2}\rangle=0$
iff such a decomposition \emph{does} exist.

It's easy to prove the first direction of the equivalence: if $\exists$$\{\wp_{1},\,\wp_{2},\,\rho_{1},\,\rho_{2}\}$
such that $\rho=\wp_{1}\rho_{1}+\wp_{2}\rho_{2}$ and $\langle x_{1}|\rho_{2}|x_{1}\rangle=\langle x_{2}|\rho_{1}|x_{2}\rangle=0$,
then $\langle x_{1}|\rho|x_{2}\rangle=0$. To show this, first note
that for any density matrix $\bar{\rho}$ and $\forall\,\{x,x'\},$
if $\langle x|\bar{\rho}|x\rangle=0$ then $\langle x|\bar{\rho}|x'\rangle=0,$
where $\langle x|x'\rangle=\delta_{x,x'}$. Since by assumption $\langle x_{1}|\rho_{2}|x_{1}\rangle=\langle x_{2}|\rho_{1}|x_{2}\rangle=0$,
then $\langle x_{1}|\rho|x_{2}\rangle=\sum_{i}\wp_{i}\langle x_{1}|\rho_{i}|x_{2}\rangle=0$.

The converse can also be proved. We use the facts that any $\rho$
can always be written as the reduced density matrix of an enlarged
pure state, where the system of interest (call it $A$) is entangled
with an ancilla $B$, i.e, \begin{equation}
\rho=Tr_{B}\{|\Psi\rangle_{AB}\langle\Psi|_{AB}\}\label{eq:app1}\end{equation}
 and that any bipartite pure state can always be written in the Schmidt
decomposition \cite{Ekert1995}\begin{equation}
|\Psi\rangle_{AB}=\sum_{i}\sqrt{\eta_{i}}|\psi_{i}\rangle|\phi_{i}^{B}\rangle.\label{eq:purification}\end{equation}
where $\{|\psi_{i}\rangle\}$ and $\{|\phi_{i}^{B}\rangle\}$ are
orthonormal and $\eta_{i}\in[0,1].$ The superscript $B$ denotes
the states of the ancilla and the absence of a superscript denotes
the states of the system of interest, $A$. We decompose each pure
state $|\psi_{i}\rangle$ that appears in the Schmidt decomposition
in the basis of eigenstates of $\hat{x}$ as $|\psi_{i}\rangle=\sum_{k}c_{i,k}|x_{k}\rangle$.
By assumption $\langle x_{1}|\rho|x_{2}\rangle=0$ and therefore $\sum_{i}\eta_{i}\langle x_{1}|\psi_{i}\rangle\langle\psi_{i}|x_{2}\rangle=\sum_{i}\eta_{i}c_{i,1}c_{i,2}^{*}=0$.
We can expand $|\Psi_{AB}\rangle$ as\begin{equation}
|\Psi_{AB}\rangle=|x_{1}\rangle|\tilde{1_{B}}\rangle+|x_{2}\rangle|\tilde{2_{B}}\rangle+\sum_{k>2,\, i}\sqrt{\eta_{i}}c_{i,k}|x_{k}\rangle|\phi_{i}^{B}\rangle,\label{eq:purification expansion}\end{equation}
where we define the (unnormalized) $|\tilde{1_{B}}\rangle\equiv\sum_{i}\sqrt{\eta_{i}}c_{i,1}|\phi_{i}^{B}\rangle$
and $|\tilde{2_{B}}\rangle\equiv\sum_{i}\sqrt{\eta_{i}}c_{i,2}|\phi_{i}^{B}\rangle$.
The inner product of these two vectors is $\langle\tilde{1_{B}}|\tilde{2_{B}}\rangle=\sum_{i}\eta_{i}c_{i,1}c_{i,2}^{*}.$
But as shown above $\sum_{i}\eta_{i}c_{i,1}c_{i,2}^{*}=0$, so $|\tilde{1_{B}}\rangle$
and $|\tilde{2_{B}}\rangle$ are orthogonal. We can therefore define
an orthonormal basis with the (normalized) $|1_{B}\rangle=|\tilde{1_{B}}\rangle/\sqrt{\sum_{i}\eta_{i}|c_{i,1}|^{2}}$
and $|2_{B}\rangle=|\tilde{2_{B}}\rangle/\sqrt{\sum_{i}\eta_{i}|c_{i,2}|^{2}},$
plus additional $|j_{B}\rangle$with $3\leq j\leq D$, where $D$
is the dimension of subsystem $B$'s Hilbert space. Taking the trace
of $\rho_{AB}=|\Psi_{AB}\rangle\langle\Psi_{AB}|$ therefore yields\begin{eqnarray}
\rho & = & Tr_{B}\{\rho_{AB}\}\nonumber \\
 & = & \left\langle 1_{B}\right|\rho_{AB}\left|1_{B}\right\rangle +\left\langle 2_{B}\right|\rho_{AB}\left|2_{B}\right\rangle \nonumber \\
 &  & +\sum_{j>2}\left\langle j_{B}\right|\rho_{AB}\left|j_{B}\right\rangle .\label{eq:traceB}\end{eqnarray}
Now referring to expansion \eqref{eq:purification expansion}, we
see that $\langle1_{B}|\rho_{AB}|1_{B}\rangle=\sum_{i}\eta_{i}|c_{i,1}|^{2}|x_{1}\rangle\langle x_{1}|$
and $\langle2_{B}|\rho_{AB}|2_{B}\rangle=\sum_{i}\eta_{i}|c_{i,2}|^{2}|x_{2}\rangle\langle x_{2}|.$
We then define $\rho_{1}\equiv|x_{1}\rangle\langle x_{1}|,$ $\wp_{1}\equiv\sum_{i}\eta_{i}|c_{i,1}|^{2},$
$\wp_{2}=1-\wp_{1}$ and $\rho_{2}\equiv\frac{1}{\wp_{2}}\{\sum_{i}\eta_{i}|c_{i,2}|^{2}|x_{2}\rangle\langle x_{2}|+\sum_{j>2}\langle j_{B}|\rho_{AB}|j_{B}\rangle\}.$
Obviously $\langle x_{2}|\rho_{1}|x_{2}\rangle=0,$ and by substituting
\eqref{eq:purification expansion} into $\rho_{2}$ we see that\textbf{
}$\langle x_{1}|\rho_{2}|x_{1}\rangle=0.$ Therefore $\rho$ can be
decomposed as $\rho=\wp_{1}\rho_{1}+\wp_{2}\rho_{2}$ with $\langle x_{1}|\rho_{2}|x_{1}\rangle=\langle x_{2}|\rho_{1}|x_{2}\rangle=0$
as desired.

\chapter{Proof of Theorem 2 of Chapter \ref{cha:Macro-Super}}

We wish to prove that if $\rho$ can be written as \begin{equation}
\rho=\wp_{L}\rho_{L}+\wp_{R}\rho_{R},\label{eq:mixture}\end{equation}
then \begin{equation}
\Delta_{inf}^{2}p^{A}\geq\wp_{L}\Delta_{inf,L}^{2}p^{A}+\wp_{R}\Delta_{inf,R}^{2}p^{A}\label{eq:B2}\end{equation}
where \[
\Delta_{inf,J}^{2}p^{A}=\sum_{p^{B}}P_{J}(p^{B})\Delta_{J}^{2}(p^{A}|p^{B}).\]
The subscript $J$ refers to the $\rho_{J}$ from which the probabilities
are calculated.

We have\begin{eqnarray*}
\Delta_{inf}^{2}p^{A} & = & \sum_{p^{B}}P(p^{B})\Delta^{2}(p^{A}|p^{B})\\
 & = & \sum_{p^{B}}\sum_{p^{A}}P(p^{A},p^{B})(p^{A}-\langle p^{A}|p^{B}\rangle)^{2}\\
 & = & \sum_{p^{B}}\sum_{p^{A}}\sum_{I=R,L}\wp_{I}P_{I}(p^{A},p^{B})(p^{A}-\langle p^{A}|p^{B}\rangle)^{2}\\
 & \geq & \sum_{p^{B}}\sum_{p^{A}}\sum_{I=R,L}\wp_{I}P_{I}(p^{A},p^{B})(p^{A}-\langle p^{A}|p^{B}\rangle_{I})^{2}\end{eqnarray*}
The inequality follows because $\langle p^{A}|p^{B}\rangle$ is the
mean of $P(p^{A}|p^{B})$ for the $\rho$ of \eqref{eq:mixture},
and the choice $a=\sum_{p}P(p)p=\langle p\rangle$ will minimise $\sum_{p}P(p)(p-a)^{2}$.
From this the desired result follows.

\end{document}